\documentclass{article}

\makeatletter
\newcommand{\termlabel}[2]{%
  \begingroup
  \ifmeasuring@\else
    \phantomsection
    \def\@currentlabel{#1}%
    \ltx@label{#2}%
  \fi
  \textnormal{(#1)}%
  \endgroup
}
\makeatother

\usepackage{amsthm}
\usepackage{thmtools,thm-restate}
\usepackage[margin=1in]{geometry}

\usepackage{graphicx}
\usepackage{xcolor}
\usepackage[ruled,linesnumbered,noend]{algorithm2e}
\usepackage{eqparbox}
\usepackage{float}
\usepackage{mathtools}
\usepackage{enumitem}
\usepackage{booktabs}
\usepackage{tabularx}
\usepackage{subcaption}
\usepackage[colorlinks,linkcolor=magenta,citecolor=magenta,urlcolor=magenta]{hyperref}
\usepackage{natbib}
\usepackage{complexity}
\usepackage[T1]{fontenc}
\usepackage{newtxtext}
\usepackage{newtxmath}
\usepackage{bm}
\usepackage{bbm}
\usepackage{xspace}
\usepackage{complexity}
\usepackage{amsthm}

\usepackage{physics2}
\usephysicsmodule{qtext.legacy,ab}

\newcommand{\delimit}[3]{
    \NewDocumentCommand#1{sm}{
    \IfBooleanTF##1
    {\left#2 ##2 \right#3}
    {\mathopen{#2} ##2 \mathclose{#3}}%
    }
}

\delimit\ceil\lceil\rceil
\delimit\floor\lfloor\rfloor
\delimit\ip\langle\rangle
\delimit\norm\lVert\rVert
\delimit\abs\lvert\rvert

\def\ve{{\bm{e}}}

\def\vm{{\bm{m}}}

\def\vr{{\bm{r}}}

\def\vu{{\bm{u}}}
\def\vv{{\bm{v}}}

\def\vx{{\bm{x}}}
\def\vy{{\bm{y}}}
\def\vz{{\bm{z}}}

\renewcommand\vec\bm

\def\mA{{\mathbf{A}}}

\def\mI{{\mathbf{I}}}
\def\mJ{{\mathbf{J}}}
\def\mK{{\mathbf{K}}}

\def\cR{{\mathcal{R}}}

\def\cX{{\mathcal{X}}}
\def\cY{{\mathcal{Y}}}
\def\cZ{{\mathcal{Z}}}

\renewcommand{\R}{\mathbb R}

\newcommand{\defeq}{:=}

\newcommand{\RM}{\texttt{RM}}

\newcommand{\DRM}{\texttt{DRM}}
\newcommand{\PRMplus}{\texttt{PRM}^+}
\newcommand{\RMplus}{\texttt{RM}^+}

\newcommand{\sPRMplus}{\texttt{SPRM}^+}
\newcommand{\ExRMplus}{\texttt{ExRM}^+}
\newcommand{\APRMplus}{\texttt{Alt-PRM}^+}
\newcommand{\PCFRplus}{\texttt{PCFR}^+}
\newcommand{\IREGPRMplus}{\texttt{IREG-PRM}^+}

\newcommand{\IRPRMplus}{\texttt{IR-PRM}^+}

\newcommand{\AIRPRMplus}{\texttt{Alt-IR-PRM}^+}
\newcommand{\AHPRMplus}{\texttt{Alt-HPRM}^+}
\newcommand{\ARMplus}{\texttt{Alt-RM}^+}

\newcommand{\HPRMplus}{\texttt{HPRM}^+}

\newcommand{\SHPRMplus}{\texttt{Sim-HPRM}^+}
\newcommand{\SPRMplus}{\texttt{Sim-PRM}^+}

\newcommand{\AIRPCFRplus}{\texttt{Alt-IR-PCFR}^+}
\newcommand{\AHPCFRplus}{\texttt{Alt-HPCFR}^+}
\newcommand{\ASAPCFRplus}{\texttt{Alt-SAPCFR}^+}

\newcommand{\APDCFRplus}{\texttt{Alt-PDCFR}^+}

\newcommand{\APCFRplus}{\texttt{Alt-PCFR}^+}

\newcommand{\AdOGD}{\texttt{AdOGD}}
\newcommand{\CFR}{\texttt{CFR}}
\newcommand{\CFRplus}{\texttt{CFR}^+}
\newcommand{\ADCFRplus}{\texttt{Alt-DCFR}^+}

\newcommand{\SAPCFRplus}{\texttt{SAPCFR}^+}

\usepackage[capitalize,nameinlink]{cleveref}
\hypersetup{pageanchor=false}
\theoremstyle{plain}
\newtheorem{theorem}{Theorem}[section]

\theoremstyle{definition}
\newtheorem{definition}[theorem]{Definition}
\newtheorem{remark}[theorem]{Remark}

\DontPrintSemicolon
\SetKwProg{Fn}{function}
\SetAlFnt{\relax}
\SetAlCapFnt{\relax}
\SetAlCapNameFnt{\relax}
\SetAlCapHSkip{0pt}



\title{On the Convergence of Alternating Regret Matching Algorithms}

\usepackage{authblk}
\author[1]{Anh Duc Nguyen}
\author[1]{Ioannis Anagnostides}
\author[2]{Brian Hu Zhang}
\author[1,3]{Tuomas Sandholm}

\affil[1]{Carnegie Mellon University, Computer Science Department}
\affil[2]{Massachusetts Institute of Technology}
\affil[3]{Additional affiliations: Strategy Robot, Inc., Strategic Machine, Inc., Optimized Markets, Inc.}
\affil[ ]{\texttt{\{anhducn,ianagnos,sandholm\}}\texttt{@cs.cmu.edu}, \texttt{zhangbh}\texttt{@csail.mit.edu}}

\begin{document}
\maketitle

\begin{abstract}
\noindent \textit{Alternating updates}, that is, where each player is updated in turn, tend to typically perform better in regret matching algorithms than simultaneous updates. Alternation within the regret matching family has long been a central ingredient of practical solvers for zero-sum games. Yet our theoretical understanding is lagging behind. In this paper, we address this gap by developing a systematic way of analyzing such algorithms. Our framework centers on an alternating analog of the RVU property that we introduce, which we use to obtain the optimal $T^{-1}$ average-iterate convergence within this family. In particular, we establish this for two simple variants of predictive regret matching$^+$: i) \emph{increasing-regret} predictive regret matching$^+$ and ii) \emph{half-prediction} regret matching$^+$. Interestingly, we find that half prediction is the unique prediction multiplier that satisfies our alternating RVU property, giving it a strong theoretical justification.

Going beyond average-iterate convergence, we refine our framework to obtain $T^{-1/2}$ best-iterate convergence and further establish convergence of their strategy profiles to a single Nash equilibrium, without assuming equilibrium uniqueness. We go on to establish separations between alternating and simultaneous updates, including a game in which half-prediction regret matching$^+$ converges under alternating updates but fails to converge in terms of the last iterate under simultaneous updates. This provides theoretical support for the empirical observation that alternation can stabilize learning dynamics. Finally, experiments on 15 game benchmarks demonstrate that the strong theoretical properties of the half-prediction mechanism translate to strong practical performance, being on par with state-of-the-art solvers.
\end{abstract}

\clearpage
\tableofcontents

\clearpage
\pagenumbering{arabic}
\hypersetup{pageanchor=true}

\section{Introduction}

\emph{Regret matching ($\RM$)}---introduced by \citet{Hart00:Simple}---and its variants underlie many of the most effective methods for practical equilibrium computation through self-play. In this approach, each player repeatedly updates its strategy using a regret minimizer, with the other player's strategy determining the observed utility vector. Prominent variants in the $\RM$ family are \emph{regret matching$^+$ ($\RMplus$)}~\citep{Tammelin14:Solving}, \emph{discounted regret matching ($\DRM$)}~\citep{Brown19:Solving}, and \emph{predictive regret matching$^+$ ($\PRMplus$)}~\citep{Farina21:Faster}. A central practical advantage of $\RM$, $\RMplus$, and $\PRMplus$ is that they require no learning-rate tuning and are invariant to positive rescaling of utilities. On the other hand, standard optimistic mirror-descent methods, including optimistic multiplicative weights, instead depend on a suitable choice of learning rate, whose appropriate magnitude changes with the payoff scale~\citep{Rakhlin13:Online,Farina21:Faster,Zhang25:Scale}. This distinction becomes particularly relevant within \textit{counterfactual regret minimization ($\CFR$)}, which decomposes regret minimization in an extensive-form game into local regret minimizers at individual information sets~\citep{Zinkevich07:Regret}. $\RM$ variants allow these local learners to operate without calibrating learning rates across potentially many information sets with different utility scales~\citep{Chakrabarti24:Extensive}. Moreover, empirical comparisons show that $\RM$-based solvers can substantially outperform optimistic mirror-descent baselines on benchmark extensive-form games, despite the latter's stronger worst-case convergence guarantees~\citep{Farina19:Optimistic}. 

\emph{Average-iterate convergence} is central to equilibrium computation in two-player zero-sum games, as it provides a way to obtain an approximate Nash equilibrium even when the individual strategy profiles do not stabilize~\citep{Freund99:Adaptive,Zinkevich07:Regret,Syrgkanis15:Fast}. For uniform averages, the $O(T^{-1})$ rate is worst-case optimal in its dependence on the horizon $T$, even when the hidden constant may depend on the game~\citep{Daskalakis15:Near,Zhang25:Scale}. Nonuniform averaging schemes that emphasize later strategies reduce the influence of poor early iterates and can substantially accelerate empirical exploitability reduction in $\CFR$ solvers~\citep{Brown19:Solving,Zhang26:Faster}. In practice, we also want to understand the quality of the individual profiles generated during self-play, especially because the best and last iterate can converge substantially faster than the average on benchmark games~\citep{Cai25:Last,Zhang25:Scale,Zhang26:General}. This motivates studying \emph{best-iterate convergence}, which quantifies how quickly self-play reaches a profile with small exploitability, together with \emph{last-iterate convergence}, which ensures that all sufficiently late profiles have small exploitability~\citep{Anagnostides22:On,Cai25:Last}. 

\emph{Alternating updates}, where players update their strategies one after the other, are widely used in practical equilibrium computation through self-play with regret matching methods. They were part of counterfactual regret minimization$^+$ ($\CFRplus$) from its introduction~\citep{Tammelin14:Solving} and were subsequently used in two landmark achievements in poker AI: \emph{Cepheus} near-optimally solved heads-up limit Texas hold'em~\citep{Bowling15:Heads}, while \emph{Libratus} achieved superhuman performance in heads-up no-limit Texas hold'em~\citep{Brown18:Superhuman}. Subsequent methods, including discounted and predictive $\CFR$ variants, also use alternating updates to obtain strong empirical improvements~\citep{Brown19:Solving,Farina21:Faster}. Despite this widespread use, substantial gaps remain in our understanding of their convergence. Guarantees for individual regret minimizers or simultaneous self-play do not automatically establish convergence under alternating updates~\citep{Farina19:Online,Burch19:Revisiting}. In fact, a recurring pattern in the $\RM$ and $\CFR$ literature is to evaluate alternating implementations for their strong empirical performance, while the accompanying theory establishes individual regret learner bounds or analyzes equilibrium convergence under simultaneous updates~\citep{Brown19:Solving,Farina19:Stable,Farina21:Faster,Farina23:Regret,Chakrabarti24:Extensive,Xu24:Minimizing,Zhang25:Scale,Meng26:Faster}. These analyses do not, by themselves, establish convergence of the alternating dynamics evaluated in practice. Despite guarantees for some alternating variants~\citep{Burch19:Revisiting, Sullins26:IR}, a systematic understanding of their worst-case convergence remains incomplete. This gap is particularly striking for alternating $\PRMplus$ ($\APRMplus$), whose CFR instantiation $\APCFRplus$ has achieved state-of-the-art empirical performance across the majority of benchmark games~\citep{Farina21:Faster,Chakrabarti24:Extensive,Zhang25:Scale,Zhang26:Faster} and is the equilibrium solver within \emph{Obscuro}, the first superhuman AI for Fog of War chess~\citep{Zhang26:General}. Yet no general guarantee is known for its average-, best-, or last-iterate convergence, even in finite two-player zero-sum normal-form games~\citep{Cai25:Last}.

Our goals are to address this gap and develop a systematic analysis of alternating updates for the $\RM$ family that distinguishes limitations of existing proof techniques from essential requirements for convergence, and guides the design of methods that retain the family's practical strengths while admitting strong theoretical guarantees. We revisit the challenges involved in establishing convergence guarantees of $\PRMplus$: i) decreases in regret vector norms (which runs counter to the recent idea of achieving convergence in certain simultaneous-update algorithms using nondecreasing regret vector norms~\citep{Zhang25:Scale}) and ii) strategy instability~\citep{Farina23:Regret} (simultaneous updates can oscillate back and forth). We focus on \textit{increasing-regret predictive regret matching$^+$ ($\IRPRMplus$)} of \citet{Zhang25:Scale}, which enforces nondecreasing regret vector norms, and an algorithm we introduce, \textit{half-prediction regret matching$^+$ ($\HPRMplus$)}, which simply halves the prediction term of $\PRMplus$. The latter algorithm belongs to the family proposed by \citet{Meng26:Faster}. We also investigate whether alternating updates merely accelerate convergence or can determine whether the strategy profile approaches equilibrium at all (the latter is suggested by the experiments of simultaneous and alternating $\PRMplus$ of \citet{Cai25:Last}).

\subsection{Our Contributions}

We develop a framework for analyzing the average-, best- and last-iterate convergence of alternating regret matching algorithms, establishing the strong convergence guarantees together with strong empirical performance on benchmarks. In more detail, we summarize our main contributions below.

\begin{itemize}
    \item \textbf{A systematic analysis of alternating self-play.}
    We introduce the \textit{alternating regret bounded by variation in utilities bound (Alt-RVU)} and its refinement, \textit{refined Alt-RVU}, together with a common analysis template for establishing these bounds. Alt-RVU yields $O(T^{-1})$ average-iterate convergence, while the refinement yields an $O(T^{-1/2})$ best-iterate rate and a last-iterate convergence to the Nash set (Subsection~\ref{sec:alt_rvu} and Section~\ref{sec:best_last_nfg}). 
    We also show that certain regret minimizers can fail the standard (simultaneous) RVU of \citep{Syrgkanis15:Fast} while having the optimal $O(T^{-1})$ average-iterate convergence. Hence, standard RVU cannot replace Alt-RVU in our analysis (Subsection \ref{sec:necessity_alt_rvu}). 
    As a first application, we establish both new Alt-RVU bounds for the increasing regret predictive regret matching algorithm $\IRPRMplus$ of \citet{Zhang25:Scale}, showing how norm monotonicity and norm-preserving prediction supply the required control (Subsection~\ref{sec:ir_alt_rvu} and Section~\ref{sec:best_last_nfg}).

    \item \textbf{Fast convergence without norm monotonicity.}
    We show that the regret-vector-norm monotonicity used to establish these bounds for $\IRPRMplus$ is, in fact, not necessary. Under alternating self-play, we prove with our Alt-RVU framework that half-prediction regret matching $\HPRMplus$ achieves its $O(T^{-1})$ average-iterate convergence, an $O(T^{-1/2})$ best-iterate rate, and last-iterate convergence to a single Nash equilibrium despite allowing its regret vector norm to decrease, while existing analysis for this family~\citep{Meng26:Faster} does not establish any such convergence guarantees (Subsection~\ref{sec:alt_rvu_hprm} and Section~\ref{sec:best_last_nfg}). Our key insight is to introduce and analyze a nondecreasing \emph{corrected squared norm} that controls the cumulative effect of norm decreases without modifying the learner's updates. We thereby retain the strong guarantees while avoiding the norm-preserving scalar computation at every iteration required by the IR mechanism in  $\IRPRMplus$ (Subsection~\ref{sec:alt_rvu_hprm}). We further show that, among variants of $\PRMplus$ that scale the prediction term by a fixed nonnegative multiplier, only half prediction satisfies Alt-RVU on every bounded utility sequence in every finite dimension (Section~\ref{sec:no_lambda_except_half}).

    \item \textbf{An analytical separation between alternating and simultaneous updates.} While \citet{Cai25:Last} provide numerical evidence that alternating and simultaneous $\PRMplus$ differ in last-iterate convergence, we establish such a distinction analytically for $\HPRMplus$. On an explicit matrix game, under the standard initialization, $\AHPRMplus$ converges to a single Nash equilibrium, whereas $\SHPRMplus$ has a duality gap asymptotically bounded away from zero. By holding the game, learner, and initialization fixed, this separation shows that update order alone can determine whether the strategy profiles approach equilibrium, beyond improving the convergence rate of their averages (Section~\ref{sec:separation_alt_sim}).

    \item \textbf{Practical gains from half prediction.}
    We evaluate the $\CFR$ instantiation of $\AHPRMplus$ on 15 extensive-form game benchmarks, where the resulting $\AHPCFRplus$ algorithm remains competitive with leading solvers across diverse games. In particular, it improves over $\APCFRplus$ in most comparisons of the iterations and learner wall-clock time needed to reach target exploitability. It also outperforms $\AIRPCFRplus$ on most benchmarks in both measures, showing that its benefits extend beyond avoiding the IR scalar computation to improving the progress made per iteration (Section~\ref{sec:exp}). 
\end{itemize}

\subsection{Other Related Research}

In this section we discuss additional related research.

\paragraph{Concurrent work.}
\citet{Sullins26:IR} independently establish $O(T^{-1})$ average-iterate, $O(T^{-1/2})$ best-iterate rates, and convergence to the Nash set for $\IRPRMplus$ under both simultaneous and alternating updates. Our $\IRPRMplus$ rates under alternating self-play overlap with theirs, but we additionally establish convergence of $\AIRPRMplus$ to a single Nash equilibrium without assuming uniqueness (Theorem~\ref{thm:nfg_best_last_convergence}). Our analytical approach differs substantially from theirs, which relies on controlling the coupled growth of the players' regret vector norms. Specifically, our Alt-RVU framework and its refinement instead formulate common conditions on individual regret learners, separating the convergence argument for alternating self-play from the algorithm-specific arguments establishing these conditions. These conditions also hold for $\HPRMplus$ despite decreases in the regret vector norm (\textit{i.e.}, the lack of the increasing-regret mechanism). Thus, our analysis extends beyond increasing-regret methods and shows that norm monotonicity is sufficient but not necessary for these convergence guarantees.

\paragraph{Regret matching family.}
Connections to online optimization have provided a basis for developing new regret-matching methods. Within Blackwell approachability, \citet{Farina21:Faster} recover $\RM$ and $\RMplus$ from follow-the-regularized-leader and online mirror descent, respectively, and derive predictive variants through their optimistic counterparts. Other work strengthens convergence guarantees through stabilization and extragradient updates~\citep{Farina23:Regret,Cai25:Last}, or through the increasing-regret mechanism~\citep{Zhang25:Scale}. Within $\CFR$, complementary developments include discounting past regrets~\citep{Brown19:Solving}, combining prediction with weighted regret accumulation~\citep{Xu24:Minimizing}, using asymmetric step sizes for regret accumulation and prediction~\citep{Meng26:Faster}, and introducing time-varying schedules for discounting and strategy averaging~\citep{Zhang26:Faster}.

\paragraph{Alternating updates.}
Alternating update order has also been studied extensively for gradient methods in unconstrained bilinear games, revealing differences in stability and convergence between simultaneous and alternating dynamics~\citep{Gidel19:Negative,Bailey20:Finite,Zhang20:Convergence}. On constrained domains, analyses of alternating mirror descent and projected gradient descent--ascent establish improved average-iterate convergence guarantees under their respective assumptions~\citep{Wibisono22:Alternating,Katona26:Symplectic,Nan26:AltGDA}. From an online-learning perspective, alternating regret provides a way to account for the update schedule and obtain stronger adversarial guarantees in online linear and convex optimization~\citep{Cevher23:Alternation,Hait25:Alternating, Tao26:Optimal}. These improvements are algorithm-dependent: \citet{Hait25:Alternating} also prove $\Omega(\sqrt T)$ alternating-regret lower bounds for $\PRMplus$ and optimistic gradient methods.

\section{Preliminaries on Zero-Sum Games}

For an integer $d\ge1$, let $[d] \defeq \{1,\ldots,d\}$ and let $\Delta(d)\defeq \{\vx \in \R^d_{\ge0}: \ip{\vec1,\vx} = 1\}$ denote the probability simplex. We consider a two-player zero-sum game with payoff matrix $\mA\in\R^{m\times n}$ and strategy spaces $\cX=\Delta(m)$ and $\cY=\Delta(n)$. The row player chooses $\vx\in\cX$ to maximize $\ip{\vx,\mA\vy}$, while the column player chooses $\vy\in\cY$ to minimize it. At a strategy profile $(\vx,\vy)\in\cX\times\cY$, their utility vectors are $\mA\vy$ and $-\mA^\top\vx$, respectively.

We measure suboptimality in the standard way as the duality gap: 
\begin{align}
\operatorname{Gap}(\vx,\vy) \defeq \max_{\vx'\in\cX}\ip{\vx',\mA\vy} -\min_{\vy'\in\cY}\ip{\vy',\mA^\top\vx} =\max_{i\in[m]}(\mA\vy)_i -\min_{j\in[n]}(\mA^\top\vx)_j.
\label{eq:gap}
\end{align}
The gap is always nonnegative and vanishes exactly at Nash equilibria. We denote the Nash equilibrium set by $\cZ^*$. For $T\ge1$, we define the uniformly averaged strategies
\begin{align*}
\bar{\vx}^{(T)}\defeq\frac{1}{T}\sum_{t=1}^T\vx^{(t)},\; \bar{\vy}^{(T)}\defeq\frac{1}{T}\sum_{t=1}^T\vy^{(t)}.
\end{align*}
We measure average-, best-, and last-iterate convergence by
\begin{align*}
\operatorname{Gap}(\bar{\vx}^{(T)},\bar{\vy}^{(T)}), \; \ \min_{1\le t\le T}\operatorname{Gap}(\vx^{(t)},\vy^{(t)}), \; \ \operatorname{Gap}(\vx^{(T)},\vy^{(T)}),
\end{align*}
respectively. Each notion requires the corresponding quantity to vanish as $T\to\infty$. Since $\cX\times\cY$ is compact and the duality gap is continuous, vanishing last-iterate gap is equivalent to the distance of $(\vx^{(T)},\vy^{(T)})$ to $\cZ^*$ tending to zero. This does not, by itself, imply the stronger property that the iterates converge to a single Nash equilibrium, namely that there exists $(\vx^*,\vy^*)\in\cZ^*$ such that $(\vx^{(T)},\vy^{(T)})\to(\vx^*,\vy^*)$.

\section{Regret Matching: Old and New}

In this work, we study optimistic regret minimizers on the simplex $\Delta(d)$, where $d$ is the number of actions. At time $t$, a learner plays a strategy $\vx^{(t)} \in \Delta(d)$ and then observes a utility vector $\vu^{(t)} \in \R^d$. The regret after $T$ observations is
\begin{align*}
    \operatorname{Reg}(T) \defeq \max_{\vx\in\Delta(d)}\sum_{t=1}^T\ip{\vu^{(t)},\vx-\vx^{(t)}}.
\end{align*}
We let $\tilde\vr^{(t)}\in\R^d_{\ge0}$ be the {\em regret vector} after the first $t-1$ observations. The prediction step forms the {\em prediction regret vector} $\vr^{(t)}$, which may have negative coordinates. The corresponding strategy is
\begin{align}
    \vx^{(t)} = \frac{[\vr^{(t)}]_+}{\|[\vr^{(t)}]_+\|_1},
    \label{eq:setup_normalization}
\end{align}
where $[\cdot]_+$ denotes the coordinatewise positive part. If the denominator is zero, we set $\vx^{(t)}=\vec1/d$.

We denote the prediction and observation routines by \textsc{NextStrategy} and \textsc{ObserveUtility}, respectively. The former takes a utility prediction $\vm^{(t)}\in\R^d$ and returns the strategy $\vx^{(t)}$; the latter receives the observed utility vector $\vu^{(t)}$ and updates the maintained regret vector to $\tilde\vr^{(t+1)}$. Unless stated otherwise, we use the standard initialization
\begin{align*}
\tilde\vr^{(1)}=\vr^{(1)}=\vm^{(1)}=\vec0, \; \vx^{(1)}=\vec1/d,
\end{align*}
and use the most recently observed utility as the next prediction, that is, $\vm^{(t+1)}=\vu^{(t)}$ for $t\ge1$.

\paragraph{Prediction.}
Predictive regret matching$^+$ ($\PRMplus$)~\citep{Farina21:Faster} updates its regret vector according to
\begin{align}
    \tilde\vr^{(t+1)} = [\tilde\vr^{(t)}+\vu^{(t)}-\ip{\vu^{(t)},\vx^{(t)}}\vec1]_+.
    \label{eq:reg_vec_update_prm}
\end{align}
The vector $\vu^{(t)}-\ip{\vu^{(t)},\vx^{(t)}}\vec1$ is the {\em instantaneous regret vector}. With the last observed utility as the prediction, the prediction regret vector update is
\begin{align}
    \vr^{(t+1)} = \tilde\vr^{(t+1)}+\vu^{(t)}-\ip{\vu^{(t)},\vx^{(t)}} \vec1.
    \label{eq:prediction_update_prm}
\end{align}
The next strategy $\vx^{(t+1)}$ is obtained by applying the normalization rule~\eqref{eq:setup_normalization} to $\vr^{(t+1)}$.

\paragraph{Scaled-prediction.}
We consider a general family of regret minimizers which we call \emph{scaled-prediction regret matching$^+$}. The family is  parameterized by a fixed scalar $\lambda \in \R_{\ge 0}$. Each algorithm in this family uses the regret vector update~\eqref{eq:reg_vec_update_prm} and forms the scaled-prediction regret vector
\begin{align*}
    \vr^{(t+1)} = \tilde\vr^{(t+1)} + \lambda (\vu^{(t)} - \ip{\vu^{(t)}, \vx^{(t)}} \vec1).
\end{align*}
The canonical choices $\lambda=0$ and $\lambda=1$ recover $\RMplus$~\citep{Tammelin14:Solving} and $\PRMplus$, respectively.

For $\lambda\in(0,1]$, these updates coincide with the local asymmetric updates of \citet{Meng26:Faster} when their parameter is fixed at $\alpha=\lambda^{-1}-1$. In particular, $\lambda=1/3$ recovers the local regret minimizer used by their simple asymmetric $\PCFRplus$ ($\SAPCFRplus$), which fixes $\alpha=2$. For every fixed $\lambda\in(0,1]$, \citet[Theorem 4.1]{Meng26:Faster}
shows that $\operatorname{Reg}(T)=O(\sqrt T)$ on bounded utility sequences. This individual regret learner guarantee also applies under alternating updates, but do not by themselves establish average-iterate, best-iterate, or last-iterate convergence to Nash equilibrium under alternating self-play.

\paragraph{Half-prediction.}
In this paper, we mainly focus on the choice $\lambda=1/2$, which gives
\begin{align}
    \vr^{(t+1)} = \tilde\vr^{(t+1)} + \frac12 (\vu^{(t)} - \ip{\vu^{(t)},\vx^{(t)}} \vec1 ).
    \label{eq:setup_hprm}
\end{align}
We call this specialization {\em half-prediction regret matching$^+$}, abbreviated as $\HPRMplus$, as it halves the prediction while keeping the regret vector update \eqref{eq:reg_vec_update_prm} identical to $\PRMplus$. Subsection~\ref{sec:alt_rvu_hprm} and Section~\ref{sec:no_lambda_except_half} will establish its theoretical guarantees and examine the role of the choice $\lambda=1/2$ within the scaled-prediction family.

\paragraph{Increasing regret.}
Increasing-regret predictive regret matching$^+$ ($\IRPRMplus$)~\citep{Zhang25:Scale} forms its prediction regret vector using a scalar shift that preserves the Euclidean norm after taking the positive part. When $\tilde\vr^{(t)}\ne\vec0$, we let $\gamma^{(t)}\in\R$ be the unique scalar satisfying
\begin{align}
    \norm{[\tilde\vr^{(t)}+\vm^{(t)}-\gamma^{(t)}\vec1]_+}_2
    = \norm{\tilde\vr^{(t)}}_2,
    \label{eq:prediction_scalar}
\end{align}
and form the prediction regret vector as
\begin{align*}
    \vr^{(t)}=\tilde\vr^{(t)}+\vm^{(t)}-\gamma^{(t)}\vec1.
\end{align*}
The strategy $\vx^{(t)}$ is then obtained by applying the normalization rule~\eqref{eq:setup_normalization}. After observing $\vu^{(t)}$, the regret vector is updated according to
\begin{align}
    \tilde\vr^{(t+1)}
    = [\tilde\vr^{(t)}+\vu^{(t)}
    -(\gamma^{(t)}+\ip{\vu^{(t)}-\vm^{(t)},\vx^{(t)}})\vec1]_+.
    \label{eq:correction}
\end{align}
When $\tilde\vr^{(t)}=\vec0$, we override the utility prediction by setting $\vm^{(t)}=\vec0$, set $\vr^{(t)}=\vec0$ and $\gamma^{(t)}=0$, and use the uniform strategy $\vx^{(t)}=\vec1/d$. After observing $\vu^{(t)}$, we apply~\eqref{eq:correction} with these values. These updates ensure that the Euclidean norm of the regret vector is nondecreasing:
\begin{align*}
\norm{\tilde\vr^{(t+1)}}_2 \ge \norm{\tilde\vr^{(t)}}_2.
\end{align*}
The extragradient specialization $\IREGPRMplus$ of \citet{Zhang25:Scale} instead uses predictions obtained from an additional utility evaluation. We do not study this specialization here, as we focus on alternating updates with last-utility predictions.

\section{Alternating vs Simultaneous Updates}
\label{sec:simultaneous-alternating}

In zero-sum games, self-play updates can be conducted simultaneously or by alternating between the agents. Let $\cR_{\cX}$ and $\cR_{\cY}$ be the learners of the row player and column player, respectively, both using the same regret minimizer defined above. Algorithms~\ref{alg:sim} and~\ref{alg:alt} specify the simultaneous and row-first alternating updates, denoted by the prefixes \texttt{Sim-} and \texttt{Alt-}, respectively. Simultaneous updates compute $\vx^{(t)}$ and $\vy^{(t)}$ using the respective utility predictions $\mA\vy^{(t-1)}$ and $-\mA^\top\vx^{(t-1)}$, before observing $\mA\vy^{(t)}$ and $-\mA^\top\vx^{(t)}$. Under row-first alternating updates, the row player likewise computes $\vx^{(t)}$ using $\mA\vy^{(t-1)}$, but the column player first observes $-\mA^\top\vx^{(t)}$ while playing $\vy^{(t-1)}$, updates its regret vector, and uses this same utility as the prediction for computing $\vy^{(t)}$. The row player then observes $\mA\vy^{(t)}$ while playing $\vx^{(t)}$ and updates its regret vector. We focus on row-first alternating updates, hereafter referred to as alternating updates, since the column-first case follows readily by exchanging the players and replacing $\mA$ with $-\mA^\top$.

Learner indices count local rounds: the normalization rule~\eqref{eq:setup_normalization} maps $\vr_{\cX}^{(t)}$ to $\vx^{(t)}$ under both schedules, and $\vr_{\cY}^{(t)}$ to $\vy^{(t)}$ under simultaneous updates or $\vy^{(t-1)}$ under alternating updates. The final line of Algorithm~\ref{alg:alt} supplies $\vx^{(T+1)}$ for alternating regret (Subsection~\ref{sec:alt_rvu}) without additional feedback. Under both schedules, the averages use only the output profiles $(\vx^{(t)},\vy^{(t)})$ for $1\le t\le T$.

\begingroup
\SetKwInput{Input}{Input}
\SetNlSty{textnormal}{}{}
\SetAlgoNlRelativeSize{-1}

\noindent
\begin{minipage}[t]{0.49\linewidth}
\vspace{0pt}
\begin{algorithm}[H]
\caption{Simultaneous Updates}
\label{alg:sim}
\Input{Matrix $\mA$; horizon $T$; learners $\cR_{\cX}, \cR_{\cY}$}

$\vu_{\cX}^{(0)}\gets\vec0$\;
$\vu_{\cY}^{(0)}\gets\vec0$\;

\For{$t=1,2,\ldots,T$}{
    $\vx^{(t)}
    \gets\cR_{\cX}.\text{\textsc{NextStrategy}}
    (\vu_{\cX}^{(t-1)})$\;

    $\vy^{(t)}
    \gets\cR_{\cY}.\text{\textsc{NextStrategy}}
    (\vu_{\cY}^{(t-1)})$\;

    $\vu_{\cX}^{(t)}
    \gets\mA\vy^{(t)}$\;

    $\vu_{\cY}^{(t)}
    \gets-\mA^\top\vx^{(t)}$\;

    $\cR_{\cX}.\text{\textsc{ObserveUtility}}
    (\vu_{\cX}^{(t)})$\;

    $\cR_{\cY}.\text{\textsc{ObserveUtility}}
    (\vu_{\cY}^{(t)})$\;
}
\end{algorithm}
\end{minipage}\hfill%
\begin{minipage}[t]{0.49\linewidth}
\vspace{0pt}
\begin{algorithm}[H]
\caption{Row-first Alternating Updates}
\label{alg:alt}
\Input{Matrix $\mA$; horizon $T$; learners $\cR_{\cX}, \cR_{\cY}$}

$\vu_{\cX}^{(0)}\gets\vec0$\;
$\vu_{\cY}^{(0)}\gets\vec0$\;

$\vy^{(0)}
\gets\cR_{\cY}.\text{\textsc{NextStrategy}}
(\vu_{\cY}^{(0)})$\;

\For{$t=1,2,\ldots,T$}{
    $\vx^{(t)}
    \gets\cR_{\cX}.\text{\textsc{NextStrategy}}
    (\vu_{\cX}^{(t-1)})$\;

    $\vu_{\cY}^{(t)}
    \gets-\mA^\top\vx^{(t)}$\;

    $\cR_{\cY}.\text{\textsc{ObserveUtility}}
    (\vu_{\cY}^{(t)})$\;

    $\vy^{(t)}
    \gets\cR_{\cY}.\text{\textsc{NextStrategy}}
    (\vu_{\cY}^{(t)})$\;

    $\vu_{\cX}^{(t)}
    \gets\mA\vy^{(t)}$\;

    $\cR_{\cX}.\text{\textsc{ObserveUtility}}
    (\vu_{\cX}^{(t)})$\;
}
$\vx^{(T+1)}
\gets\cR_{\cX}.\text{\textsc{NextStrategy}}
(\vu_{\cX}^{(T)})$\;
\end{algorithm}
\end{minipage}
\par

\endgroup

\section{Average-Iterate Convergence}

Average-iterate convergence quantifies how quickly self-play produces an approximate Nash equilibrium, even when the individual strategy profiles have not stabilized. Alternating methods in the regret matching family often outperform their simultaneous counterparts for practical self-play~\citep{Brown19:Solving, Farina21:Faster}, yet a systematic understanding of their convergence remains largely incomplete. For $\ARMplus$, a suboptimal $O(T^{-1/2})$ average-iterate guarantee is known~\citep{Burch19:Revisiting}, while for $\APRMplus$, an average-iterate convergence remains unresolved~\citep{Cai25:Last}. By comparison, simultaneous optimistic and extragradient variants in this family admit stronger guarantees, including the optimal $O(T^{-1})$ average-iterate rates~\citep{Farina23:Regret,Zhang25:Scale}. These gaps between theory and practice motivate us to develop a systematic analysis of fast average-iterate convergence under alternating updates.

We first adapt standard analytical tools for simultaneous updates to the alternating setting, providing a clear route to $O(T^{-1})$ average-iterate convergence. While some modifications may appear to be mere bookkeeping, our later separation results show why they are needed: the existing (simultaneous) bounds can fail for learners whose alternating self-play nevertheless converges (Subsection~\ref{sec:necessity_alt_rvu}). The estimates we develop for average-iterate convergence in this section also provide the starting point for the refined analysis of best- and last-iterate convergence in Section~\ref{sec:best_last_nfg}.

We begin by recalling the {\em Regret bounded by Variation in Utilities} ({\em RVU}) bound~\citep{Syrgkanis15:Fast}, which underlies the standard analysis of simultaneous optimistic methods. For the following definitions, we fix a bounded utility domain in $\R^d$ and a pair of primal-dual norms $(| \cdot |,| \cdot |_\ast)$.

\begin{definition}[RVU]
\label{def:rvu}
A learner satisfies the RVU bound if there exist constants $\alpha,\beta,\gamma>0$ such that, for every utility sequence $(\vu^{(t)})_{t\ge1}$ and every $T\ge1$,
\begin{align}
    \operatorname{Reg}(T) \le \alpha + \beta \sum_{t=2}^T \|\vu^{(t)} - \vu^{(t-1)}\|_\ast^2 - \gamma \sum_{t=2}^T \|\vx^{(t)} - \vx^{(t-1)}\|^2.
    \label{eq:rvu}
\end{align}
\end{definition}
Under simultaneous updates, when the constants allow the negative strategy movement terms to absorb the utility variation induced by the opponents' movements, summing the two players' RVU bounds yields $O(T^{-1})$ average-iterate convergence: the players' regrets sum to $T\operatorname{Gap}(\bar{\vx}^{(T)},\bar{\vy}^{(T)})\ge0$~\citep{Syrgkanis15:Fast}.

Under alternating updates, however,
\begin{align*}
    \operatorname{Reg}_{\cal X} (T) + \operatorname{Reg}_{\cal Y} (T)  = T \operatorname{Gap}(\bar{\vx}^{(T)}, \bar{\vy}^{(T)}) - \sum_{t=1}^T (\vx^{(t)})^\top \mA (\vy^{(t)} -  \vy^{(t-1)}), 
\end{align*}
where the last term sum has no fixed sign and does not generally telescope. This extra term arises because the row player observes $\mA\vy^{(t)}$ at $\vx^{(t)}$, whereas the column player observes $-\mA^\top\vx^{(t)}$ at $\vy^{(t-1)}$. An individual RVU bound can still hold under alternating feedback, but summing the two players' bounds no longer gives the usual cancellation. Thus, in addition to establishing individual regret bounds, a key challenge in analyzing alternating updates is to control the \textit{interaction} between the players’ updates.

\subsection{Alternating RVU and Analysis Template}
\label{sec:alt_rvu}

Our goal is to adapt the analysis to this alternating update order. Our first step is to consider the recent idea of alternating regret \citep{Cevher23:Alternation, Hait25:Alternating}, which is more natural to the structure of alternating updates:
\begin{align*}
    \operatorname{Alt-Reg}_{\cal X} (T) = \max_{\vz \in \cal X} \sum_{t=1}^T \ip{\mA \vy^{(t)}, 2 \vz - \vx^{(t)} - \vx^{(t+1)}}, \; \operatorname{Alt-Reg}_{\cal Y} (T) = \max_{\vz \in \cal Y} \sum_{t=1}^T \ip{- \mA^T \vx^{(t)}, 2 \vz - \vy^{(t-1)} - \vy^{(t)}}.
\end{align*}
Intuitively, in the alternating update scheme, both $\vx^{(t)}$ and $\vx^{(t+1)}$ are played against $\vy^{(t)}$. Alternating regret for the $\vx$ player takes into account both of these strategies, while standard (simultaneous) regret only considers $\vx^{(t)}$ against $\vy^{(t)}$. The same goes for the $\vy$ player, as both $\vy^{(t-1)}$ and $\vy^{(t)}$ are played against $\vx^{(t)}$. Moreover, the sum of two players' alternating regrets restores the desired cancellation
\begin{align}
    \operatorname{Alt-Reg}_{\cal X} (T) + \operatorname{Alt-Reg}_{\cal Y} (T) = 2T \operatorname{Gap}(\bar{\vx}^{(T)}, \bar{\vy}^{(T)}) + (\vx^{(1)})^\top \mA \vy^{(0)} - (\vx^{(T+1)})^\top \mA \vy^{(T)},
    \label{eq:alt_payoff_telescope}
\end{align}
since the endpoint difference is at most $\operatorname{range}(\mA)$.

Using alternating regret, we introduce the following Alt-RVU bound. Unlike the uniform requirement in the RVU definition~(\ref{def:rvu}), its constants may depend on the utility sequence, but must remain valid for every horizon $T$. This dependence also arises in the zero-initialized convergence analysis of~\citet[Section~5.1]{Zhang25:Scale}: in both their and our later analyses, the positive constant of the negative strategy movement term is obtained from the first nonzero regret vector.
\begin{definition}[Alt-RVU]
\label{def:alt_rvu}
A learner satisfies the Alt-RVU bound if, for every utility sequence $(\vu^{(t)})_{t\ge1}$, there exist constants $\alpha,\beta,\gamma>0$ such that, for every $T\ge1$,
\begin{align}
    \operatorname{Alt-Reg}(T) \le \alpha + \beta \sum_{t=2}^T \|\vu^{(t)} - \vu^{(t-1)}\|_\ast^2 - \gamma \sum_{t=1}^T \|\vx^{(t+1)} - \vx^{(t)}\|^2.
    \label{eq:alt_rvu}
\end{align}
\end{definition}
Compared with standard RVU (Definition~\ref{def:rvu}), Alt-RVU not only bounds alternating regret but also pairs utility variation with the subsequent strategy movement, rather than with movement over the same pair of consecutive iterations. This shift in indexing precisely reflects the alternating update order. Using Alt-RVU, we obtain the following guarantees for alternating updates through an argument analogous to the standard RVU analysis for simultaneous updates.
\begin{restatable}{proposition}{altRvuGap}
     Suppose that in a normal-form game with alternating update scheme, for a pair of primal-dual norms $(\| \cdot \|, \| \cdot \|_{\ast})$, both players satisfy the above Alt-RVU bound with $(\alpha_{\cal X}, \beta_{\cal X}, \gamma_{\cal X}), (\alpha_{\cal Y}, \beta_{\cal Y}, \gamma_{\cal Y}) \in \R^3_{>0}$ such that $\|\mA\|^2 \beta_{\cal X} < \gamma_{\cal Y}$ and $\|\mA\|^2 \beta_{\cal Y} < \gamma_{\cal X}$, where $\|\mA\| = \sup_{\|\vx\| = \|\vy\| = 1} |\vx^\top \mA \vy |$ is the induced bilinear norm. Then, both players' squared strategy movement sum and both players' utility variation sum are finite, and we obtain the average-iterate convergence
    \begin{align*}
        \operatorname{Gap} (\bar{\vx}^{(T)}, \bar{\vy}^{(T)}) \le \frac{\alpha_{\cal X} + \alpha_{\cal Y} + \operatorname{range}(\mA)}{2 T}.
    \end{align*}
\end{restatable}
\paragraph{Analysis Template.} We now develop a common approach to establishing Alt-RVU for the regret-matching learners considered in this work. Throughout the following analyses, we consider a learner on $\Delta(d)$, where $d$ is the number of actions (\textit{i.e.}, $d=m$ for the row player and $d=n$ for the column player), with the standard initialization (first strategies are uniform and regret vectors are zero) and last-utility predictions (\textit{i.e.}, the prediction is the previous utility). 
Lastly, we only consider the pair of Euclidean norms for all the Alt-RVU bounds in this work. 

For the individual-learner estimates below, we assume
$\|\vu^{(t)}\|_2\le\|\mA\|_2$ for every $t\ge1$.
Apart from this bound, these estimates impose no restriction on how the utility sequence is generated. The same proofs apply to any bounded utility sequence by replacing $\|\mA\|_2$ in the individual-learner estimates with $\sup_{t\ge1}\|\vu^{(t)}\|_2$. In self-play, the displayed bound holds for both players because $\vu_{\cX}^{(t)}=\mA\vy^{(t)}$ and $\vu_{\cY}^{(t)}=-\mA^\top\vx^{(t)}$, with $\|\vx^{(t)}\|_2,\|\vy^{(t)}\|_2\le1$. For notational convenience, we also set $\tilde\vr^{(0)}=\vec0$.

To start with, we have the following general bound on the alternating regret. 
\begin{restatable}{lemma}{altRegBoundSimple}
\label{lem:alt_reg_bound_simple}
Consider a learner on $\Delta(d)$ with $\tilde\vr^{(1)}=\vec0$ whose regret vectors satisfy
\begin{align}
    \tilde\vr^{(t+1)} \ge \tilde\vr^{(t)}+\vu^{(t)}-\ip{\vu^{(t)},\vx^{(t)}}\vec1
    \label{eq:ineq_tilde_r}
\end{align}
coordinate-wise. Then, for every $T\ge1$, this learner satisfies
\begin{align}
    \operatorname{Alt-Reg}(T) \le 2 \|\tilde \vr^{T+1}\|_2 - \sum_{t=1}^T \ip{\vu^{(t)}, \vx^{(t+1)}-\vx^{(t)}}.
\label{eq:alt_reg_bound_simple}
\end{align}
\end{restatable}
For a specific no-regret learner with alternating updates, this lemma identifies the two key components of our later analyses: an upper bound on the {\em terminal regret vector norm} $\|\tilde\vr^{(T+1)}\|_2$ and a lower bound on
the {\em cumulative utility improvement} $\sum_{t=1}^T \ip{\vu^{(t)}, \vx^{(t+1)}-\vx^{(t)}}$. Specifically, these upper and lower bounds must depend on the utility variation and strategy movement terms in Alt-RVU \eqref{eq:alt_rvu}.
 
We also define the first nonzero regret vector index
\begin{align}
    t_0 := \min\{t \ge 1 : \tilde \vr^{(t)} \neq 0\},
    \label{eq:t_0}
\end{align}
with $t_0=\infty$ if this set is empty. If $t_0=\infty$, the strategy remains uniform and every observed utility vector has equal coordinates. Thus, Alt-RVU holds trivially. We therefore assume $t_0<\infty$ in the subsequent analysis. 

At $t=t_0-1$, the strategy is still uniform, and all three learners we consider in this paper have the same regret-vector update:
\begin{align*}
    \tilde \vr^{(t_0)} = \left[\vu^{(t_0-1)} - \frac{\ip{\vu^{(t_0-1)}, \vec1}}{d} \vec1\right]_+.
\end{align*}
Consequently, the first nonzero regret vector satisfies
\begin{align}
    0 < \|\tilde \vr^{(t_0)}\|_2^2 \le \left\|\vu^{(t_0-1)} - \frac{\ip{\vu^{(t_0-1)}, \vec1}}{d} \vec1\right\|_2^2 = \|\vu^{(t_0-1)}\|_2^2 - \frac{\ip{\vu^{(t_0-1)}, \vec1}^2}{d} \le \|\mA\|_2^2.
    \label{eq:first_nonzero_regret_bound}
\end{align}
This provides the initialization bound used in our subsequent estimates of the terminal regret vector norm.

For the learners considered here, the first utility improvement is positive, as we verify in their respective analyses. Since the strategy does not change before iteration $t_0-1$, we obtain
\begin{align}
    \sum_{t=1}^{t_0-1} \ip{\vu^{(t)}, \vx^{(t+1)}-\vx^{(t)}} = \ip{\vu^{(t_0-1)}, \vx^{(t_0)}-\vx^{(t_0-1)}} > 0.
    \label{eq:t_0_1_sum}
\end{align}
For $T \ge t_0$, we can therefore drop this initial contribution when applying \eqref{eq:alt_reg_bound_simple} and focus on the utility improvements from $t_0$ onward.

\paragraph{Obstacles that arise when applying the analysis template to $\APRMplus$.} As discussed in the introduction, $\APRMplus$ is among the best-performing methods in practice \citep{Farina21:Faster, Chakrabarti24:Extensive}, yet its average-iterate convergence remains unresolved. We would therefore want to apply the analysis template to $\APRMplus$, seeking to explain its practical success through Alt-RVU. However, that does not work. Although we cannot establish Alt-RVU for $\APRMplus$, we identify the obstacles to obtaining such a bound.

We begin by bounding the terminal regret vector norm.
\begin{restatable}{lemma}{terminalTildeRBoundPrm}
\label{lem:terminal_tilde_r_bound_prm} 
 For every $T\ge1$, a $\PRMplus$ learner satisfies
    \begin{align}
        \max_{1 \leq s \leq T+1} \| \tilde \vr^{(s)}\|_2^2 \le \|\mA\|_2^2 + 2 (1+\sqrt{d})^2 \sum_{t=2}^T \| \vu^{(t)} - \vu^{(t-1)} \|^2_2  + 2d \|\mA\|_2^2 \sum_{t=1}^T \| \vx^{(t+1)} - \vx^{(t)} \|^2_2.
        \label{eq:tilde_r_s_bound}
    \end{align}
\end{restatable}
We next bound the utility improvement in terms of the regret vectors.
\begin{restatable}{lemma}{utilImprovementPrm}
\label{eq:util_improvement_prm}
    At time step $t \ge t_0$, a $\PRMplus$ learner satisfies
\begin{align*}
    \ip{\vu^{(t)}, \vx^{(t+1)}-\vx^{(t)}} \geq \frac{ \|\tilde \vr^{(t+1)}\|_2^2 - \|\tilde \vr^{(t)}\|_2^2 + \|\tilde \vr^{(t+1)}-\tilde \vr^{(t)}\|_2^2 + 2\|[\vr^{(t+1)}]_+-\tilde \vr^{(t+1)}\|_2^2}{2\|[\vr^{(t+1)}]_+\|_1}.
\end{align*}
\end{restatable}
For $T \ge t_0$, summing the above bound and separating the change in the squared regret norm from the two squared differences, we obtain
\begin{align}
        \label{eq:prm_decomp}
    \sum_{t=t_0}^T \ip{\vu^{(t)}, \vx^{(t+1)} - \vx^{(t)}} \ge \underbrace{\sum_{t=t_0}^T \frac{\|\tilde \vr^{(t+1)}\|_2^2 - \|\tilde \vr^{(t)}\|_2^2}{2\|[\vr^{(t+1)}]_+\|_1}}_{\termlabel{I}{term:alt_B}} + \underbrace{\sum_{t=t_0}^T \frac{\|\tilde \vr^{(t+1)} - \tilde \vr^{(t)}\|_2^2 + 2\|[\vr^{(t+1)}]_+ - \tilde \vr^{(t+1)}\|_2^2}{2\|[\vr^{(t+1)}]_+\|_1}}_{\termlabel{II}{term:alt_C}}.
\end{align}
We can show \eqref{term:alt_C} controls a weighted sum of squared strategy movements, up to a fixed initial term. 
\begin{restatable}{lemma}{movementBoundPrm}
    For any $T \ge t_0$, we have
\begin{align*}
    \sum_{t=t_0}^{T} \frac{\| \tilde \vr^{(t+1)} - \tilde \vr^{(t)} \|_2^2 + 2 \| [\vr^{(t+1)}]_+ - \tilde \vr^{(t+1)} \|_2^2}{2 \| [\vr^{(t+1)}]_+ \|_1}
&\ge \frac1{6d} \sum_{t=t_0}^{T} \max \{ \| [\vr^{(t)}]_+ \|_1, \| [\vr^{(t+1)}]_+ \|_1 \} \| \vx^{(t+1)} - \vx^{(t)} \|_2^2 \\
&- \frac{\| [\vr^{(t_0)}]_+ - \tilde \vr^{(t_0)} \|_2^2}{2 \| [\vr^{(t_0)}]_+ \|_1}.
\end{align*}
\end{restatable}
Two obstacles remain. The strategy movement weights $\max \{ \| [\vr^{(t)}]_+ \|_1, \| [\vr^{(t+1)}]_+ \|_1 \}$ in the preceding bound depend on the prediction regret vectors, so this estimate does not by itself provide the fixed positive coefficient required by Alt-RVU. More importantly, the main problem lies in \eqref{term:alt_B}, since the summand numerator does not have a fixed sign and cannot be readily controlled by the updates of $\APRMplus$. We will illustrate such negative contributions of the summands numerically under alternating self-play in a matrix game and discuss it in further detail in Subsection \ref{sec:alt_rvu_hprm}. 

\subsection{Alt-RVU with Norm Monotonicity: $\AIRPRMplus$}
\label{sec:ir_alt_rvu}

The obstacles we identified above motivate us to consider the learning algorithm of \citet{Zhang25:Scale}. Interpreting the regret vector norm as an inverse learning rate, they seek to prevent increases in the effective learning rate by introducing the IR mechanism, which ensures $\| \tilde\vr^{(t+1)} \|_2^2 \ge \| \tilde\vr^{(t)} \|_2^2$ at every iteration $t$. Our decomposition \eqref{eq:prm_decomp} provides complementary motivations for this mechanism: norm monotonicity prevents adverse norm changes in~\eqref{term:alt_B} and, together with norm preservation during prediction, keeps the weights in~\eqref{term:alt_C} bounded away from $0$ for $t\ge t_0$. We next derive the corresponding utility-improvement bounds for $\IRPRMplus$ and show how these properties yield Alt-RVU, and thus that $\IRPRMplus$ has $O(T^{-1})$ average-iterate convergence.

Recall that under IR, the update for the regret vector $\tilde \vr^{(t+1)}$ becomes \eqref{eq:correction}. We can rewrite it as
\begin{align*}
    \tilde \vr^{(t+1)}=\left[\tilde \vr^{(t)} + \vu^{(t)}-\langle \vu^{(t)}, \vx^{(t)} \rangle \mathbf1+c^{(t)} \mathbf1\right]_+,
\end{align*}
where $c^{(t)} = \ip{\vm^{(t)}, \vx^{(t)}} - \gamma^{(t)}$ highlights the new addition from the $\PRMplus$ regret vector update \eqref{eq:reg_vec_update_prm}. \citet[Proof of Theorem 3.2]{Zhang25:Scale} showed that $c^{(t)} \geq 0$ for any $t$. While they did not explicitly use $c^{(t)}$ in the analysis, we will fully utilize this scalar to obtain tight bounds. 

Since $c^{(t)} \geq 0$, for $\IRPRMplus$ we obtain a similar inequality to \eqref{eq:ineq_tilde_r}: 
\begin{align*}
    \tilde{\vr}^{(t+1)} = \left[\tilde \vr^{(t)} + \vu^{(t)}-\langle \vu^{(t)}, \vx^{(t)} \rangle \mathbf1+c^{(t)} \mathbf1\right]_+ \ge \tilde{\vr}^{(t)} + \vu^{(t)} - \langle \vu^{(t)}, \vx^{(t)} \rangle \mathbf1.
\end{align*}

We now bound the utility improvement from updating the strategy
\begin{restatable}{lemma}{utilImprovementIr}
    At every time step $t\ge1$, an $\IRPRMplus$ learner satisfies
    \begin{align}
        \ip{\vu^{(t)}, \vx^{(t+1)} - \vx^{(t)}}  \ge \frac{1}{4d} \| \tilde \vr^{(t + 1)} \|_2 \|\vx^{(t+1)} - \vx^{(t)}\|_2^2  + c^{(t + 1)} -  c^{(t)}.
        \label{eq:util_improvement_ir}
    \end{align}
\end{restatable}
The next step is to bound the terminal regret vector norm. Applying the norm estimate of \citet[Eq.~(5)]{Zhang25:Scale} over iterations \(t_0,\ldots,s-1\), we obtain
\begin{align*}
    \|\tilde{\vr}^{(s)}\|_2^2 &\le \|\tilde{\vr}^{(t_0)}\|_2^2 + \sum_{t=t_0}^{s-1} \|\vu^{(t)} - \vu^{(t-1)} - \langle \vu^{(t)} - \vu^{(t-1)}, \vx^{(t)}\rangle \vec1 \|_2^2 \\
    &\le \| \mA \|_2^2 + \sum_{t=t_0}^{s-1} \|\vu^{(t)} - \vu^{(t-1)} - \langle \vu^{(t)} - \vu^{(t-1)}, \vx^{(t)}\rangle \vec1 \|_2^2,
\end{align*}
where the second inequality comes from \eqref{eq:first_nonzero_regret_bound}. Since $\|\vx^{(t)}\|_2 \le 1$, we have
\begingroup
\small
\begin{align*}
    \|\vu^{(t)} - \vu^{(t-1)} - \langle \vu^{(t)} - \vu^{(t-1)}, \vx^{(t)}\rangle \vec1 \|_2 \le \|\vu^{(t)} - \vu^{(t-1)}\|_2 + \sqrt d  |\langle \vu^{(t)} - \vu^{(t-1)}, \vx^{(t)}\rangle | \le (1+\sqrt d)\|\vu^{(t)} - \vu^{(t-1)}\|_2.
\end{align*}
\endgroup
We hence obtain the bound for the terminal regret vector norm: 
\begin{align}
    \max_{1 \le s \le T+1} \|\tilde \vr^{(s)}\|_2^2 \le \|\mA\|_2^2 + (1 + \sqrt d)^2 \sum_{ t= t_0}^{T} \|\vu^{(t)} - \vu^{(t-1)}\|_2^2 \le \|\mA\|_2^2 + (1 + \sqrt d)^2 \sum_{t=2}^{T} \|\vu^{(t)} - \vu^{(t-1)}\|_2^2.
    \label{eq:terminal_tilde_r_bound_ir}
\end{align}
With that, we can now complete the Alt-RVU bound for $\IRPRMplus$ as follows.
\begin{restatable}[Alt-RVU for $\IRPRMplus$]{proposition}{altRvuIr}
    \label{prop:alt_rvu_IR}
    For any $\beta > 0$, a $\IRPRMplus$ learner satisfies the Alt-RVU bound for $\alpha = 2 \|\mA \|_2 + (1+\sqrt{d})^2/\beta$ and $\gamma = \| \tilde \vr^{(t_0)}\|_2 / (4d)$. Therefore, in every finite two-player zero-sum normal-form game, alternating self-play with $\IRPRMplus$ has $O(T^{-1})$ average-iterate convergence.
\end{restatable}

The IR mechanism thus provides a sufficient way to obtain the control required by Alt-RVU. We next investigate whether enforcing norm monotonicity is necessary or whether Alt-RVU can be obtained without enforcing nondecreasing regret vector norms.

\subsection{Alt-RVU without Norm Monotonicity: $\AHPRMplus$}
\label{sec:alt_rvu_hprm}

The preceding subsection establishes Alt-RVU for $\IRPRMplus$ by exploiting the nondecreasing regret vector norm enforced by the IR mechanism. We now show that $\HPRMplus$ satisfies Alt-RVU without requiring this monotonicity. The motivation for allowing norm decreases comes from $\APRMplus$ self-play, which exhibits fast empirical average-iterate convergence despite such decreases on the following matrix game.
\begin{align}
    \mA = \begin{pmatrix}
200 &-200 &0\\
-200 &202 &-1\\
0 &-1 &1/2
\end{pmatrix}.
\label{eg:prm_norm_decrease}
\end{align}

\begin{figure}[!h]
    \centering
    \includegraphics[width=\linewidth]{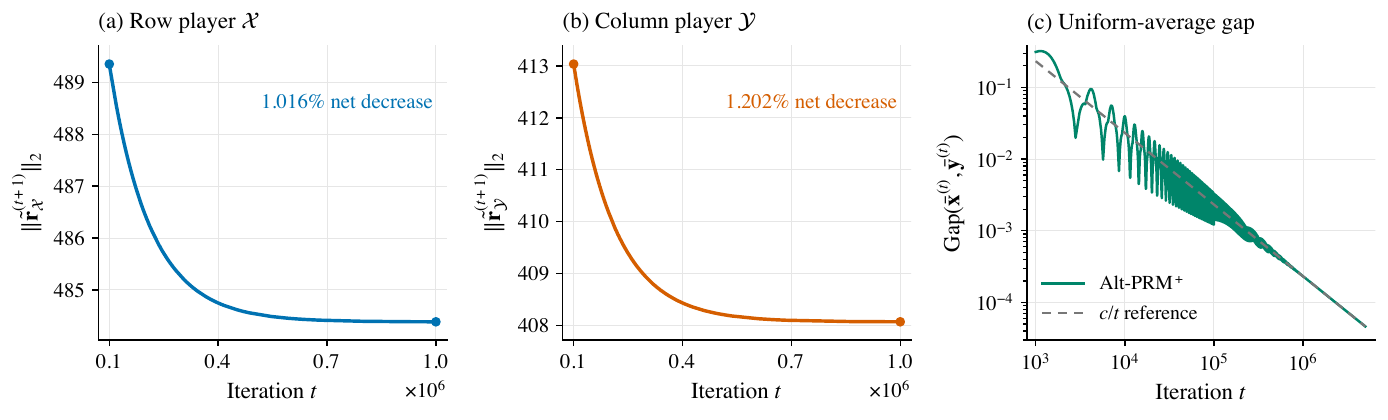}
    \caption{$\APRMplus$ on the normal-form game \eqref{eg:prm_norm_decrease}.}
    \label{fig:prm_norm_decrease}
\end{figure}

Figure~\ref{fig:prm_norm_decrease} shows both players' regret vector norms decreasing over iterations $10^5$ to $10^6$, producing negative contributions to term~\eqref{term:alt_B}. Nevertheless, in the rightmost plot, the average-iterate gap empirically decays approximately as $T^{-1}$. This behavior motivates controlling the cumulative effect of norm changes without requiring every change to be nonnegative. Moreover, the IR mechanism incurs additional computational overhead by solving for the norm-preserving scalar shift $\gamma$ in~\eqref{eq:prediction_scalar} per iteration. These considerations motivate us to seek an algorithm that not only permits norm decreases but also avoids this scalar computation while retaining the Alt-RVU guarantee. In fact, as we will show later in this paper, an algorithm that this relaxation enables ($\HPRMplus$) significantly outperforms $\IRPRMplus$ in practice and they have the same theoretical convergence guarantees under alternating self-play.

We begin by relating changes in the squared regret vector norm to consecutive regret vector increments.
\begin{restatable}{lemma}{halfBoundRTilde}
\label{lem:half_bound_r_tilde}
    For every $t \geq 1$, a $\HPRMplus$ learner satisfies
    \begin{align}
        \|\tilde \vr^{(t+1)}\|_2^2 - \|\tilde \vr^{(t)}\|_2^2 \geq \frac12 \|\tilde \vr^{(t+1)} - \tilde \vr^{(t)}\|_2^2 - \frac12 \|\tilde \vr^{(t)} - \tilde \vr^{(t -1)}\|_2^2.
        \label{eq:half_bound_r_tilde}
    \end{align}
\end{restatable}
This bound gives us the idea of proposing the following \emph{corrected squared norm} 
\begin{align*}
    f^{(t)} = \|\tilde \vr^{(t)}\|_2^2 - \frac12 \|\tilde \vr^{(t)} - \tilde \vr^{(t -1)}\|_2^2,
\end{align*}
because it implies that 
\begin{align*}
     f^{(t + 1)} -  f^{(t)} \ge 0.
\end{align*}

The corrected squared norm $f^{(t)}$ remains nondecreasing even when the regret vector norm decreases. An analogous correction is nondecreasing for every $0\le\lambda\le1/2$, but half prediction additionally eliminates a positive movement contribution from the terminal regret-norm bound, allowing us to combine the two estimates to establish Alt-RVU. Appendix~\ref{sec:why_half_prediction} derives the correction for general $\lambda$ and explains this distinction. 

The key idea is that since $f^{(t)}$ is nondecreasing, it provides us more control than the regret vector norm $\|\tilde \vr^{(t)}\|_2^2$. Our analysis can shift to $f^{(t)}$ if its increments also control the utility improvement. We can, indeed, bound the utility improvement with $f^{(t + 1)} - f^{(t)}$.
\begin{restatable}{lemma}{utilImprovementHprm}
    At every time step $t\ge t_0-1$, a $\HPRMplus$ learner satisfies
    \begin{align}
        \langle\vu^{(t)},\vx^{(t+1)}-\vx^{(t)}\rangle
         \ge \frac{\|\tilde{\vr}^{(t+1)} -\tilde{\vr}^{(t)}\|_2^2}{4\|[\vr^{(t+1)}]_+\|_1} - \frac{f^{(t + 1)} - f^{(t)}}{2 \|[\vr^{(t+1)}]_+\|_1}.
         \label{eq:util_improvement_hprm}
    \end{align}
\end{restatable}
Following our Alt-RVU analysis template, we now bound the terminal regret vector norm. Compared with the $\PRMplus$ estimate, this bound also involves regret-vector increments. Their coefficient can be chosen to be arbitrarily small, which will allow us to control their contribution using the utility-improvement bound.
\begin{restatable}{lemma}{terminalTildeRBoundHalf}
    For every $\varepsilon>0$ and $T\ge1$, a $\HPRMplus$ learner satisfies
    \begingroup
    \small
    \begin{align}
        \max_{1 \le s \le T+1} \|\tilde \vr^{(s)}\|_2^2 \le \|\mA\|_2^2 + \left(2+\frac{1}{2\varepsilon}\right) \left[(1+\sqrt d)^2 \sum_{t=2}^{T} \|\vu^{(t)}-\vu^{(t-1)}\|_2^2 + d\|\mA\|_2^2 \sum_{t=1}^{T} \|\vx^{(t+1)}-\vx^{(t)}\|_2^2\right] + \varepsilon \sum_{t=2}^{T} \|\tilde \vr^{(t)}-\tilde \vr^{(t-1)}\|_2^2.
        \label{eq:tilde_r_s_bound_half}
    \end{align}
    \endgroup
\end{restatable}
We now obtain the following.
\begin{restatable}[Alt-RVU for $\AHPRMplus$]{proposition}{altRvuHprm}
    Set $\gamma = \sqrt{f^{(t_0)}}/(64 \sqrt{2} (1 + \sqrt{d})^2) >0$. For every $\beta > 0$, there exists $\alpha > 0$ so that $\AHPRMplus$ satisfies the Alt-RVU bound. Therefore, in every finite two-player zero-sum normal-form game, alternating self-play with $\HPRMplus$ has $O(T^{-1})$ average-iterate convergence.
\end{restatable}

\begin{figure}[!h]
    \centering
    \includegraphics[width=\linewidth]{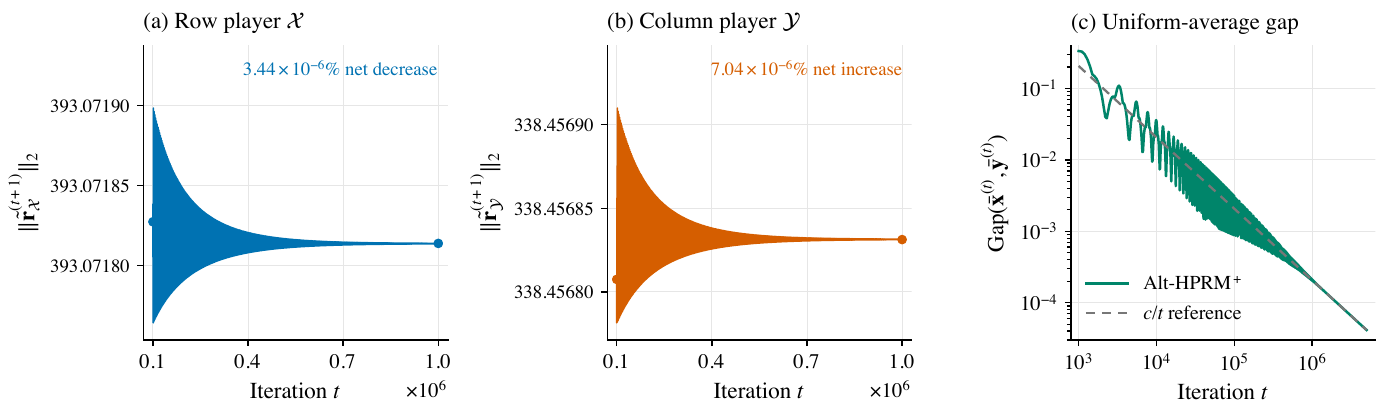}
    \caption{$\AHPRMplus$ on the normal-form game \eqref{eg:prm_norm_decrease}.}
    \label{fig:hprm_norm_decrease}
\end{figure}

This proposition thus gives $\AHPRMplus$ the same $O(T^{-1})$ average-iterate guarantee as $\AIRPRMplus$ without enforcing nondecreasing regret vector norms. Indeed, such norm decreases occur on the same game~\eqref{eg:prm_norm_decrease} for $\AHPRMplus$ self-play, where Figure~\ref{fig:hprm_norm_decrease} shows small decreases in both players' regret vector norms up to iterations $10^6$. By shifting the analysis to the nondecreasing $f^{(t)}$, we control the cumulative effect of regret vector norm decreases and retain the simpler half-prediction update, which avoids the additional scalar computation required by the IR mechanism. The benefits extend beyond the lower per-iteration cost: Section~\ref{sec:exp} provides experiments in which the $\CFR$ instantiation of $\AHPRMplus$ outperforms that of $\AIRPRMplus$ on most benchmarks, achieving faster exploitability reduction in terms of both iteration count and learner wall-clock time. 

\subsection{Insufficiency of Standard RVU}
\label{sec:necessity_alt_rvu}

In the above subsections, we developed Alt-RVU to establish average-iterate convergence guarantees for $\AIRPRMplus$ and $\AHPRMplus$. A natural question is whether these guarantees for alternating self-play could instead be obtained by establishing a standard RVU bound (Definition~\ref{def:rvu}) for each learner. We show in this subsection that standard RVU is too restrictive to serve as a learner-level condition covering all optimistic methods that converge under alternating self-play. Indeed, $\HPRMplus$ fails the standard RVU bound despite satisfying Alt-RVU and achieving the optimal $O(T^{-1})$ average-iterate convergence rate under alternating self-play. This result highlights the need to go beyond standard RVU to encompass such alternating methods, and our Alt-RVU framework and accompanying analysis provide a systematic extension that does so.

We now show that a standard RVU bound is impossible for $\AHPRMplus$, even on a single bounded utility sequence.
\begin{restatable}{proposition}{halfStandardRvuFailure}
\label{prop:half_standard_rvu_failure}
There is a utility sequence $(\vu^{(t)})_{t \ge 1} \subset [-1,1]^2$, which can be generated by an opponent under alternating updates, on which a $\HPRMplus$ learner satisfies, for every $T \ge 1$,
\begin{align}
    0 \le \operatorname{Reg}(T) \le 3, \; \sum_{t=2}^{\infty} \|\vu^{(t)} - \vu^{(t-1)}\|_2^2 < \infty, \; \sum_{t=2}^{\infty} \|\vx^{(t)} - \vx^{(t-1)}\|_2^2 = \infty.
    \label{eq:half_standard_rvu_failure}
\end{align}
Consequently, the standard RVU bound fails on this sequence for every choice of horizon-independent positive constants, even when they may depend on the entire utility sequence.
\end{restatable}

The distinction between ordinary and alternating regret is noteworthy here. Since $\HPRMplus$ satisfies Alt-RVU, the same finite utility variation and divergent strategy movement imply
$\operatorname{Alt-Reg}(T)\to-\infty$ on this sequence. From the definitions, we have the identity
\begin{align*}
    \operatorname{Alt-Reg}(T)
    = 2\operatorname{Reg}(T)
    - \sum_{t=1}^{T}
    \ip{\vu^{(t)},\vx^{(t+1)}-\vx^{(t)}}.
\end{align*}
This equality implies that the cumulative utility improvement satisfies $\sum_{t=1}^{T}\ip{\vu^{(t)},\vx^{(t+1)}-\vx^{(t)}}\to+\infty$, while $0\le\operatorname{Reg}(T)\le3$. Alternating regret accounts for the gains from updating $\vx^{(t)}$ to $\vx^{(t+1)}$ against the same utility vector $\vu^{(t)}$, whereas ordinary regret evaluates only the strategy $\vx^{(t)}$ played before the update. On this sequence, these accumulated gains drive $\operatorname{Alt-Reg}(T)\to-\infty$, as Alt-RVU requires when utility variation is finite but $\sum_{t=1}^{T}\|\vx^{(t+1)}-\vx^{(t)}\|_2^2\to\infty$. Since $\operatorname{Reg}(T)\ge0$, no choice of horizon-independent $\alpha,\beta,\gamma>0$ can make ordinary regret satisfy the corresponding bound, whose right-hand side tends to $-\infty$. Changing the regret quantity is therefore essential here, rather than merely adjusting the constants in standard RVU.
\begin{remark} 
    The Alt-RVU framework itself is not restricted to the regret matching family. Its definition involves only utilities, strategies, and alternating regret, so its convergence implications apply to other alternating learners whenever the required bounds and coefficient conditions hold. We can prove similar Alt-RVU-based guarantees for other classes of algorithms, for example, $O(T^{-1})$ average-iterate convergence under alternating self-play for the \textit{Adaptive Optimistic Gradient Descent ($\AdOGD$)} algorithm of \citet{Zhang25:Scale}. We omit the routine details and focus on regret matching algorithms here.
\end{remark}

\section{Alt-RVU Failure of Scaled-Prediction Regret Matching}
\label{sec:no_lambda_except_half}

The preceding analysis establishes the Alt-RVU bound for the half-prediction, \textit{i.e.}, $\lambda=1/2$, but leaves open whether other  choices of $\lambda$ can admit the same bound. Now, we show that they cannot: for every fixed $\lambda\ge0$ with $\lambda\ne1/2$, we construct a bounded infinite utility sequence on which Alt-RVU fails. 

\citet[Theorem~3.1]{Farina23:Regret} construct finite utility sequences on which the strategies generated by $\RMplus$ and $\PRMplus$ repeatedly alternate between $(1/2,1/2)$ and $(0,1)$. Normalization can produce such movements from small changes in the prediction regret vectors when their positive parts have small norms. For $\PRMplus$, this difficulty also appears in our lower bound on~\eqref{term:alt_C}, which weights squared strategy movement by $\max\{\|[\vr^{(t)}]_+\|_1,\|[\vr^{(t+1)}]_+\|_1\}$. Without a positive lower bound on these weights, this estimate does not provide a fixed positive coefficient on unweighted strategy movement. Their construction also rules out standard RVU bounds with constants chosen uniformly across utility sequences. Rescaling their geometrically growing sequences as in \citet[Remark~B.1]{Farina23:Regret} gives finite utility sequences in $[-1,1]^2$ with uniformly bounded ordinary regret and squared utility variation, while cumulative squared strategy movement grows linearly. Extending each sequence by zero utilities therefore implies that, for each of $\RMplus$ and $\PRMplus$,
\begin{align*}
    \forall\,\alpha,\beta,\gamma>0,\; \exists\,(\vu^{(t)})_{t\ge1}\subset[-1,1]^2,\; \exists\,T \ge 1 : \operatorname{Reg}(T) > \alpha + \beta \sum_{t=2}^{T} \|\vu^{(t)}-\vu^{(t-1)}\|_2^2 - \gamma \sum_{t=2}^{T} \|\vx^{(t)}-\vx^{(t-1)}\|_2^2.
\end{align*}

We seek an analogous failure result for Alt-RVU, whose definition allows the constants to depend on the utility sequence. Establishing such a result therefore requires more than replacing ordinary regret by alternating regret and changing the indices. We must construct a single bounded infinite utility sequence on which the Alt-RVU bound fails for every choice of horizon-independent positive constants, that is,
\begin{align*}
    \exists\,(\vu^{(t)})_{t\ge1}\subset[-1,1]^d,\; \forall\,\alpha,\beta,\gamma>0,\; \exists\,T \ge 1 : \operatorname{Alt-Reg}(T) > \alpha + \beta \sum_{t=2}^{T} \|\vu^{(t)}-\vu^{(t-1)}\|_2^2 - \gamma \sum_{t=1}^{T} \|\vx^{(t+1)}-\vx^{(t)}\|_2^2.
\end{align*}

\begin{figure}[!h]
    \centering
    \includegraphics[width=\linewidth]{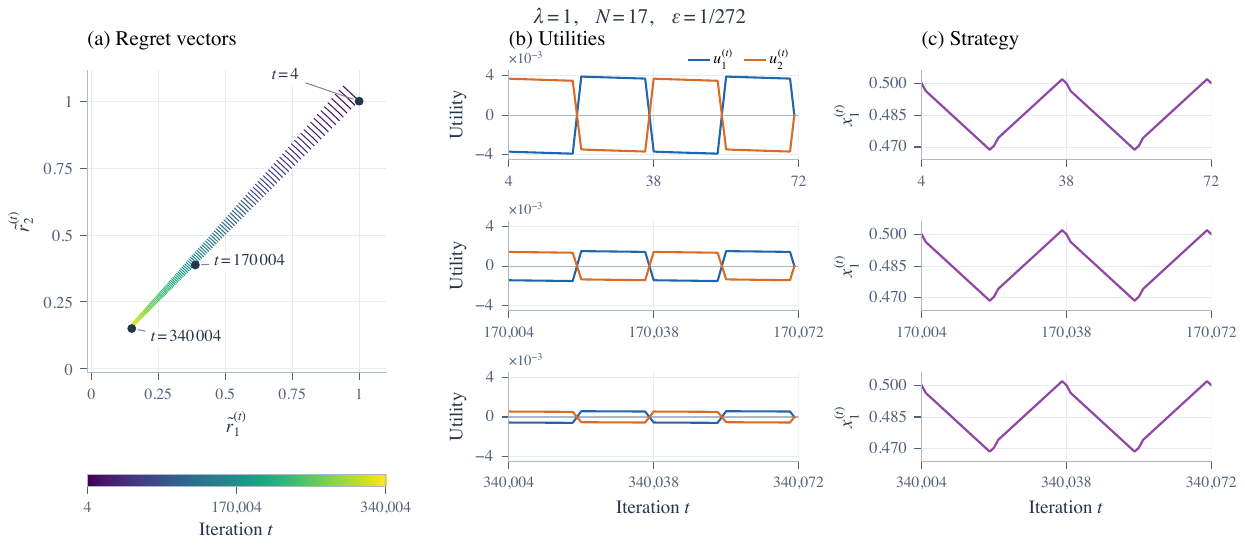}
    \caption{The construction of Proposition~\ref{prop:failure_above_half_explicit} with 2 actions.}
    \label{fig:lambda_equal_1}
\end{figure}

We construct one such utility sequence for every $\lambda \ne 1/2$. For $\lambda>1/2$, we make the regret vector norms decrease geometrically while the strategies repeat. 
\begin{restatable}{proposition}{altRvuFailureAboveHalf}
\label{prop:failure_above_half_explicit}
For every fixed $\lambda > 1/2$, there is a utility sequence $(\vu^{(t)})_{t \ge 1} \subset [-1,1]^2$, on which the scaled-prediction regret matching$^+$ learner satisfies
\begin{align}
    \sum_{t=1}^{\infty}\|\vu^{(t)}\|_2<\infty, \; \ \sup_{T\ge1}|\operatorname{Alt-Reg}(T)|<\infty, \; \ \sum_{t=1}^T\|\vx^{(t+1)}-\vx^{(t)}\|_2^2=\Theta(T).
    \label{eq:failure_above_half}
\end{align}
Consequently, Alt-RVU fails on this sequence for every choice of horizon-independent constants.
\end{restatable}
We visualize this construction in Figure~\ref{fig:lambda_equal_1}. Intuitively, we construct successive blocks with similar sizes that multiply the regret vectors $\tilde{\vr}^{(t)}$ and utilities $\vu^{(t)}$ by the same factor $\rho\in(0,1)$ at the same positions across different blocks. Normalization removes this factor, so the strategies $\vx^{(t)}$ cycle.

Taking $\lambda=1$ in Proposition~\ref{prop:failure_above_half_explicit} shows that $\PRMplus$ fails Alt-RVU. This construction clarifies why controlling the norm decrease in \eqref{term:alt_B} alone is insufficient. Both~\eqref{term:alt_B} and~\eqref{term:alt_C} remain bounded on this sequence, since their contributions over successive blocks decay geometrically. Thus, the negative contribution of~\eqref{term:alt_B} can be absorbed into a horizon-independent constant.

The remaining difficulty is that our lower bound on~\eqref{term:alt_C} supplies a strategy-movement coefficient proportional to the prediction regret vector norms:
\begin{align*}
    \frac1{6d}\max\{\|[\vr^{(t)}]_+\|_1,\|[\vr^{(t+1)}]_+\|_1\} \rightarrow 0.
\end{align*}
These coefficients decay geometrically across blocks, while the same nonzero strategy movements repeat. Consequently, the weighted movement sum remains bounded even though the unweighted sum grows linearly. Since~\eqref{term:alt_C} itself remains bounded, this cannot be repaired by sharpening its lower bound. In particular, no fixed $\gamma>0$ and $C<\infty$ can make
\begin{align*}
    \sum_{t=t_0}^{T} \frac{\|\tilde\vr^{(t+1)}-\tilde\vr^{(t)}\|_2^2 + 2\|[\vr^{(t+1)}]_+-\tilde\vr^{(t+1)}\|_2^2}{2\|[\vr^{(t+1)}]_+\|_1} \ge \gamma \sum_{t=t_0}^{T}\|\vx^{(t+1)}-\vx^{(t)}\|_2^2 - C
\end{align*}
hold for every $T \ge t_0$, because the left-hand side remains bounded while the right-hand side grows linearly with $T$.

For $0\le\lambda<1/2$, we instead construct alternating regret of order $\sqrt T$ with only logarithmic squared utility variation. This rules out even the upper bound obtained by removing the negative movement term from Alt-RVU. 
\begin{restatable}{proposition}{altRvuFailureBelowHalf}
\label{prop:failure_below_half_explicit}
Fix $0\le\lambda<1/2$ and an integer $d\ge4$ satisfying $d>((1+2\lambda)/(1-2\lambda))^2$. There is a utility sequence $(\vu^{(t)})_{t\ge1}\subset[-1,1]^d$ on which the scaled-prediction regret matching$^+$ learner satisfies
\begin{align}
    \operatorname{Alt-Reg}(T) = \Theta(\sqrt T), \; \sum_{t=2}^{T}\|\vu^{(t)} - \vu^{(t-1)}\|_2^2 = O(\log T).
    \label{eq:failure_below_half}
\end{align}
Consequently, Alt-RVU fails on this sequence for every choice of horizon-independent constants.
\end{restatable}

Taking $\lambda=1/3$ in Proposition~\ref{prop:failure_below_half_explicit} shows that the local learner underlying the $\SAPCFRplus$ algorithm of \citet{Meng26:Faster} fails Alt-RVU in every dimension $d\ge26$. Together with the positive result in Subsection~\ref{sec:alt_rvu_hprm}, Propositions~\ref{prop:failure_above_half_explicit} and~\ref{prop:failure_below_half_explicit} identify $\lambda=1/2$ as the \emph{unique} fixed multiplier for which Alt-RVU holds on every bounded utility sequence in every finite dimension.

Alt-RVU is a sufficient criterion for $O(T^{-1})$ average-iterate convergence, and its failure does not rule out slower convergence guarantees for other multipliers. Any $O(T^{-1})$ guarantee for another fixed multiplier across all finite games must exploit additional structure of self-play or use an analysis beyond our general Alt-RVU analysis. In Subsection~\ref{sec:separation_last_avg}, we complement these theoretical results with numerical separations in average- and last-iterate behavior under alternating self-play, comparing half prediction with selected multipliers.

\section{Best- and Last-Iterate Convergence}
\label{sec:best_last_nfg}

The preceding discussion provided the optimal average-iterate convergence $O(T^{-1})$ for $\AIRPRMplus$ and $\AHPRMplus$. Average-iterate convergence, however, does not fully capture the practical performance of self-play, as averaging can retain the influence of poor early strategies and converge more slowly than the last iterate in standard benchmarks~\citep{Cai25:Last,Zhang25:Scale}. Our experiments in Section~\ref{sec:exp} also exhibit this behavior, reinforcing the importance of also studying best- and last-iterate convergence.

Now, to study best- and last-iterate convergence, we must focus on the Nash gap at the $t$-th iteration. This Nash gap is exactly the sum of the players’ best-response errors at that iteration:
\begin{align*}
    \operatorname{Gap}(\vx^{(t)},\vy^{(t)}) = \max_{\vz \in \cX} \langle \mA \vy^{(t)}, \vz - \vx^{(t)} \rangle + \max_{\vz \in \cY} \langle -\mA^\top\vx^{(t)}, \vz - \vy^{(t)} \rangle.
\end{align*}
Here, one might try to bound the two terms on the RHS. However, these terms do not directly reflect the alternating update structure discussed in Subsection~\ref{sec:alt_rvu}. Specifically, $\vy^{(t)}$ is played against both $\vx^{(t)}$ and $\vx^{(t + 1)}$, whereas the first term involves only $\vx^{(t)}$. Likewise, $\vx^{(t)}$ is played against both $\vy^{(t - 1)}$ and $\vy^{(t)}$, whereas the second term involves only $\vy^{(t)}$.

Our key insight is to bound the Nash gap at the $t$-th iteration with terms that are more natural to alternating update structure. We noticed that
\begin{align*}
    \max_{\vz \in \cX} \langle \mA \vy^{(t)}, \vz - \vx^{(t+1)} \rangle + \max_{\vz \in \cX} \langle \mA \vy^{(t)}, \vz - \vx^{(t)} \rangle = \max_{\vz \in \cX} \langle \mA \vy^{(t)}, 2 \vz - \vx^{(t)} - \vx^{(t + 1)} \rangle.
\end{align*}
Analogously,
\begin{align*}
    \max_{\vz \in \cY} \langle -\mA^\top\vx^{(t)}, \vz - \vy^{(t - 1)} \rangle + \max_{\vz \in \cY} \langle -\mA^\top\vx^{(t)}, \vz - \vy^{(t)} \rangle = \max_{\vz \in \cY} \langle -\mA^\top\vx^{(t)}, 2 \vz - \vy^{(t)} - \vy^{(t - 1)} \rangle.
\end{align*}
With these identities, we can bound the Nash gap at the $t$-th iteration as follows.
\begin{align*}
    \operatorname{Gap}(\vx^{(t)},\vy^{(t)}) &\le \operatorname{Gap}(\vx^{(t)},\vy^{(t)}) + \max_{\vz \in \cX} \langle \mA \vy^{(t)}, \vz - \vx^{(t+1)} \rangle + \max_{\vz \in \cY} \langle -\mA^\top\vx^{(t)}, \vz - \vy^{(t - 1)} \rangle\\
    &= \max_{\vz \in \cX} \langle \mA \vy^{(t)}, 2 \vz - \vx^{(t)} - \vx^{(t + 1)} \rangle + \max_{\vz \in \cY} \langle -\mA^\top\vx^{(t)}, 2 \vz - \vy^{(t)} - \vy^{(t - 1)} \rangle. 
\end{align*}
This bound identifies the quantity missing from Alt-RVU. Its expression is the same as the contribution of one observation to alternating regret, but the comparator $\vz$ is now maximized separately at each iteration. A bound on alternating regret hence does not directly control these individual maxima.

For best-iterate convergence, we seek at least one iteration with a small Nash gap. For last-iterate convergence to the Nash set, we need the gap to become small at every sufficiently late iteration. A bound on the sum of squared errors is able to provide both. Specifically, we have
\begin{align}
    \sum_{t=1}^{T} \operatorname{Gap}(\vx^{(t)}, \vy^{(t)})^2
    \le 2 \sum_{t=1}^{T} \left[\max_{\vz \in \cX} \ip{\mA\vy^{(t)}, 2\vz - \vx^{(t)} - \vx^{(t+1)}}\right]^2 + 2 \sum_{t=1}^{T} \left[\max_{\vz \in \cY} \ip{-\mA^\top\vx^{(t)}, 2\vz - \vy^{(t-1)} - \vy^{(t)}}\right]^2.
    \label{eq:square_gap_bound}
\end{align}
It therefore suffices to bound each of the two sums on the right hand side independently of $T$. This motivates us to refine the Alt-RVU bound as follows.
\begin{definition}[Refined Alt-RVU]\footnote{This parallels the analysis of \citet[Proposition~A.13]{Anagnostides22:On}, as they retain squared distances between the played and auxiliary optimistic mirror descent iterates in a refined RVU bound and use smoothness of the regularizer to control $\max_{\vz\in\Delta(d)}\ip{\vu^{(t)},\vz-\vx^{(t)}}$. In our settings, we instead control $\max_{\vz\in\Delta(d)}\ip{\vu^{(t)},2\vz-\vx^{(t)}-\vx^{(t+1)}}$ due to the alternating structure through $\|\tilde\vr^{(t+1)}-\tilde\vr^{(t)}\|_2$.}
\label{def:nfg_refined_rvu}
A learner satisfies the refined Alt-RVU bound on a bounded utility sequence if there are constants $\alpha, \beta, \gamma, \zeta >0$, independent of $T$, such that, for every $T\ge1$,
\begin{align}
\operatorname{Alt-Reg}(T) + \zeta \sqrt{\sum_{t=1}^{T} \left[ \max_{\vz \in \Delta(d)} \left \langle \vu^{(t)}, 2\vz- \vx^{(t)} - \vx^{(t+1)}\right\rangle \right]^2} \le\alpha+\beta\sum_{t=2}^T\|\vu^{(t)}-\vu^{(t-1)}\|_2^2-\gamma\sum_{t=1}^T\|\vx^{(t+1)}-\vx^{(t)}\|_2^2.
\label{eq:refined_rvu}
\end{align}
\end{definition}

The additional second term does not replace alternating regret. Indeed, maximizing separately at each iteration and applying Cauchy-Schwarz, we obtain
\begin{align*}
    \operatorname{Alt-Reg}(T) \le \sum_{t=1}^T \max_{\vz \in \Delta(d)} \ip{\vu^{(t)}, 2\vz - \vx^{(t)} - \vx^{(t+1)}} \le \sqrt{T}\sqrt{\sum_{t=1}^T \left[\max_{\vz \in \Delta(d)} \ip{\vu^{(t)}, 2\vz - \vx^{(t)} - \vx^{(t+1)}}\right]^2}.
\end{align*}
Thus, bounding the square-root term independently of $T$ gives only an $O(T^{1/2})$ upper bound on alternating regret, which yields an $O(T^{-1/2})$ average-iterate guarantee rather than $O(T^{-1})$. Conversely, bounded alternating regret does not control the best-response errors at individual iterations. Retaining both terms allows us to establish all the desired guarantees.

We now show how this refined Alt-RVU yields best- and last-iterate convergence to the Nash set.
\begin{restatable}{proposition}{refinedAltRvuConvergence}
\label{prop:refined_alt_rvu_convergence}
    Suppose both players in a finite two-player zero-sum normal-form game satisfy the refined Alt-RVU bound under alternating updates, with common constants $\alpha, \beta, \gamma, \zeta > 0$ satisfying $\beta\|\mA\|_2^2 \le \gamma$. Then, we have 
    \begin{align*}
        \sum_{t=1}^{\infty} \operatorname{Gap}(\vx^{(t)}, \vy^{(t)})^2 \le \left(\frac{2\alpha + \operatorname{range}(\mA)}{\zeta}\right)^2.
    \end{align*}
    Consequently, for every $T \ge 1$, we obtain
    \begin{align*}
        \min_{1 \le t \le T} \operatorname{Gap}(\vx^{(t)}, \vy^{(t)}) \le \frac{2\alpha + \operatorname{range}(\mA)}{\zeta\sqrt T},
    \end{align*}
    which implies $O(T^{-1/2})$ best-iterate rate, and the last iterates converge to the Nash set $\cZ^*$.
\end{restatable}
We first relate these best-response errors to regret vector increments.
\begin{restatable}{lemma}{refineAltRvuErrors}
\label{lem:refine_alt_rvu_errors}
$\IRPRMplus$ and $\HPRMplus$ learners satisfy, for every $t \ge 1$,
\begin{align}
    \max_{\vz \in \Delta(d)} \ip{\vu^{(t)}, 2\vz - \vx^{(t)} - \vx^{(t+1)}} \le \sqrt{5}\|\tilde\vr^{(t+1)} - \tilde\vr^{(t)}\|_2.
    \label{eq:errors}
\end{align}
For $\IRPRMplus$, the constant $\sqrt{5}$ can be improved to $2$.
\end{restatable}
The preceding lemma relates the additional best-response errors to regret vector increments. We now show that the negative terms retained in the Alt-RVU analysis also control these increments, yielding the refined Alt-RVU bound.
\begin{restatable}{proposition}{refinedAltRvuAlgorithms}
\label{prop:refined_alt_rvu_algorithms}
Both $\IRPRMplus$ and $\HPRMplus$ satisfy refined Alt-RVU. Specifically, for every bounded utility sequence, there exist $\gamma,\zeta>0$ such that, for every $\beta>0$, there exists $\alpha>0$ for which \eqref{eq:refined_rvu} holds for every $T\ge1$. These constants are independent of $T$.
\end{restatable}
Proposition~\ref{prop:refined_alt_rvu_algorithms} allows us to choose common constants for the two players satisfying $\beta\|\mA\|_2^2<\gamma$. Proposition~\ref{prop:refined_alt_rvu_convergence} then yields an $O(T^{-1/2})$ best-iterate rate and convergence of the last iterates to the Nash set $\cZ^*$. 

Convergence to a single equilibrium, however, requires more than refined Alt-RVU: vanishing exploitability and square-summable strategy movements do not rule out multiple equilibrium accumulation points. We establish this stronger conclusion by further exploiting the structure of the $\IRPRMplus$ and $\HPRMplus$ updates, showing that convergence of a subsequence to a Nash equilibrium forces the entire strategy sequence to converge to the same equilibrium.
\begin{restatable}{theorem}{nfgBestConvergence}
    \label{thm:nfg_best_last_convergence}
    In every finite two-player zero-sum normal-form game, $\AIRPRMplus$ and $\AHPRMplus$ satisfy 
    \begin{align*}
        \min_{1 \le t \le T} \operatorname{Gap}(\vx^{(t)}, \vy^{(t)}) = O(T^{-1/2}) \text{ and } (\vx^{(t)}, \vy^{(t)}) \rightarrow (\vx^{*}, \vy^{*}),
    \end{align*}
    where $(\vx^{*}, \vy^{*}) \in \cal Z^\ast$ is a Nash equilibrium. 
\end{restatable}
\begin{remark}
\citet{Cai25:Last} establish similar convergence to a single equilibrium for $\ExRMplus$ and smooth $\sPRMplus$ under simultaneous updates, but both methods require payoff-dependent step sizes. For IR variants, \citet{Sullins26:IR} and \citet{Li26:Why} establish convergence to the Nash set for $\IRPRMplus$ and $\IREGPRMplus$, respectively, leaving convergence to a single equilibrium unresolved when the game does not have a unique equilibrium. Theorem~\ref{thm:nfg_best_last_convergence} establishes this stronger conclusion for the parameter-free $\AIRPRMplus$ and $\AHPRMplus$ algorithms.
\end{remark}

\section{Separation of Alternating and Simultaneous Updates}
\label{sec:separation_alt_sim}

In two-player zero-sum games, alternating mirror descent admits $O(T^{-4/5})$ average-iterate convergence guarantees under certain regularity assumptions, improving on the classical $O(T^{-1/2})$ guarantee for simultaneous mirror descent~\citep{Katona26:Symplectic}. Similarly, projected alternating gradient descent--ascent achieves an $O(T^{-1})$ average-iterate rate in finite matrix games, compared with the classical $O(T^{-1/2})$ guarantee for simultaneous updates~\citep{Nan26:AltGDA}. For regret-matching methods, \citet[Section~3]{Cai25:Last} provide numerical evidence that alternation can affect convergence itself. On a $3\times3$ matrix game, their experiments suggest last-iterate nonconvergence of $\SPRMplus$, whereas $\APRMplus$ appears to converge to equilibrium. These observations motivate a stronger separation than an improvement in average-iterate rates, but leave open whether both behaviors can be established analytically. 

We establish such an analytical separation for $\HPRMplus$ by combining its alternating convergence guarantee with the following simultaneous nonconvergence result.
\begin{restatable}{theorem}{AltSimSeparation}
\label{thm:alt_sim_separation}
Consider the normal-form game
\begin{align}
    \mA=
    \begin{pmatrix}
        0 &-1 &1 &4\\
        1 &0 &-1 &3\\
        -1 &1 &0 &3\\
        -4 &-3 &-3 &0
    \end{pmatrix}.
    \label{eq:separation_matrix}
\end{align}
Under standard initialization, $\SHPRMplus$ satisfies
\begin{align*}
    \liminf_{t\to\infty}
    \operatorname{Gap}(\vx^{(t)},\vy^{(t)})
    \ge \frac1{\sqrt3},
\end{align*}
while $\AHPRMplus$ converges to the unique Nash equilibrium $\vx^*=\vy^*=(1/3,1/3,1/3,0)$ (Theorem~\ref{thm:nfg_best_last_convergence}).
\end{restatable}
We constructed the example above by starting from three-action rock--paper--scissors (RPS), whose skew symmetry keeps the two simultaneous learners synchronized, with $\vx^{(t)}=\vy^{(t)}$ and zero realized utility. However, standard initialization would place them exactly at the uniform equilibrium, leaving no nontrivial dynamics to analyze. We therefore added a strictly dominated fourth action and assigned unequal payoffs $4,3,3$ against it, choosing the fourth row to preserve skew symmetry. Under uniform initialization, the RPS payoffs cancel, so the unequal payoffs $4,3,3$ against the fourth action produce $\vx^{(2)}=\vy^{(2)}=(2/5,3/10,3/10,0)$ after the first simultaneous update. The fourth action receives negative utility throughout the subsequent simultaneous trajectory and is never played again. Its only role is hence to generate a nonuniform starting point for the RPS dynamics through the prescribed updates themselves. Since both players have identical regret vectors and strategies, we analyze one learner's subsequent updates within the three-action RPS subgame and prove that its strategy remains bounded away from equilibrium. A visualization for this separation is shown in Figure~\ref{fig:separation} in the Appendix.

\section{Experiments}
\label{sec:exp}

In this section, we first empirically examine how prediction scaling affects average- and last-iterate behavior in normal-form games, then evaluate whether half prediction improves the practical performance of $\CFR$ on standard extensive-form benchmark games.

\subsection{Numerical Separations under Prediction Scaling}
\label{sec:separation_last_avg}

The preceding results in Section \ref{sec:no_lambda_except_half} identify $\lambda=1/2$ as the unique fixed nonnegative prediction multiplier satisfying Alt-RVU on every bounded utility sequence in every finite dimension, and therefore rule out both Alt-RVU and its refinement for all other fixed nonnegative prediction multipliers. We now examine whether prediction scaling produces separations in actual alternating self-play.

\begin{figure}[!h]
    \centering
    \includegraphics[width=\linewidth]{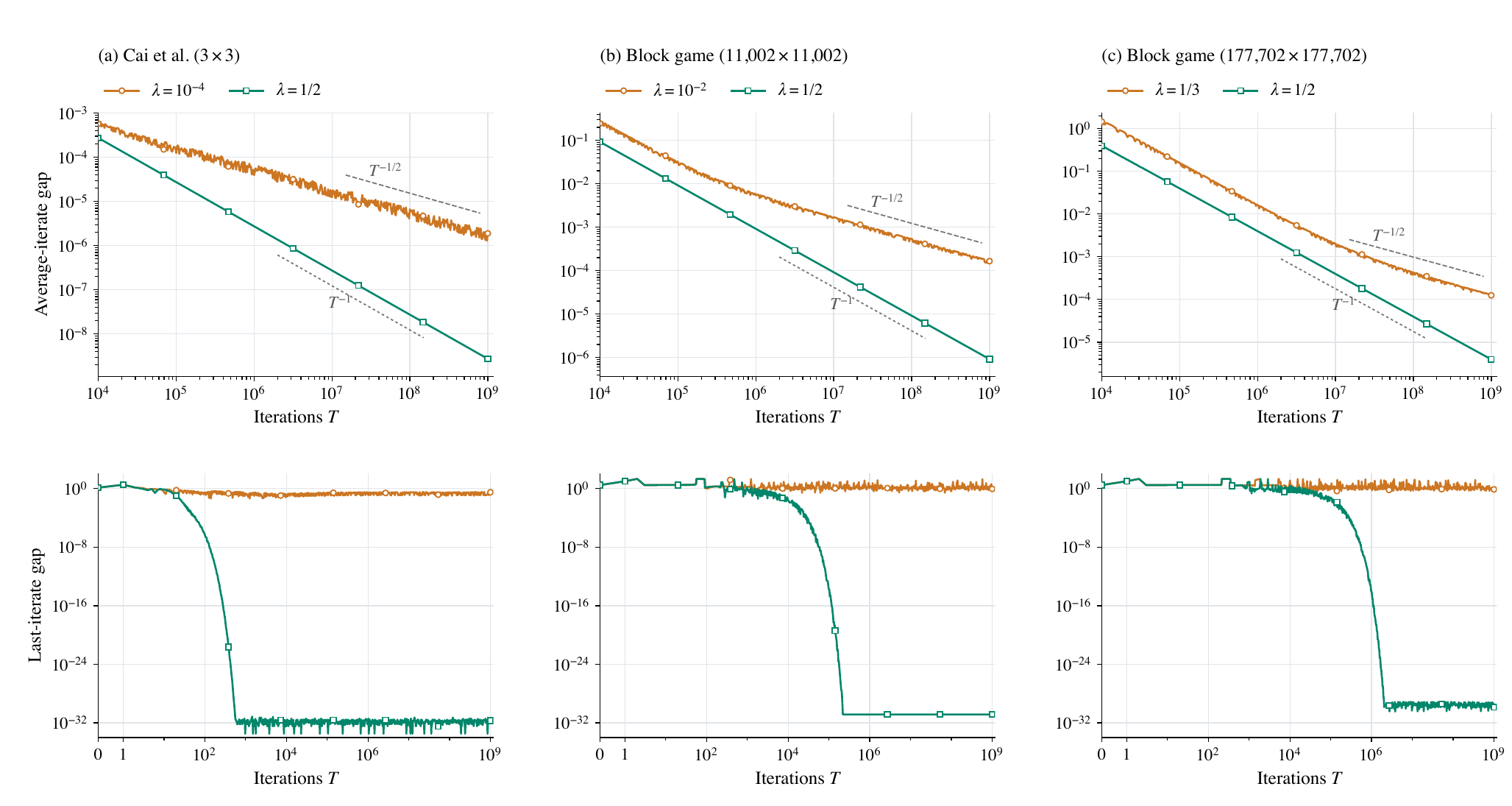}
    \caption{Average-iterate and last-iterate gaps under alternating self-play. Half prediction is compared with $\lambda=10^{-4}$ on the $3\times3$ game of \citet{Cai25:Last} (with the payoff matrix negated to match our maximizing-row convention), $\lambda=10^{-2}$ on a $11,002 \times 11,002$ block game, and $\lambda=1/3$ on a $177,702  \times 177,702$ block game. The gray lines indicate the reference slopes $T^{-1/2}$ and $T^{-1}$.}
    \label{fig:average_last_separation}
\end{figure}

Across the three games in Figure~\ref{fig:average_last_separation}, the choices $\lambda=10^{-4}$, $\lambda=10^{-2}$, and $\lambda=1/3$, respectively, exhibit uniform average-iterate gap decay approximately proportional to $T^{-1/2}$ at late iterations, whereas half prediction exhibits $T^{-1}$ decay on the same games. The last-iterate gaps reveal a further separation: for the tested non-half multipliers, they remain large and oscillatory through $10^9$ iterations, while half prediction reduces them to the numerical precision limit.

The comparison with $\lambda=1/3$ is particularly noteworthy because this is the prediction multiplier used by the local learner underlying $\SAPCFRplus$~\citep{Meng26:Faster}. Their analysis does not establish convergence under alternating self-play, and our numerical results suggest a suboptimal $T^{-1/2}$ average-iterate rate and possible last-iterate nonconvergence for this prediction choice on the third game, in contrast to our convergence guarantees for half prediction.

That being said, these experiments do not establish asymptotic lower bounds or last-iterate nonconvergence. Our numerical comparison here parallels the $3\times3$ example of \citet[Section~3]{Cai25:Last}, who report numerical evidence of last-iterate nonconvergence for $\SPRMplus$ and convergence for $\APRMplus$ on the same game. The three games in Figure~\ref{fig:average_last_separation} concern only the tested multipliers below $1/2$; whether analogous separations occur for $\lambda>1/2$ under alternating self-play remains unresolved, despite the failure of Alt-RVU in that range. The two block games share a simple $3\times3$ integer kernel and differ only in their action multiplicities, which we selected through heuristic numerical search. Appendix~\ref{sec:separation_avg} gives further details on the construction and numerical validation.

\subsection{Performance on Extensive-Form Benchmarks}
\label{sec:benchmark}
We conducted experiments on sizeable, standard extensive-form benchmark games. Our algorithms for normal-form games are incorporated into extensive form using the standard counterfactual regret minimization (CFR) framework~\citep{Zinkevich07:Regret, Farina19:Efficient}, so in the names of the algorithms, we replace $\RM$ with $\CFR$ as is standard practice. We compare the algorithms $\APCFRplus$, $\AHPCFRplus$, $\ASAPCFRplus$ \citep{Meng26:Faster}, $\AIRPCFRplus$, $\APDCFRplus$~\citep{Xu24:Minimizing}, and $\ADCFRplus$~\citep{Brown19:Solving, Xu24:Minimizing}. Here, $\ASAPCFRplus$ uses local learners from the same scaled-prediction family as $\AHPCFRplus$, but with $\lambda=1/3$. The baselines $\APDCFRplus$ and $\ADCFRplus$ are based on the family of discounted regret matching methods (with the constant $\alpha$ set as suggested in those papers). Table~\ref{tab:game-sizes} in the appendix lists the 15 benchmark games and their iteration budgets. All methods use alternating updates, with one iteration consisting of one update by each player. We evaluate the resulting strategy profile after both updates. Further details about the experiments are in Appendix~\ref{sec:exp_details}. 

In this section, $\vx^{(t)}$ and $\vy^{(t)}$ denote realization plans. We report the last iterate, the uniform average, and the HS30 weighted average~\citep{Zhang26:Faster}. For a fixed budget $T$ on the number of iterations, the latter is
\begin{align}
    \bar{\vx}_{\mathrm{HS30}}^{(t)} &\defeq \frac{\sum_{s=1}^{t}w_s^{(t)}\vx^{(s)}} {\sum_{s=1}^{t}w_s^{(t)}}, \; \bar{\vy}_{\mathrm{HS30}}^{(t)} \defeq \frac{\sum_{s=1}^{t}w_s^{(t)}\vy^{(s)}} {\sum_{s=1}^{t}w_s^{(t)}}, \text{ where } w_s^{(t)} \defeq \prod_{k=s+1}^{t} \left(\frac{k-1}{k}\right)^{30-5k/T}.
    \label{eq:hs30_average}
\end{align}
All of the methods use the same averaging rules for computing average strategies. Those rules do not affect their regret-updating rules.

We measure exploitability as half the duality gap, with best responses taken over the extensive-form strategy spaces:
\begin{align*}
    E_{\mathrm{last}}(t) \defeq \frac12\operatorname{Gap}(\vx^{(t)},\vy^{(t)}), \; \ E_{\mathrm{avg}}(t) \defeq \frac12\operatorname{Gap} (\bar{\vx}^{(t)}, \bar{\vy}^{(t)}), \; \ E_{\mathrm{HS30}}(t) \defeq \frac12\operatorname{Gap} (\bar{\vx}_{\mathrm{HS30}}^{(t)}, \bar{\vy}_{\mathrm{HS30}}^{(t)}).
\end{align*}

Across most benchmarks, the HS30 average closely tracks the last iterate and usually achieves lower exploitability. Both generally outperform the uniform average by a substantial margin. We therefore present the last-iterate and HS30 results against iteration count in Figures~\ref{fig:Last_iter} and~\ref{fig:HS30_iter}, respectively, and defer the average-iterate results to Figure~\ref{fig:standard_iter} in Appendix~\ref{sec:exp_add}. Figures~\ref{fig:last_runtime}, \ref{fig:hs_runtime}, and~\ref{fig:unweighted_runtime} in the same appendix report the last-iterate, HS30, average-iterate convergence against learner wall-clock time.

\begin{figure}[!h]
    \centering
    \includegraphics[width=0.75\linewidth]{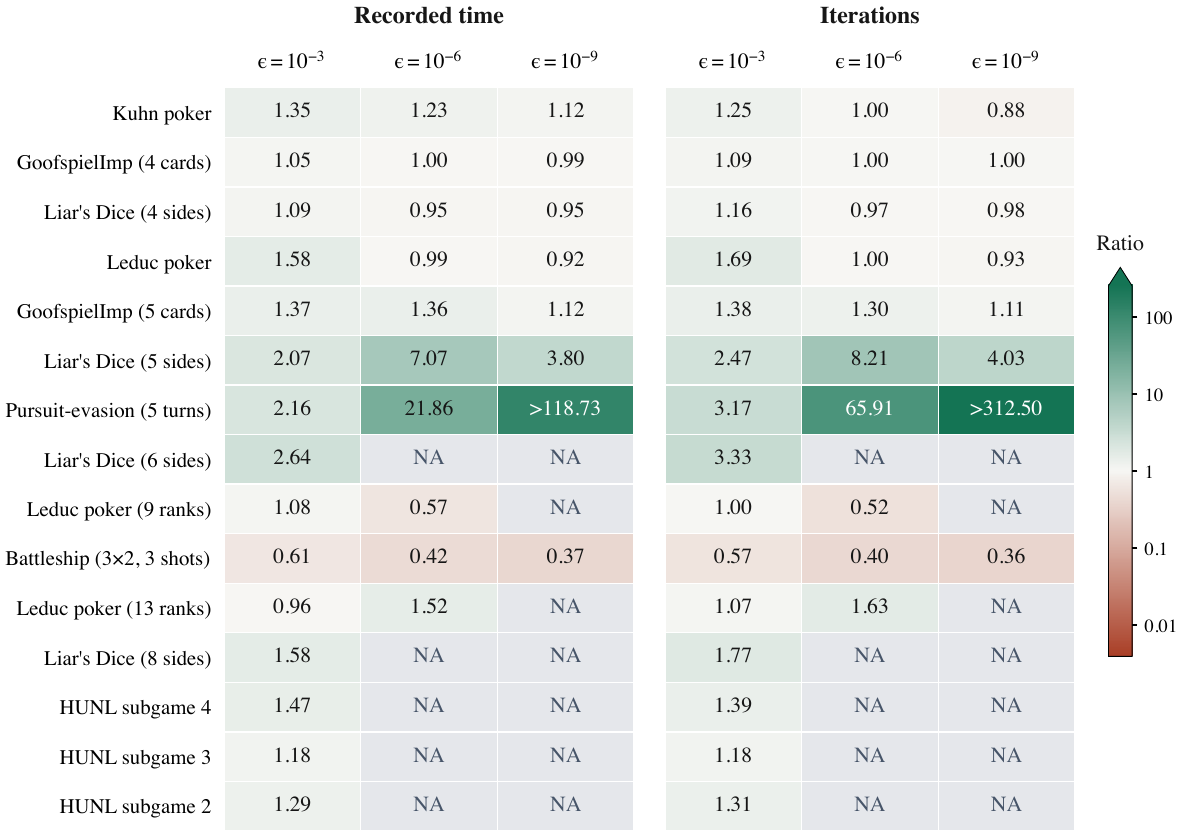}
    \caption{Runtime and iteration ratios ($\APCFRplus/\AHPCFRplus$) to reach target exploitability. Values larger than $1$ mean that $\AHPCFRplus$ outperforms $\APCFRplus$.}    
    \label{fig:heatmap}
\end{figure}

\begin{figure}[!h]
    \centering
    \includegraphics[width=0.9\linewidth]{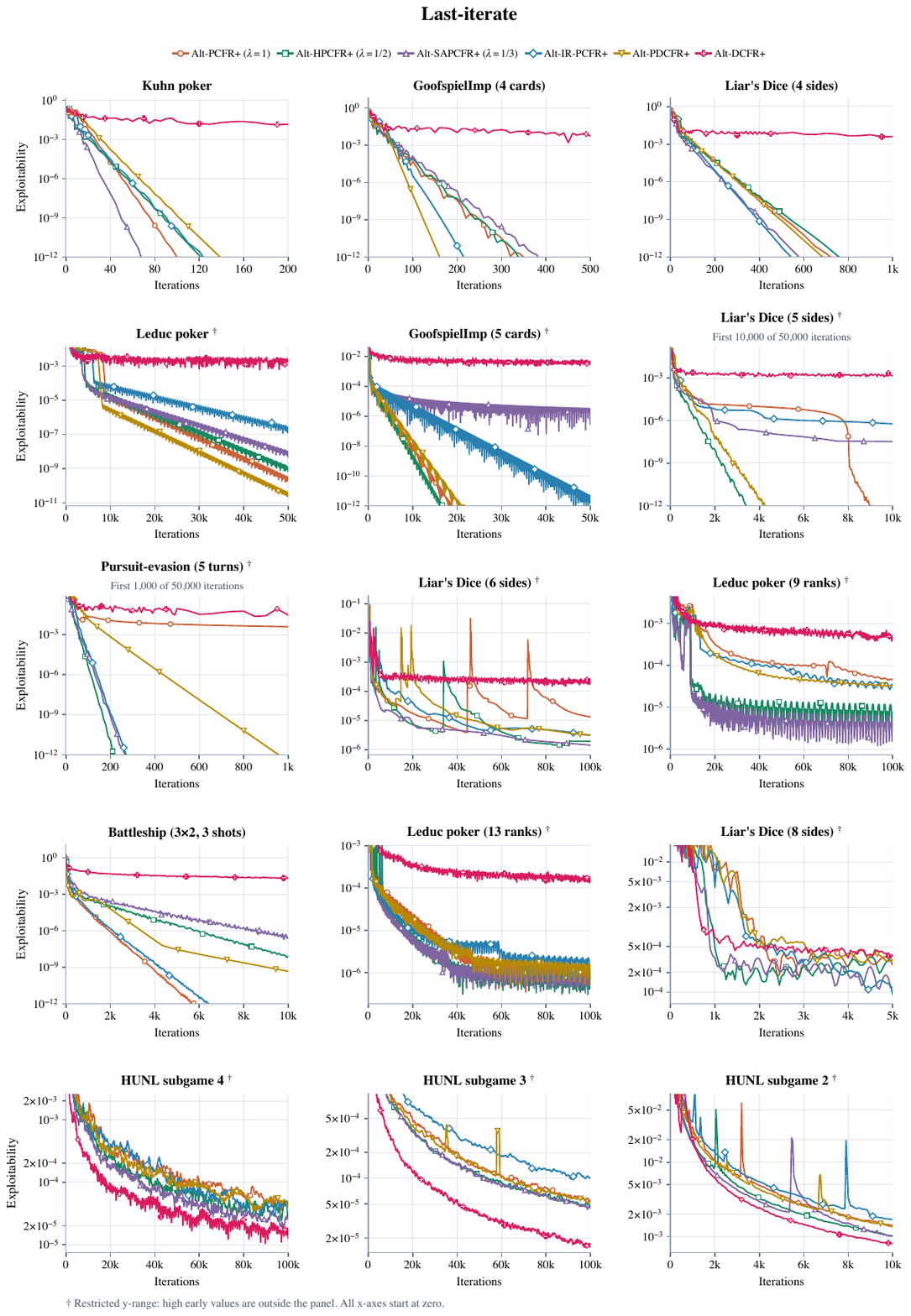}
    \caption{Last-iterate convergence against iteration count.}
    \label{fig:Last_iter}
\end{figure}

\paragraph{$\AHPCFRplus$ vs $\APCFRplus$.} We compare $\AHPCFRplus$ with $\APCFRplus$, which has achieved state-of-the-art empirical performance on the majority of standard non-poker benchmarks~\citep{Farina21:Faster,Chakrabarti24:Extensive,Zhang26:Faster}. For each game, we compare algorithms by the first saved iteration at which either the HS30 average or last iterate reaches exploitability at most $\epsilon$, where $\epsilon\in\{10^{-3}, 10^{-6},10^{-9}\}$. We consider the first saved checkpoint that satisfies
\begin{align}
    \min\left\{E_{\mathrm{HS30}}(t),E_{\mathrm{last}}(t)\right\} \leq \epsilon.
    \label{eq:criterion}
\end{align}
We report the ratio of the iteration count, and separately the recorded runtime, of $\APCFRplus$ to that of $\AHPCFRplus$ for reaching each $\epsilon$ threshold in Figure~\ref{fig:heatmap}. A ratio greater than one shows that $\AHPCFRplus$ outperforms $\APCFRplus$. If neither algorithm reaches the threshold, we report NA for not applicable. Across most games and thresholds, $\AHPCFRplus$ is competitive with $\APCFRplus$ and often reaches the target with substantially fewer iterations and less runtime. The noticeable exceptions are Battleship ($3\times2$, 3 shots) and, at $\epsilon=10^{-6}$, Leduc poker (9 ranks). Across the 33 game-threshold pairs for which at least one algorithm reaches the target, $\AHPCFRplus$ achieves win–loss–draw counts of $20:6:7$ in iteration count and $22:6:5$ in recorded runtime, treating actual ratios that round to $1.0$ at one decimal place as draws and counting sole threshold attainment as a win. Moreover, among the $7$ games where neither algorithm reaches at least one target (\textit{i.e.}, NA), $\AHPCFRplus$ achieves lower exploitability at the end of the run in $6$ games under the same criterion \eqref{eq:criterion}. The only exception is Leduc poker (9 ranks).

\paragraph{All 6 algorithms comparisons.} We compare all six algorithms using the same criterion above \eqref{eq:criterion}, ranking them by recorded runtime and, separately, iteration count to reach the exploitability $\epsilon\in\{10^{-3}, 10^{-6},10^{-9}\}$. We retain the 33 game-threshold pairs for which at least three algorithms reach the target. To handle ties, we group algorithms relative to the fastest remaining algorithm whenever their actual cost ratios round to $1.0$ at one decimal place, and assign ranks such as $1,1,3$. Algorithms that do not reach the target receive no placement. Table~\ref{tab:top_placements} reports the number of top-one and top-three placements.  We include all tied algorithms in the top-three count, even if more than three qualify, and report separately how often each algorithm finishes first alone or ties for first place.

\begin{table*}[t]
    \centering
    \setlength{\tabcolsep}{8pt}
    \begin{tabular}{lcccc}
        \toprule
        & \multicolumn{2}{c}{Recorded runtime}
        & \multicolumn{2}{c}{Iteration count} \\
        \cmidrule(lr){2-3}\cmidrule(lr){4-5}
        Algorithm & Top 1 & Top 3 & Top 1 & Top 3 \\
        \midrule
        $\AHPCFRplus$
        & \textbf{14} (11 sole, 3 ties) & \textbf{30}
        & \textbf{10} (9 sole, 1 tie) & \textbf{28} \\
        $\ASAPCFRplus$
        & 10 (7 sole, 3 ties) & 24
        & \textbf{10} (6 sole, 4 ties) & 22 \\
        $\APCFRplus$
        & 5 (4 sole, 1 tie) & 21
        & 3 (3 sole, 0 ties) & 19 \\
        $\APDCFRplus$
        & 3 (3 sole, 0 ties) & 11
        & 6 (6 sole, 0 ties) & 19 \\
        $\AIRPCFRplus$
        & 0 (0 sole, 0 ties) & 9
        & 2 (0 sole, 2 ties) & 17 \\
        $\ADCFRplus$
        & 5 (5 sole, 0 ties) & 8
        & 6 (5 sole, 1 tie) & 8 \\
        \bottomrule
    \end{tabular}
    \caption{Top-one and top-three placements across the 33 game-threshold comparisons.}
    \label{tab:top_placements}
\end{table*}

Regarding $\AIRPCFRplus$, its performance is weaker in recorded runtime than in iteration count, consistent with the additional per-iteration computations required by the IR mechanism. Meanwhile, the discounted method $\ADCFRplus$ remains the fastest method on the tested HUNL subgames, consistent with previous findings that show that the DCFR family of algorithms is particularly well suited for poker games~\citep{Farina21:Faster,Xu24:Minimizing}. However, its lowest number of top-three placements in both recorded runtime and iteration count indicates less consistent and robust performance across the broader benchmark suite. The predictive variant $\APDCFRplus$ from the same discounted family achieves more top-three placements than $\ADCFRplus$, but remains less consistently competitive in recorded runtime than the scaled-prediction family under our evaluation protocol.


Within the scaled-prediction family, $\APCFRplus$ remains consistently competitive, placing in the top three frequently in both recorded runtime and iteration count. This agrees with its strong empirical performance reported in the literature~\citep{Farina21:Faster,Chakrabarti24:Extensive,Zhang26:Faster}. Nevertheless, $\AHPCFRplus$ and $\ASAPCFRplus$ achieve more first-place and top-three finishes in both measures, showing that scaling the prediction can further improve speed and consistency across the tested benchmarks. Although $\ASAPCFRplus$ is competitive with $\AHPCFRplus$, the analysis of \citet{Meng26:Faster} does not establish convergence for its local learner under alternating self-play, even in normal-form games. Moreover, our numerical results in Subsection~\ref{sec:separation_last_avg} suggest a suboptimal $T^{-1/2}$ average-iterate rate and possible last-iterate nonconvergence for its prediction multiplier $\lambda=1/3$. Overall, $\AHPCFRplus$ achieves the strongest aggregate performance under our evaluation protocol. Half prediction is also, to our knowledge, the only choice among fixed positive prediction multipliers with established convergence guarantees for alternating self-play in every finite two-player zero-sum normal-form game.

\begin{figure}[!h]
    \centering
    \includegraphics[width=0.9 \linewidth]{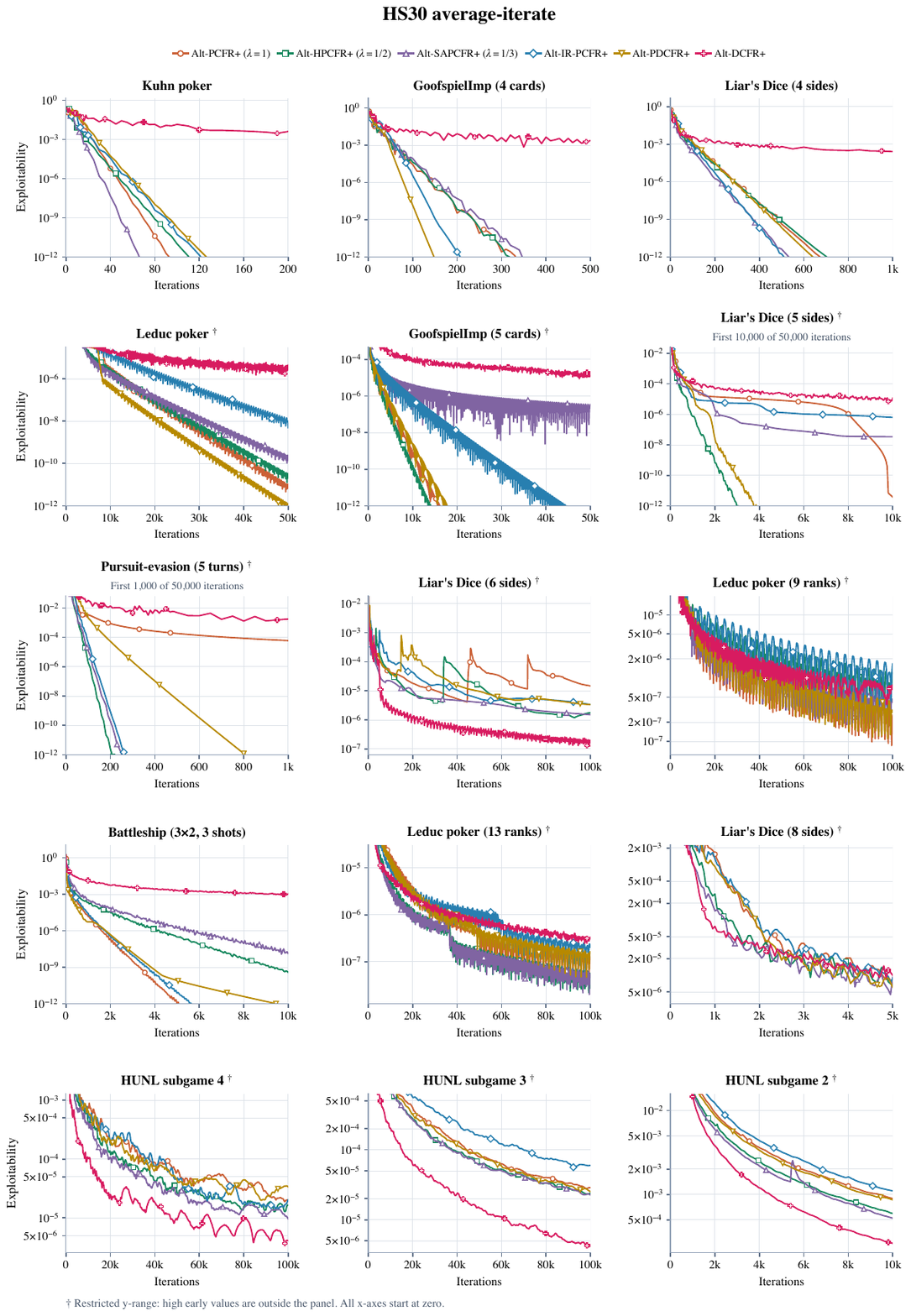}
    \caption{HS30 average convergence against iteration count.}
    \label{fig:HS30_iter}
\end{figure}

\section{Conclusions}
We developed Alt-RVU and its refinement to connect conditions on individual learners with the convergence of alternating self-play. For $\AIRPRMplus$ and $\AHPRMplus$, this framework yields $O(T^{-1})$ average-iterate and $O(T^{-1/2})$ best-iterate rates in every finite two-player zero-sum normal-form game. Further analysis of their updates establishes convergence to a single Nash equilibrium without assuming uniqueness.

Our analysis of half prediction shows that fast convergence can be obtained by controlling the cumulative effect of regret-norm decreases without preventing those decreases through additional computation. Our impossibility results further identify half prediction as the unique fixed nonnegative multiplier satisfying Alt-RVU on every bounded utility sequence in every finite dimension. 

Our separations show that both the regret quantity and the update schedule matter. The $\HPRMplus$ learner fails the standard RVU bound despite achieving the optimal $O(T^{-1})$ average-iterate convergence rate under alternating self-play, demonstrating the importance of our Alt-RVU. Separately, on the same matrix game with the same standard initialization, alternating half prediction converges to equilibrium while simultaneous half prediction maintains a duality gap asymptotically bounded away from zero. Thus, update order can determine whether the individual strategy profiles approach equilibrium at all.

Half prediction also achieves strong practical performance. Its $\CFR$ instantiation has the strongest aggregate performance among the six tested solvers across 15 extensive-form benchmarks under our evaluation protocol. Complementary numerical examples show slower average-iterate decay and persistent last-iterate oscillations for selected smaller prediction multipliers, suggesting that prediction scaling can affect more than rate constants. 
\section{AI Disclosure}

The construction in Proposition \ref{prop:failure_above_half_explicit} was inspired by an idea from a chat with GPT 6 Astra in which the AI suggested a construction that allows the regret vectors and utilities to shrink while the strategies oscillate. GPT-6 Astra also assisted in developing the numerical search procedure for the $3\times3$ kernel matrix used in the example games in Section \ref{sec:separation_last_avg}, following the authors' heuristic. The same AI was also used to optimize the code for the experiments.

\section{Acknowledgments}
The work presented in this paper was supported by Vannevar Bush Faculty Fellowship ONR N00014-23-1-2876 and National Science Foundation grant RI-2312342. Any opinions, findings, conclusions, or recommendations expressed in this material are those of the author(s) and do not necessarily reflect the views of the funding agencies.

\newpage
\bibliographystyle{plainnat}
\bibliography{dairefs}

\appendix
\newpage


\section{Omitted Proofs}

\subsection{Proofs from Subsection \ref{sec:alt_rvu}}
\altRvuGap*
\begin{proof}
    Since each player's utility vector depends linearly on the opponent's strategy, the operator norm bound gives
\begin{align}
    \sum_{t=2}^T\|\vu_{\cX}^{(t)}-\vu_{\cX}^{(t-1)}\|_\ast^2 \le \|\mA\|^2\sum_{t=2}^T \|\vy^{(t)}-\vy^{(t-1)}\|^2, \: \sum_{t=2}^T\|\vu_{\cY}^{(t)}-\vu_{\cY}^{(t-1)}\|_\ast^2 \le \|\mA\|^2\sum_{t=1}^{T-1} \|\vx^{(t+1)}-\vx^{(t)}\|^2.
    \label{eq:util_bound_from_strat}
\end{align}
We can sum the Alt-RVU bounds of the two players and use the identity
\eqref{eq:alt_payoff_telescope} to obtain
\begin{align*}
&2T\operatorname{Gap}(\bar\vx^{(T)},\bar\vy^{(T)}) + (\gamma_{\cX} - \|\mA\|^2\beta_{\cY}) \sum_{t=1}^T\|\vx^{(t+1)}-\vx^{(t)}\|^2 + (\gamma_{\cY}-\|\mA\|^2\beta_{\cX}) \sum_{t=1}^T\|\vy^{(t)}-\vy^{(t-1)}\|^2 \\
&\le \alpha_{\cX} + \alpha_{\cY} + \operatorname{range}(\mA).
\end{align*}
Every LHS term is nonnegative due the condition  $\|\mA\|^2 \beta_{\cal X} < \gamma_{\cal Y}$ and $\|\mA\|^2 \beta_{\cal Y} < \gamma_{\cal X}$. We can drop the nonnegative terms to obtain the desired average-iterate convergence
\begin{align*}
        \operatorname{Gap} (\bar{\vx}^{(T)}, \bar{\vy}^{(T)}) \le \frac{\alpha_{\cal X} + \alpha_{\cal Y} + \operatorname{range}(\mA)}{2 T}.
    \end{align*}
We can also drop the $2T\operatorname{Gap}(\bar\vx^{(T)},\bar\vy^{(T)})$ term to obtain
\begin{align*}
    \sum_{t=1}^T \|\vx^{(t+1)}-\vx^{(t)}\|^2 + \sum_{t=1}^T\|\vy^{(t)}-\vy^{(t-1)}\|^2 \le \frac{\alpha_{\cX}+\alpha_{\cY} + \operatorname{range}(\mA)}{\min\{\gamma_{\cX}-\|\mA\|^2\beta_{\cY}, \gamma_{\cY}-\|\mA\|^2\beta_{\cX}\}},
\end{align*}
which implies that the squared strategy movement sum is finite. This also implies from the bound \eqref{eq:util_bound_from_strat} that the squared utility variation sum is finite. 
\end{proof}

\altRegBoundSimple*
\begin{proof}
Inner producting both sides of \eqref{eq:ineq_tilde_r} with $\vz$, we obtain
\begin{align*}
    \ip{\tilde{\vr}^{(t+1)}, \vz} \ge \ip{\tilde{\vr}^{(t)}, \vz} + \ip{\vu^{(t)}, \vz} - \langle \vu^{(t)}, \vx^{(t)} \rangle \ip{\mathbf1, \vz} = \ip{\tilde{\vr}^{(t)}, \vz} + \ip{\vu^{(t)}, \vz} - \langle \vu^{(t)}, \vx^{(t)} \rangle,
\end{align*}
which implies
\begin{align*}
    \ip{\tilde{\vr}^{(t+1)} - \tilde{\vr}^{(t)}, \vz} \ge \ip{\vu^{(t)}, \vz - \vx^{(t)}}.
\end{align*}
Consequently, each summand in alternating regret obeys
\begin{align*}
    \langle \vu^{(t)}, 2\vz - \vx^{(t)} - \vx^{(t+1)} \rangle \le 2\langle\tilde{\vr}^{(t+1)}-\tilde{\vr}^{(t)}, \vz\rangle -\langle \vu^{(t)}, \vx^{(t+1)} - \vx^{(t)}\rangle.
\end{align*}
For the initialization $\tilde{\vr}^{(1)}=0$, we can sum both sides from $1$ to $T$ to get
\begin{align*}
    \sum_{t=1}^T \langle \vu^{(t)}, 2\vz - \vx^{(t)} - \vx^{(t+1)} \rangle \le 2\langle\tilde{\vr}^{(T+1)}, \vz\rangle - \sum_{t=1}^T \langle \vu^{(t)}, \vx^{(t+1)} - \vx^{(t)}\rangle \le 2 \|\tilde{\vr}^{(T+1)} \|_2 - \sum_{t=1}^T \langle \vu^{(t)}, \vx^{(t+1)} - \vx^{(t)}\rangle,
\end{align*}
where the last inequality follows from the fact that $\|\vz\|_2 \le 1$. Maximizing over $\vz\in\Delta(d)$, we obtain the desired inequality.
\end{proof}

\utilImprovementPrm*
\begin{proof}
From $\PRMplus$ updates \eqref{eq:reg_vec_update_prm} and \eqref{eq:prediction_update_prm}, we have the following two identities
\begin{align*}
    \ip{\vu^{(t)}-\ip{\vu^{(t)},\vx^{(t)}}\vec1,\tilde\vr^{(t+1)}} &= \ip{\tilde\vr^{(t+1)}-\tilde\vr^{(t)},\tilde \vr^{(t+1)}}, \\
    \vu^{(t)} - \langle \vu^{(t)}, \vx^{(t)} \rangle \vec1 &=([\vr^{(t+1)}]_+ -\tilde{\vr}^{(t+1)})
+(\vr^{(t+1)}-[\vr^{(t+1)}]_+).
\end{align*}
where the first equality holds because, on coordinates where \(\tilde{\vr}^{(t+1)}\) is positive, the regret vector update is untruncated; on the remaining coordinates, both inner products contribute zero. For the first expression, we can expand its RHS to get
\begin{align}
    \ip{\vu^{(t)}-\ip{\vu^{(t)},\vx^{(t)}}\vec1,\tilde\vr^{(t+1)}} = \frac{\|\tilde\vr^{(t+1)}\|_2^2-\|\tilde\vr^{(t)}\|_2^2+\|\tilde\vr^{(t+1)}-\tilde\vr^{(t)}\|_2^2}{2}.
    \label{eq:useful_identity}
\end{align}
For the second expression, we can take the inner product with \([\vr^{(t+1)}]_+-\tilde{\vr}^{(t+1)}\) for both sides to obtain
\begin{align*}
&\langle \vu^{(t)} - \langle \vu^{(t)}, \vx^{(t)}\rangle \vec1,[\vr^{(t+1)}]_+ - \tilde{\vr}^{(t+1)} \rangle \\
&= \|[\vr^{(t+1)}]_+ - \tilde{\vr}^{(t+1)}\|_2^2 + \langle \vr^{(t+1)}-[\vr^{(t+1)}]_+,[\vr^{(t+1)}]_+ \rangle - \langle \vr^{(t+1)} - [\vr^{(t+1)}]_+, \tilde{\vr}^{(t+1)} \rangle \\
&= \|[\vr^{(t+1)}]_+ - \tilde{\vr}^{(t+1)}\|_2^2 - \langle \vr^{(t+1)} - [\vr^{(t+1)}]_+, \tilde{\vr}^{(t+1)} \rangle \\
&\ge \|[\vr^{(t+1)}]_+-\tilde\vr^{(t+1)}\|_2^2,
\end{align*}
where the inequality follows from the fact that  $\vr^{(t+1)}-[\vr^{(t+1)}]_+\le\vec0$ and $\tilde\vr^{(t+1)}\ge\vec0$.

Adding this inequality to the above identity for $\ip{\vu^{(t)}-\ip{\vu^{(t)},\vx^{(t)}}\vec1,\tilde\vr^{(t+1)}}$, we obtain
\begin{align*}
    \ip{\vu^{(t)} - \ip{\vu^{(t)}, \vx^{(t)}}\vec1,[\vr^{(t+1)}]_+} \ge \frac{\|\tilde\vr^{(t+1)}\|_2^2 -\|\tilde\vr^{(t)}\|_2^2 + \|\tilde\vr^{(t+1)}-\tilde\vr^{(t)}\|_2^2}{2} + \|[\vr^{(t+1)}]_+ -\tilde\vr^{(t+1)}\|_2^2,
\end{align*}
which implies
\begin{align*}
    \frac{\ip{\vu^{(t)} - \ip{\vu^{(t)}, \vx^{(t)}}\vec1,[\vr^{(t+1)}]_+} }{\|[\vr^{(t+1)}]_+\|_1} \ge \frac{\|\tilde\vr^{(t+1)}\|_2^2 -\|\tilde\vr^{(t)}\|_2^2 + \|\tilde\vr^{(t+1)}-\tilde\vr^{(t)}\|_2^2 + 2 \|[\vr^{(t+1)}]_+ -\tilde\vr^{(t+1)}\|_2^2}{2 \|[\vr^{(t+1)}]_+\|_1}.
\end{align*}
Since $\vx^{(t+1)} = [\vr^{(t+1)}]_+/\|[\vr^{(t+1)}]_+\|_1$, we can rewrite the LHS as
\begin{align*}
    \frac{\ip{\vu^{(t)} - \ip{\vu^{(t)}, \vx^{(t)}} \vec1, [\vr^{(t+1)}]_+}}{\|[\vr^{(t+1)}]_+\|_1} = \ip{\vu^{(t)} - \ip{\vu^{(t)}, \vx^{(t)}} \vec1, \vx^{(t + 1)}} = \ip{\vu^{(t)},\vx^{(t+1)}-\vx^{(t)}}.
\end{align*}
Hence, we obtain the desired claim.
\end{proof}
\terminalTildeRBoundPrm*
\begin{proof}
For $t \ge 2$, we use identity \eqref{eq:useful_identity} established in the proof of Lemma~\ref{eq:util_improvement_prm} to obtain
\begin{align*}
    \| \tilde \vr^{(t + 1)} \|_2^2 - \| \tilde \vr^{(t)} \|_2^2 = 2 \ip{\vu^{(t)} - \ip{\vu^{(t)}, \vx^{(t)}} \vec1, \tilde \vr^{(t + 1)} - [\vr^{(t)}]_+} - \| \tilde \vr^{(t + 1)} - \tilde \vr^{(t)} \|_2^2.
\end{align*}
Substituting $\vr^{(t)} = \tilde \vr^{(t)} + \vu^{(t - 1)} - \ip{\vu^{(t - 1)}, \vx^{(t - 1)}} \vec1$ and expanding the squared norm $\| \tilde \vr^{(t + 1)} - \tilde \vr^{(t)} \|_2^2$, we get
\begin{align*}
    \| \tilde \vr^{(t + 1)} \|_2^2 - \| \tilde \vr^{(t)} \|_2^2 &= 2 \ip{\vu^{(t)} - \vu^{(t - 1)} - ( \ip{\vu^{(t)}, \vx^{(t)}} - \ip{\vu^{(t - 1)}, \vx^{(t - 1)}} ) \vec1, \tilde \vr^{(t + 1)} - [\vr^{(t)}]_+} \\
    & - \| \tilde \vr^{(t + 1)} - [\vr^{(t)}]_+ \|_2^2 - \| [\vr^{(t)}]_+ - \tilde \vr^{(t)} \|_2^2 + 2 \ip{\vr^{(t)} - [\vr^{(t)}]_+, \tilde \vr^{(t + 1)} - [\vr^{(t)}]_+}\\
    &\le 2 \ip{\vu^{(t)} - \vu^{(t - 1)} - ( \ip{\vu^{(t)}, \vx^{(t)}} - \ip{\vu^{(t - 1)}, \vx^{(t - 1)}} ) \vec1, \tilde \vr^{(t + 1)} - [\vr^{(t)}]_+} \\
    &- \| \tilde \vr^{(t + 1)} - [\vr^{(t)}]_+ \|_2^2 - \| [\vr^{(t)}]_+ - \tilde \vr^{(t)} \|_2^2 \\
    &\le \left\| \vu^{(t)} - \vu^{(t - 1)} - ( \ip{\vu^{(t)}, \vx^{(t)}} - \ip{\vu^{(t - 1)}, \vx^{(t - 1)}} ) \vec1 \right\|_2^2,
\end{align*}
where the first inequality follows from $\ip{\vr^{(t)} - [\vr^{(t)}]_+, \tilde \vr^{(t + 1)} - [\vr^{(t)}]_+} = \ip{\vr^{(t)} - [\vr^{(t)}]_+, \tilde \vr^{(t + 1)}} \le 0$, and the second inequality follows by completing the square and discarding $- \| [\vr^{(t)}]_+ - \tilde \vr^{(t)} \|_2^2$.

Using $\| \vx^{(t)} \|_2 \le 1$ and $\| \vu^{(t - 1)} \|_2 \le \| \mA \|_2$, we further obtain
\begin{align*}
    &\left\| \vu^{(t)} - \vu^{(t - 1)} - \bigl( \ip{\vu^{(t)}, \vx^{(t)}} - \ip{\vu^{(t - 1)}, \vx^{(t - 1)}} \bigr) \vec1 \right\|_2
    \\
    &\le \| \vu^{(t)} - \vu^{(t - 1)} \|_2 + \sqrt d \, \left| \ip{\vu^{(t)} - \vu^{(t - 1)}, \vx^{(t)}} \right| + \sqrt d \, \left| \ip{\vu^{(t - 1)}, \vx^{(t)} - \vx^{(t - 1)}} \right| \\
    &\le (1 + \sqrt d) \| \vu^{(t)} - \vu^{(t - 1)} \|_2 + \sqrt d \, \| \mA \|_2 \| \vx^{(t)} - \vx^{(t - 1)} \|_2.
\end{align*}
This implies that
\begin{align*}
    \| \tilde \vr^{(t + 1)} \|_2^2 - \| \tilde \vr^{(t)} \|_2^2 &\le \left( (1 + \sqrt d) \| \vu^{(t)} - \vu^{(t - 1)} \|_2 + \sqrt d \, \| \mA \|_2 \| \vx^{(t)} - \vx^{(t - 1)} \|_2 \right)^2 \\
     &\le 2 (1 + \sqrt d)^2 \| \vu^{(t)} - \vu^{(t - 1)} \|_2^2 + 2d \| \mA \|_2^2 \| \vx^{(t)} - \vx^{(t - 1)} \|_2^2. 
\end{align*}
At initialization, $\tilde \vr^{(1)} = \vec0$ and $\vx^{(1)} = \vec1 / d$, so
\begin{align*}
    \| \tilde \vr^{(2)} \|_2^2 \le \left\| \vu^{(1)} - \frac{\ip{\vu^{(1)}, \vec1}}{d} \vec1 \right\|_2^2 = \| \vu^{(1)} \|_2^2 - \frac{\ip{\vu^{(1)}, \vec1}^2}{d} \le \| \mA \|_2^2.
\end{align*}
For each $2 \le s \le T + 1$, we now sum the norm increments:
\begin{align*}
    \| \tilde \vr^{(s)} \|_2^2
    &= \| \tilde \vr^{(2)} \|_2^2 + \sum_{t = 2}^{s - 1} \left( \| \tilde \vr^{(t + 1)} \|_2^2 - \| \tilde \vr^{(t)} \|_2^2 \right) \\
    &\le \| \mA \|_2^2 + 2 (1 + \sqrt d)^2 \sum_{t = 2}^{s - 1} \| \vu^{(t)} - \vu^{(t - 1)} \|_2^2 + 2d \| \mA \|_2^2 \sum_{t = 1}^{s - 2} \| \vx^{(t + 1)} - \vx^{(t)} \|_2^2 \\
    &\le \| \mA \|_2^2 + 2 (1 + \sqrt d)^2 \sum_{t = 2}^{T} \| \vu^{(t)} - \vu^{(t - 1)} \|_2^2 + 2d \| \mA \|_2^2 \sum_{t = 1}^{T} \| \vx^{(t + 1)} - \vx^{(t)} \|_2^2,
\end{align*}
where we reindex the strategy-movement sum the first inequality, and extend the nonnegative sums for the second. The resulting bound is independent of $s$ and also holds for $s = 1$, since $\tilde \vr^{(1)} = \vec0$. Taking the maximum over $1 \le s \le T + 1$ gives us the desired claim.
\end{proof}
\movementBoundPrm*
\begin{proof}
We first show the normalization bound
\begin{align}
    \max\{\|[\vr^{(t)}]_+\|_1,\|[\vr^{(t+1)}]_+\|_1\}\|\vx^{(t+1)}-\vx^{(t)}\|_2 \le \sqrt d\,\|[\vr^{(t+1)}]_+-[\vr^{(t)}]_+\|_2.
    \label{eq:normalization_bound}
\end{align}
Indeed, if $\|[\vr^{(t+1)}]_+\|_1\ge\|[\vr^{(t)}]_+\|_1$, we can write the LHS as
\begin{align*}
    \|[\vr^{(t+1)}]_+\|_1(\vx^{(t+1)}-\vx^{(t)}) &= \|[\vr^{(t+1)}]_+\|_1 \left( \frac{[\vr^{(t+1)}]_+}{\|[\vr^{(t+1)}]_+\|_1} - \vx^{(t)} \right) \\
    &= [\vr^{(t+1)}]_+ - \|[\vr^{(t+1)}]_+\|_1 \vx^{(t)}\\
    &= (I - \vx^{(t)} \vec1^\top)[\vr^{(t+1)}]_+.
\end{align*}
Since $[\vr^{(t)}]_+ = \|[\vr^{(t)}]_+\|_1 \vx^{(t)}$, we note that
\begin{align*}
    (I - \vx^{(t)} \vec1^\top) [\vr^{(t)}]_+ = [\vr^{(t)}]_+ - \vx^{(t)} \| [\vr^{(t)}]_+ \|_1 = [\vr^{(t)}]_+ - [\vr^{(t)}]_+ = 0.
\end{align*}
As $\|I-\vx^{(t)}\vec1^\top\|_2\le\sqrt d$, we obtain the above normalization bound. The other case is symmetric.

We now decompose the prediction difference as
\begin{align*}
    [\vr^{(t + 1)}]_+ - [\vr^{(t)}]_+ = ( [\vr^{(t + 1)}]_+ - \tilde \vr^{(t + 1)} ) + ( \tilde \vr^{(t + 1)} - \tilde \vr^{(t)} ) + ( \tilde \vr^{(t)} - [\vr^{(t)}]_+ ).
\end{align*}
Applying the normalization bound and then Cauchy--Schwarz, we obtain
\begin{align*}
    \frac{\max \{ \| [\vr^{(t)}]_+ \|_1, \| [\vr^{(t + 1)}]_+ \|_1 \} \| \vx^{(t + 1)} - \vx^{(t)} \|_2^2}{6d} 
    &\le \frac{\| [\vr^{(t + 1)}]_+ - [\vr^{(t)}]_+ \|_2^2}{6 \max \{ \| [\vr^{(t)}]_+ \|_1, \| [\vr^{(t + 1)}]_+ \|_1 \}} \\
    &\le \frac{\| \tilde \vr^{(t + 1)} - \tilde \vr^{(t)} \|_2^2 + \| [\vr^{(t + 1)}]_+ - \tilde \vr^{(t + 1)} \|_2^2 + \| [\vr^{(t)}]_+ - \tilde \vr^{(t)} \|_2^2}{2 \max \{ \| [\vr^{(t)}]_+ \|_1, \| [\vr^{(t + 1)}]_+ \|_1 \}} \\
    &\le \frac{\| \tilde \vr^{(t + 1)} - \tilde \vr^{(t)} \|_2^2 + \| [\vr^{(t + 1)}]_+ - \tilde \vr^{(t + 1)} \|_2^2}{2 \| [\vr^{(t + 1)}]_+ \|_1} + \frac{\| [\vr^{(t)}]_+ - \tilde \vr^{(t)} \|_2^2}{2 \| [\vr^{(t)}]_+ \|_1},
\end{align*}
where we replace the common denominator by the corresponding smaller norms for the last inequality.

We can sum the last term for $t=t_0, \ldots, T$ and rearrange
\begin{align*}
    \sum_{t=t_0}^{T} \frac{\| [\vr^{(t)}]_+ - \tilde\vr^{(t)} \|_2^2}{2\|[\vr^{(t)}]_+\|_1} &= \sum_{t=t_0}^{T} \frac{\|[\vr^{(t+1)}]_+ - \tilde\vr^{(t+1)}\|_2^2}{2\|[\vr^{(t+1)}]_+\|_1} + \frac{\|[\vr^{(t_0)}]_+ - \tilde\vr^{(t_0)}\|_2^2}{2 \|[\vr^{(t_0)}]_+\|_1} - \frac{\|[\vr^{(T+1)}]_+ - \tilde\vr^{(T+1)}\|_2^2}{2 \|[\vr^{(T+1)}]_+\|_1} \\
    &\le \sum_{t=t_0}^{T} \frac{\|[\vr^{(t+1)}]_+ - \tilde\vr^{(t+1)}\|_2^2}{2\|[\vr^{(t+1)}]_+\|_1} + \frac{\|[\vr^{(t_0)}]_+ - \tilde\vr^{(t_0)}\|_2^2}{2 \|[\vr^{(t_0)}]_+\|_1}.
\end{align*}
Substituting this identity and discarding the final nonpositive term, we obtain the claimed inequality.
\end{proof}
\subsection{Proofs from Subsection \ref{sec:ir_alt_rvu}}
\utilImprovementIr*
\begin{proof}
Suppose first that $\tilde\vr^{(t+1)} \neq \vec0$. The IR regret vector update gives
\begin{align*}
    \tilde\vr^{(t+1)} \ge \tilde\vr^{(t)} + \vu^{(t)} - \ip{\vu^{(t)}, \vx^{(t)}} \vec1 + c^{(t)} \vec1.
\end{align*}
Taking the inner product with $[\vr^{(t)}]_+$, we obtain
\begin{align*}
    \ip{\tilde\vr^{(t+1)}, [\vr^{(t)}]_+}
    &\ge \ip{\tilde\vr^{(t)}, [\vr^{(t)}]_+} + \ip{\vu^{(t)} - \ip{\vu^{(t)}, \vx^{(t)}} \vec1, [\vr^{(t)}]_+} + c^{(t)} \|[\vr^{(t)}]_+\|_1 \\
    &= \ip{\vr^{(t)}, [\vr^{(t)}]_+} + \ip{\vu^{(t)} - \ip{\vu^{(t)}, \vx^{(t)}} \vec1, [\vr^{(t)}]_+},
\end{align*}
where the equality follows from the definitions of $\vr^{(t)}$ and $c^{(t)}$. Since $[\vr^{(t)}]_+= \|[\vr^{(t)}]_+\|_1 \vx^{(t)}$, we can get
\begin{align}
    \ip{\vu^{(t)} - \ip{\vu^{(t)}, \vx^{(t)}} \vec1, [\vr^{(t)}]_+}
    &= \|[\vr^{(t)}]_+\|_1 (\ip{\vu^{(t)}, \vx^{(t)}} - \ip{\vu^{(t)}, \vx^{(t)}} \ip{\vec1, \vx^{(t)}}) \notag \\
    &= \|[\vr^{(t)}]_+\|_1 (\ip{\vu^{(t)}, \vx^{(t)}} - \ip{\vu^{(t)}, \vx^{(t)}}) \notag \\
    &= 0.
    \label{eq:centered_payoff_inner_prod}
\end{align}
Hence, we obtain
\begin{align*}
        \ip{\tilde\vr^{(t+1)}, [\vr^{(t)}]_+} \ge  \ip{\vr^{(t)}, [\vr^{(t)}]_+}  = \|[\vr^{(t)}]_+\|_2^2 = \|\tilde\vr^{(t)}\|_2^2.
\end{align*}
With this inequality, we get
\begin{align}
    \|\tilde\vr^{(t+1)} - [\vr^{(t)}]_+\|_2^2
    &= \|\tilde\vr^{(t+1)}\|_2^2 + \|\tilde\vr^{(t)}\|_2^2 - 2 \ip{\tilde\vr^{(t+1)}, [\vr^{(t)}]_+} \notag \\
    &\le \|\tilde\vr^{(t+1)}\|_2^2 - \|\tilde\vr^{(t)}\|_2^2.
    \label{eq:ir_bound_1}
\end{align}
From the regret vector update, we have the identity
\begin{align*}
    \ip{\vu^{(t)} - \ip{\vu^{(t)}, \vx^{(t)}} \vec1 + c^{(t)} \vec1, \tilde\vr^{(t+1)}} = \ip{\tilde\vr^{(t+1)} - \tilde\vr^{(t)}, \tilde\vr^{(t+1)}}.
\end{align*}
The next prediction satisfies $\vr^{(t+1)} - \tilde\vr^{(t+1)} = \vu^{(t)} - \ip{\vu^{(t)}, \vx^{(t+1)}} \vec1 + c^{(t+1)} \vec1$. Subtracting its inner product with $\tilde\vr^{(t+1)}$ from the preceding identity, we obtain
\begin{align*}
    \|\tilde\vr^{(t+1)}\|_1 (\ip{\vu^{(t)}, \vx^{(t+1)} - \vx^{(t)}} + c^{(t)} - c^{(t+1)})
    &= \ip{\tilde\vr^{(t+1)} - \tilde\vr^{(t)}, \tilde\vr^{(t+1)}} + \ip{\tilde\vr^{(t+1)} - \vr^{(t+1)}, \tilde\vr^{(t+1)}} \\
    &\ge \frac{\|\tilde\vr^{(t+1)}\|_2^2 - \|\tilde\vr^{(t)}\|_2^2 + \|\tilde\vr^{(t+1)} - \tilde\vr^{(t)}\|_2^2 + \|\tilde\vr^{(t+1)} - [\vr^{(t+1)}]_+\|_2^2}{2} \\
    &\overset{\eqref{eq:ir_bound_1}}{\ge}\frac{\|\tilde\vr^{(t+1)} - [\vr^{(t)}]_+\|_2^2 + \|\tilde\vr^{(t+1)} - [\vr^{(t+1)}]_+\|_2^2}{2}.
\end{align*}
The first inequality follows from $\vr^{(t+1)} \le [\vr^{(t+1)}]_+$, and $\|[\vr^{(t+1)}]_+\|_2 = \|\tilde\vr^{(t+1)}\|_2$; for the second, we use \eqref{eq:ir_bound_1} bound and discard $\|\tilde\vr^{(t+1)} - \tilde\vr^{(t)}\|_2^2$. We can divide $\|\tilde\vr^{(t+1)}\|_1$ on both sides to obtain 
\begin{align*}
     \ip{\vu^{(t)}, \vx^{(t+1)} - \vx^{(t)}} + c^{(t)} - c^{(t+1)} \ge \frac{\|\tilde\vr^{(t+1)} - [\vr^{(t)}]_+\|_2^2 + \|\tilde\vr^{(t+1)} - [\vr^{(t+1)}]_+\|_2^2}{2 \|\tilde\vr^{(t+1)}\|_1}.
\end{align*}
Applying the normalization bound \eqref{eq:normalization_bound} to $\tilde\vr^{(t+1)}$ and $[\vr^{(s)}]_+$, we obtain
\begin{align*}
    \left\| \tilde\vr^{(t+1)} - \|\tilde\vr^{(t+1)}\|_1 \vx^{(s)} \right\|_2
    &= \|\tilde\vr^{(t+1)}\|_1 \left\|\frac{\tilde\vr^{(t+1)}}{\|\tilde\vr^{(t+1)}\|_1} - \vx^{(s)}\right\|_2 \\
    &\le \max\{\|\tilde\vr^{(t+1)}\|_1, \|[\vr^{(s)}]_+\|_1\} \left\|\frac{\tilde\vr^{(t+1)}}{\|\tilde\vr^{(t+1)}\|_1} - \vx^{(s)}\right\|_2 \\
    &\le \sqrt d \, \|\tilde\vr^{(t+1)} - [\vr^{(s)}]_+\|_2.
\end{align*}
For $s = t$ and $s = t+1$, these bounds control the two terms in the identity
\begin{align*}
    \|\tilde\vr^{(t+1)}\|_1 (\vx^{(t+1)} - \vx^{(t)}) = (\tilde\vr^{(t+1)} - \|\tilde\vr^{(t+1)}\|_1 \vx^{(t)}) - (\tilde\vr^{(t+1)} - \|\tilde\vr^{(t+1)}\|_1 \vx^{(t+1)}).
\end{align*}
Taking squared norms and applying the squared triangle inequality, we obtain
\begin{align*}
    \|\tilde\vr^{(t+1)}\|_1^2 \|\vx^{(t+1)} - \vx^{(t)}\|_2^2
    &\le 2 \|\tilde\vr^{(t+1)} - \|\tilde\vr^{(t+1)}\|_1 \vx^{(t)}\|_2^2 + 2 \|\tilde\vr^{(t+1)} - \|\tilde\vr^{(t+1)}\|_1 \vx^{(t+1)}\|_2^2 \\
    &\le 2d (\|\tilde\vr^{(t+1)} - [\vr^{(t)}]_+\|_2^2 + \|\tilde\vr^{(t+1)} - [\vr^{(t+1)}]_+\|_2^2).
\end{align*}
where the second inequality applies the two normalization bounds after squaring them. Dividing by $2d$ and substituting into the utility-improvement bound, we get
\begin{align*}
    \ip{\vu^{(t)}, \vx^{(t+1)} - \vx^{(t)}} + c^{(t)} - c^{(t+1)}
    &\ge \frac{\|\tilde\vr^{(t+1)}\|_1^2 \|\vx^{(t+1)} - \vx^{(t)}\|_2^2}{4d \|\tilde\vr^{(t+1)}\|_1} \\
    &= \frac{\|\tilde\vr^{(t+1)}\|_1}{4d} \|\vx^{(t+1)} - \vx^{(t)}\|_2^2 \\
    &\ge \frac{\|\tilde\vr^{(t+1)}\|_2}{4d} \|\vx^{(t+1)} - \vx^{(t)}\|_2^2.
\end{align*}
Rearranging, we obtain the desired bound
\begin{align*}
    \ip{\vu^{(t)}, \vx^{(t+1)} - \vx^{(t)}} \ge \frac{\|\tilde\vr^{(t+1)}\|_2}{4d} \|\vx^{(t+1)} - \vx^{(t)}\|_2^2 + c^{(t+1)} - c^{(t)}.
\end{align*}
If $\tilde\vr^{(t+1)} = \vec0$, the norm-increase bound proved above implies $\tilde\vr^{(t)} = \vec0$. Both strategies are then the uniform default strategy and $c^{(t)} = c^{(t+1)} = 0$, so both sides of the claimed inequality are zero.
\end{proof}
\altRvuIr*
\begin{proof}
We have just shown that
\begin{align*}
    \|\tilde\vr^{(T+1)}\|_2^2 \le \|\mA\|_2^2 + (1 + \sqrt d)^2 \sum_{t=2}^{T} \|\vu^{(t)} - \vu^{(t-1)}\|_2^2.
\end{align*}
Taking square roots and applying Young's inequality, we obtain
\begin{align*}
    2 \|\tilde\vr^{(T+1)}\|_2
    \le 2 \|\mA\|_2 + 2(1 + \sqrt d) \sqrt{\sum_{t=2}^{T} \|\vu^{(t)} - \vu^{(t-1)}\|_2^2} \le 2 \|\mA\|_2 + \frac{(1 + \sqrt d)^2}{\beta} + \beta \sum_{t=2}^{T} \|\vu^{(t)} - \vu^{(t-1)}\|_2^2.
\end{align*}
Substituting into the above alternating regret bound proves the claim with the stated $\alpha$ and $\gamma$.
\end{proof}

\subsection{The Theoretical Role of Half Prediction}
\label{sec:why_half_prediction}

To explain the choice of half prediction, we first derive a corrected squared norm for a general prediction multiplier $\lambda\ge0$. Repeating the argument underlying~\eqref{eq:half_bound_r_tilde}, the prediction step gives
\begin{align*}
    \|\tilde\vr^{(t+1)}\|_2^2 - \|\tilde\vr^{(t)}\|_2^2 \ge \|\tilde\vr^{(t+1)}-\tilde\vr^{(t)}\|_2^2 - 2\lambda \|\tilde\vr^{(t+1)}-\tilde\vr^{(t)}\|_2 \|\tilde\vr^{(t)}-\tilde\vr^{(t-1)}\|_2.
\end{align*}
Applying $2ab\le a^2+b^2$ and regrouping, we obtain
\begin{align*}
    \|\tilde\vr^{(t+1)}\|_2^2 - \|\tilde\vr^{(t)}\|_2^2 \ge \lambda \left(\|\tilde\vr^{(t+1)}-\tilde\vr^{(t)}\|_2^2 - \|\tilde\vr^{(t)}-\tilde\vr^{(t-1)}\|_2^2\right) + (1-2\lambda)\|\tilde\vr^{(t+1)}-\tilde\vr^{(t)}\|_2^2.
\end{align*}
The first term on the right is itself a difference between consecutive squared increments. We therefore subtract the corresponding correction from the squared regret vector norm and define
\begin{align*}
    f_\lambda^{(t)} = \|\tilde\vr^{(t)}\|_2^2 - \lambda \|\tilde\vr^{(t)}-\tilde\vr^{(t-1)}\|_2^2.
\end{align*}
This absorbs that difference into the left-hand side and yields
\begin{align*}
    f_\lambda^{(t+1)} - f_\lambda^{(t)} \ge (1-2\lambda)\|\tilde\vr^{(t+1)}-\tilde\vr^{(t)}\|_2^2.
\end{align*}
Thus, $f_\lambda^{(t)}$ is nondecreasing for every $0\le\lambda\le1/2$, and at $\lambda=1/2$ it coincides with the corrected squared norm used in our analysis.

This explains the correction, but monotonicity alone does not select half prediction. We also need to bound the terminal regret vector norm. For $0\le\lambda\le1/2$, the argument underlying~\eqref{eq:tilde_r_s_bound_half} leaves a coefficient $1-2\lambda+\varepsilon$ on cumulative squared regret-vector increments, where $\varepsilon>0$ is arbitrary. At half prediction, the fixed contribution $1-2\lambda$ vanishes, allowing us to choose the remaining coefficient sufficiently small to absorb this contribution using the utility-improvement bound. For smaller multipliers, a nondecreasing corrected squared norm remains available, but this cancellation does not occur. Half prediction therefore supplies both forms of control used to establish Alt-RVU.

\subsection{Proofs from Subsection \ref{sec:alt_rvu_hprm}}
\halfBoundRTilde*
\begin{proof}
For $t = 1$, the inequality follows immediately from $\tilde\vr^{(0)} = \tilde\vr^{(1)} = \vec0$. For $t \ge 2$, the regret vector and half-prediction updates \eqref{eq:setup_hprm} imply that
\begin{align*}
    [\vr^{(t)}]_+ = \left[\tilde\vr^{(t)} + \frac12 (\tilde\vr^{(t)} - \tilde\vr^{(t-1)})\right]_+.
\end{align*}
Indeed, on coordinates where $\tilde\vr^{(t)}$ is positive, the preceding regret vector update is untruncated; on the remaining coordinates, both sides are zero. Since taking the positive part is nonexpansive and $\tilde\vr^{(t)} \ge \vec0$, we obtain
\begin{align}
    \|[\vr^{(t)}]_+ - \tilde\vr^{(t)}\|_2 \le \frac12 \|\tilde\vr^{(t)} - \tilde\vr^{(t-1)}\|_2.
    \label{eq:bound_in_hprm_1}
\end{align}
The regret vector update also gives the coordinatewise inequality
\begin{align*}
    \tilde\vr^{(t+1)} - \tilde\vr^{(t)} \ge \vu^{(t)} - \ip{\vu^{(t)}, \vx^{(t)}} \vec1.
\end{align*}
Taking its inner product with $[\vr^{(t)}]_+ \ge \vec0$,
we get
\begin{align}
    \ip{\tilde\vr^{(t+1)} - \tilde\vr^{(t)}, [\vr^{(t)}]_+} \ge \ip{\vu^{(t)} - \ip{\vu^{(t)}, \vx^{(t)}} \vec1, [\vr^{(t)}]_+} = 0,
    \label{eq:bound_in_hprm_2}
\end{align}
where the equality follows from
$[\vr^{(t)}]_+ = \|[\vr^{(t)}]_+\|_1 \vx^{(t)}$,
including when $[\vr^{(t)}]_+ = \vec0$.

Expanding the difference of squared norms and applying the above bounds, we obtain the desired claim
\begin{align*}
    \|\tilde\vr^{(t+1)}\|_2^2 -\|\tilde\vr^{(t)}\|_2^2 &= \|\tilde\vr^{(t+1)} - \tilde\vr^{(t)}\|_2^2 - 2 \ip{\tilde\vr^{(t+1)} - \tilde\vr^{(t)}, - \tilde\vr^{(t)}} \\
    & \overset{\eqref{eq:bound_in_hprm_2}}{\ge}  \|\tilde\vr^{(t+1)} - \tilde\vr^{(t)}\|_2^2 - 2 \ip{\tilde\vr^{(t+1)} - \tilde\vr^{(t)}, [\vr^{(t)}]_+ - \tilde\vr^{(t)}} \\
    &\ge \|\tilde\vr^{(t+1)} - \tilde\vr^{(t)}\|_2^2
    - 2 \|\tilde\vr^{(t+1)} - \tilde\vr^{(t)}\|_2
    \|[\vr^{(t)}]_+ - \tilde\vr^{(t)}\|_2 \\
    & \overset{\eqref{eq:bound_in_hprm_1}}{\ge} \|\tilde\vr^{(t+1)} - \tilde\vr^{(t)}\|_2^2 - \|\tilde\vr^{(t+1)} - \tilde\vr^{(t)}\|_2 \|\tilde\vr^{(t)} - \tilde\vr^{(t-1)}\|_2 \\
    &\ge \frac12 \|\tilde\vr^{(t+1)} - \tilde\vr^{(t)}\|_2^2
    - \frac12 \|\tilde\vr^{(t)} - \tilde\vr^{(t-1)}\|_2^2,
\end{align*}
where the last inequality uses $\|\tilde\vr^{(t+1)} - \tilde\vr^{(t)}\|_2 \|\tilde\vr^{(t)} - \tilde\vr^{(t-1)}\|_2 \le \|\tilde\vr^{(t+1)} - \tilde\vr^{(t)}\|_2^2/2 + \|\tilde\vr^{(t)} - \tilde\vr^{(t-1)}\|_2^2/2$.
\end{proof}
\utilImprovementHprm*
\begin{proof}

The case $t=1$ arises only when $t_0=2$, so $\tilde\vr^{(2)}\ne\vec0$. Since $[\vr^{(2)}]_+ = 3\tilde\vr^{(2)}/2$, we have
\begin{align*}
    \ip{\vu^{(1)}, \vx^{(2)}-\vx^{(1)}} = \frac{\|\tilde\vr^{(2)}\|_2^2}{\|\tilde\vr^{(2)}\|_1} > 0.
\end{align*}
Moreover, $f^{(1)}=0$ and $f^{(2)}=\|\tilde\vr^{(2)}\|_2^2/2$, so the right-hand side of~\eqref{eq:util_improvement_hprm} is zero, proving the claim for $t=1$.

For $t \ge \max\{2,t_0-1\}$, the proof of Lemma~\ref{lem:half_bound_r_tilde} establishes
\begin{align*}
    \ip{\tilde\vr^{(t+1)} - \tilde\vr^{(t)}, [\vr^{(t)}]_+} \ge 0, \; \|[\vr^{(t)}]_+ - \tilde\vr^{(t)}\|_2 \le \frac12 \|\tilde\vr^{(t)} - \tilde\vr^{(t-1)}\|_2.
\end{align*}
Expanding the squared norms and applying these bounds, we obtain
\begin{align}
    \|\tilde\vr^{(t+1)}\|_2^2 - \|\tilde\vr^{(t)}\|_2^2
    &= \|\tilde\vr^{(t+1)} - [\vr^{(t)}]_+\|_2^2 - \|[\vr^{(t)}]_+ - \tilde\vr^{(t)}\|_2^2 + 2 \ip{\tilde\vr^{(t+1)} - \tilde\vr^{(t)}, [\vr^{(t)}]_+} \notag \\
    &\ge - \|[\vr^{(t)}]_+ - \tilde\vr^{(t)}\|_2^2 \notag \\
    &\ge - \frac14 \|\tilde\vr^{(t)} - \tilde\vr^{(t-1)}\|_2^2.
    \label{eq:hprm_bound}
\end{align}
We next apply the positive-part argument from Lemma~\ref{eq:util_improvement_prm} to the half-prediction update. Since $\vr^{(t+1)} - \tilde\vr^{(t+1)} = (\vu^{(t)} - \ip{\vu^{(t)}, \vx^{(t)}} \vec1)/2$, we get
\begin{align*}
    \ip{\vu^{(t)} - \ip{\vu^{(t)}, \vx^{(t)}} \vec1, [\vr^{(t+1)}]_+ - \tilde\vr^{(t+1)}} \ge 2 \|[\vr^{(t+1)}]_+ - \tilde\vr^{(t+1)}\|_2^2 \ge 0.
\end{align*}
Combining this with the identity \eqref{eq:useful_identity}, which also holds for $\HPRMplus$ because its regret vector update is unchanged, we obtain
\begin{align*}
    \ip{\vu^{(t)} - \ip{\vu^{(t)}, \vx^{(t)}} \vec1, [\vr^{(t+1)}]_+}
    &\ge \ip{\vu^{(t)} - \ip{\vu^{(t)}, \vx^{(t)}} \vec1, \tilde\vr^{(t+1)}} \\
    &= \frac{\|\tilde\vr^{(t+1)}\|_2^2 - \|\tilde\vr^{(t)}\|_2^2 + \|\tilde\vr^{(t+1)} - \tilde\vr^{(t)}\|_2^2}{2}.
\end{align*}
By the definition of $f^{(t)}$, we have
\begin{align*}
    f^{(t+1)} - f^{(t)} = \|\tilde\vr^{(t+1)}\|_2^2 - \|\tilde\vr^{(t)}\|_2^2 - \frac12 \|\tilde\vr^{(t+1)} - \tilde\vr^{(t)}\|_2^2 + \frac12 \|\tilde\vr^{(t)} - \tilde\vr^{(t-1)}\|_2^2.
\end{align*}
Consequently, we can rewrite the preceding lower bound as
\begin{align*}
    &\frac{\|\tilde\vr^{(t+1)}\|_2^2 - \|\tilde\vr^{(t)}\|_2^2 + \|\tilde\vr^{(t+1)} - \tilde\vr^{(t)}\|_2^2}{2} \\
    &= \frac14 \|\tilde\vr^{(t+1)} - \tilde\vr^{(t)}\|_2^2 - \frac12 (f^{(t+1)} - f^{(t)}) + \|\tilde\vr^{(t+1)}\|_2^2 - \|\tilde\vr^{(t)}\|_2^2 + \frac14 \|\tilde\vr^{(t)} - \tilde\vr^{(t-1)}\|_2^2 \\
    &\ge \frac14 \|\tilde\vr^{(t+1)} - \tilde\vr^{(t)}\|_2^2 - \frac12 (f^{(t+1)} - f^{(t)}),
\end{align*}
where the inequality follows from \eqref{eq:hprm_bound}. We hence have
\begin{align*}
    \ip{\vu^{(t)} - \ip{\vu^{(t)}, \vx^{(t)}} \vec1, [\vr^{(t+1)}]_+} \ge \frac14 \|\tilde\vr^{(t+1)} - \tilde\vr^{(t)}\|_2^2 - \frac12 (f^{(t+1)} - f^{(t)}),
\end{align*}
which implies that
\begin{align*}
    \frac{ \ip{\vu^{(t)} - \ip{\vu^{(t)}, \vx^{(t)}} \vec1, [\vr^{(t+1)}]_+}}{\|[\vr^{(t+1)}]_+\|_1} \ge \frac{\|\tilde\vr^{(t+1)}  - \tilde\vr^{(t)}\|_2^2}{4 \|[\vr^{(t+1)}]_+\|_1} - \frac{f^{(t+1)} - f^{(t)}}{2 \|[\vr^{(t+1)}]_+\|_1}.
\end{align*}
Since $\vx^{(t+1)} = [\vr^{(t+1)}]_+/ \|[\vr^{(t+1)}]_+\|_1$, we can obtain the LHS as 
\begin{align*}
     \frac{ \ip{\vu^{(t)} - \ip{\vu^{(t)}, \vx^{(t)}} \vec1, [\vr^{(t+1)}]_+}}{\|[\vr^{(t+1)}]_+\|_1} = \ip{\vu^{(t)} - \ip{\vu^{(t)}, \vx^{(t)}} \vec1, \vx^{(t+1)}} = \ip{\vu^{(t)}, \vx^{(t+1)} - \vx^{(t)}},
\end{align*}
so we obtained the desired claim. 
\end{proof}
\terminalTildeRBoundHalf*
\begin{proof}
For $t \ge 2$, the proof of Lemma~\ref{lem:half_bound_r_tilde} establishes
\begin{align*}
    [\vr^{(t)}]_+ = \left[\tilde\vr^{(t)} + \frac12 \bigl(\tilde\vr^{(t)} - \tilde\vr^{(t-1)}\bigr)\right]_+, \; \|[\vr^{(t)}]_+ - \tilde\vr^{(t)}\|_2 \le \frac12 \|\tilde\vr^{(t)} - \tilde\vr^{(t-1)}\|_2.
\end{align*}
Recall from the proof of Lemma~\ref{lem:terminal_tilde_r_bound_prm} that
\begin{align*}
    \ip{\vr^{(t)} - [\vr^{(t)}]_+, \tilde\vr^{(t+1)} - [\vr^{(t)}]_+}
    = \ip{\vr^{(t)} - [\vr^{(t)}]_+, \tilde\vr^{(t+1)}} \le 0.
\end{align*}
Substituting $\vr^{(t)} = \tilde\vr^{(t)} + \frac12 \bigl(\vu^{(t-1)} - \ip{\vu^{(t-1)}, \vx^{(t-1)}} \vec1\bigr)$ and rearranging, we obtain
\begin{align*}
    \frac12 \ip{\vu^{(t-1)} - \ip{\vu^{(t-1)}, \vx^{(t-1)}} \vec1, \tilde\vr^{(t+1)} - [\vr^{(t)}]_+}
    \le \ip{[\vr^{(t)}]_+ - \tilde\vr^{(t)}, \tilde\vr^{(t+1)} - [\vr^{(t)}]_+},
\end{align*}
which implies
\begin{align}
    \ip{\vu^{(t-1)} - \ip{\vu^{(t-1)}, \vx^{(t-1)}} \vec1, \tilde\vr^{(t+1)} - [\vr^{(t)}]_+}
    \le 2 \ip{[\vr^{(t)}]_+ - \tilde\vr^{(t)}, \tilde\vr^{(t+1)} - [\vr^{(t)}]_+}.
    \label{eq:hprm_useful_bound}
\end{align}
Since the regret vector update is the same, we use identity \eqref{eq:useful_identity} in the proof of Lemma~\ref{eq:util_improvement_prm} to obtain
\begin{align*}
    \|\tilde\vr^{(t+1)}\|_2^2 - \|\tilde\vr^{(t)}\|_2^2
    = 2 \ip{\vu^{(t)} - \ip{\vu^{(t)}, \vx^{(t)}} \vec1, \tilde\vr^{(t+1)} - [\vr^{(t)}]_+} - \|\tilde\vr^{(t+1)} - \tilde\vr^{(t)}\|_2^2.
\end{align*}
We first add and subtract the previous instantaneous regret vector. Applying the inequality~\eqref{eq:hprm_useful_bound}, we obtain
\begin{align*}
    \|\tilde\vr^{(t+1)}\|_2^2 - \|\tilde\vr^{(t)}\|_2^2
    &= 2 \ip{\vu^{(t)} - \vu^{(t-1)} - \left(\ip{\vu^{(t)}, \vx^{(t)}} - \ip{\vu^{(t-1)}, \vx^{(t-1)}}\right) \vec1, \tilde\vr^{(t+1)} - [\vr^{(t)}]_+} \\
    & + 2 \ip{\vu^{(t-1)} - \ip{\vu^{(t-1)}, \vx^{(t-1)}} \vec1, \tilde\vr^{(t+1)} - [\vr^{(t)}]_+} - \|\tilde\vr^{(t+1)} - \tilde\vr^{(t)}\|_2^2 \\
    &\le 2 \ip{\vu^{(t)} - \vu^{(t-1)} - \left(\ip{\vu^{(t)}, \vx^{(t)}} - \ip{\vu^{(t-1)}, \vx^{(t-1)}}\right) \vec1, \tilde\vr^{(t+1)} - [\vr^{(t)}]_+} \\
    & + 4 \ip{[\vr^{(t)}]_+ - \tilde\vr^{(t)}, \tilde\vr^{(t+1)} - [\vr^{(t)}]_+} - \|\tilde\vr^{(t+1)} - \tilde\vr^{(t)}\|_2^2.
\end{align*}
To expand the last square, we write
\begin{align*}
    \|\tilde\vr^{(t+1)} - \tilde\vr^{(t)}\|_2^2
    &= \|\tilde\vr^{(t+1)} - [\vr^{(t)}]_+\|_2^2 + \|[\vr^{(t)}]_+ - \tilde\vr^{(t)}\|_2^2 + 2 \ip{[\vr^{(t)}]_+ - \tilde\vr^{(t)}, \tilde\vr^{(t+1)} - [\vr^{(t)}]_+}.
\end{align*}
Substituting this identity, the coefficient of the cross term decreases from $4$ to $2$. We therefore obtain
\begin{align*}
    \|\tilde\vr^{(t+1)}\|_2^2 - \|\tilde\vr^{(t)}\|_2^2
    &\le 2 \ip{\vu^{(t)} - \vu^{(t-1)} - \left(\ip{\vu^{(t)}, \vx^{(t)}} - \ip{\vu^{(t-1)}, \vx^{(t-1)}}\right) \vec1 + [\vr^{(t)}]_+ - \tilde\vr^{(t)}, \tilde\vr^{(t+1)} - [\vr^{(t)}]_+} \\
    &- \|\tilde\vr^{(t+1)} - [\vr^{(t)}]_+\|_2^2 - \|[\vr^{(t)}]_+ - \tilde\vr^{(t)}\|_2^2 \\
    &= \left\|\vu^{(t)} - \vu^{(t-1)} - \left(\ip{\vu^{(t)}, \vx^{(t)}} - \ip{\vu^{(t-1)}, \vx^{(t-1)}}\right) \vec1 + [\vr^{(t)}]_+ - \tilde\vr^{(t)}\right\|_2^2 - \|[\vr^{(t)}]_+ - \tilde\vr^{(t)}\|_2^2 \\
    &- \left\|\tilde\vr^{(t+1)} - [\vr^{(t)}]_+ - \left(\vu^{(t)} - \vu^{(t-1)} - \left(\ip{\vu^{(t)}, \vx^{(t)}} - \ip{\vu^{(t-1)}, \vx^{(t-1)}}\right) \vec1 + [\vr^{(t)}]_+ - \tilde\vr^{(t)}\right)\right\|_2^2.
\end{align*}
Discarding the last nonpositive term and expanding the first square, we get
\begin{align}
    \|\tilde\vr^{(t+1)}\|_2^2 - \|\tilde\vr^{(t)}\|_2^2
    &\le \left\|\vu^{(t)} - \vu^{(t-1)} - \left(\ip{\vu^{(t)}, \vx^{(t)}} - \ip{\vu^{(t-1)}, \vx^{(t-1)}}\right) \vec1\right\|_2^2 \notag \\
    &+ 2 \ip{\vu^{(t)} - \vu^{(t-1)} - \left(\ip{\vu^{(t)}, \vx^{(t)}} - \ip{\vu^{(t-1)}, \vx^{(t-1)}}\right) \vec1, [\vr^{(t)}]_+ - \tilde\vr^{(t)}} \notag \\
    &\le \left\|\vu^{(t)} - \vu^{(t-1)} - \left(\ip{\vu^{(t)}, \vx^{(t)}} - \ip{\vu^{(t-1)}, \vx^{(t-1)}}\right) \vec1\right\|_2^2 \notag\\
    &+ 2 \left\|\vu^{(t)} - \vu^{(t-1)} - \left(\ip{\vu^{(t)}, \vx^{(t)}} - \ip{\vu^{(t-1)}, \vx^{(t-1)}}\right) \vec1\right\|_2 \|[\vr^{(t)}]_+ - \tilde\vr^{(t)}\|_2 \notag \\
    &\le \left\|\vu^{(t)} - \vu^{(t-1)} - \left(\ip{\vu^{(t)}, \vx^{(t)}} - \ip{\vu^{(t-1)}, \vx^{(t-1)}}\right) \vec1\right\|_2^2 \notag \\
    &+ \left\|\vu^{(t)} - \vu^{(t-1)} - \left(\ip{\vu^{(t)}, \vx^{(t)}} - \ip{\vu^{(t-1)}, \vx^{(t-1)}}\right) \vec1\right\|_2 \|\tilde\vr^{(t)} - \tilde\vr^{(t-1)}\|_2,
    \label{eq:hprm_bound_2}
\end{align}
where we use Cauchy--Schwarz in the second inequality and the bound $\|[\vr^{(t)}]_+ - \tilde\vr^{(t)}\|_2 \le \frac12 \|\tilde\vr^{(t)} - \tilde\vr^{(t-1)}\|_2$ in the last inequality.

The utility-variation estimate already established in the proof of Lemma~\ref{lem:terminal_tilde_r_bound_prm} gives
\begin{align*}
    &\sum_{t=2}^{T} \left\|\vu^{(t)} - \vu^{(t-1)} - \bigl(\ip{\vu^{(t)}, \vx^{(t)}} - \ip{\vu^{(t-1)}, \vx^{(t-1)}}\bigr) \vec1\right\|_2^2 \\
    & \le 2(1+\sqrt d)^2 \sum_{t=2}^{T} \|\vu^{(t)} - \vu^{(t-1)}\|_2^2 + 2d\|\mA\|_2^2 \sum_{t=1}^{T} \|\vx^{(t+1)} - \vx^{(t)}\|_2^2.
\end{align*}

Fix $\varepsilon>0$. Applying Young's inequality to the product in \eqref{eq:hprm_bound_2}, and using the initialization bound, we obtain, for every $2 \le s \le T+1$,
\begin{align*}
    \|\tilde \vr^{(s)}\|_2^2 &= \|\tilde \vr^{(2)}\|_2^2 + \sum_{t=2}^{s-1} \left(\|\tilde \vr^{(t+1)}\|_2^2 - \|\tilde \vr^{(t)}\|_2^2\right) \\
    &\le \|\mA\|_2^2 + \left(1+\frac{1}{4\varepsilon}\right) \sum_{t=2}^{s-1} \left\|\vu^{(t)}-\vu^{(t-1)}-\left(\ip{\vu^{(t)}, \vx^{(t)}}-\ip{\vu^{(t-1)}, \vx^{(t-1)}}\right)\vec1\right\|_2^2 + \varepsilon \sum_{t=2}^{s-1} \|\tilde \vr^{(t)}-\tilde \vr^{(t-1)}\|_2^2 \\
    &\le \|\mA\|_2^2 + \left(2+\frac{1}{2\varepsilon}\right) \left[(1+\sqrt d)^2 \sum_{t=2}^{T} \|\vu^{(t)}-\vu^{(t-1)}\|_2^2 + d\|\mA\|_2^2 \sum_{t=1}^{T} \|\vx^{(t+1)}-\vx^{(t)}\|_2^2 \right] + \varepsilon \sum_{t=2}^{T} \|\tilde \vr^{(t)}-\tilde \vr^{(t-1)}\|_2^2.
\end{align*}
The last inequality follows by extending the nonnegative sums to $T$ and applying the preceding utility-variation estimate. Since $\tilde \vr^{(1)}=\vec0$, taking the maximum over $1 \le s \le T+1$ proves the claim.
\end{proof}
\altRvuHprm*
\begin{proof}
If $t_0$ does not exist, the instantaneous regret vectors and strategy movements are zero, so the conclusion holds with any positive constants. Otherwise, $f^{(t_0)} = \|\tilde\vr^{(t_0)}\|_2^2/2 > 0$. It suffices to consider $T \ge t_0-1$, since alternating regret and strategy movement are zero before that time.

We first bound the prediction norms from below. By \eqref{eq:bound_in_hprm_1}, for every $t \ge t_0$,
\begin{align*}
    f^{(t)}
    &= \|\tilde\vr^{(t)}\|_2^2 - \frac12 \|\tilde\vr^{(t)} - \tilde\vr^{(t-1)}\|_2^2 \le \|\tilde\vr^{(t)}\|_2^2 - 2 \|[\vr^{(t)}]_+ - \tilde\vr^{(t)}\|_2^2 = 2 \|[\vr^{(t)}]_+\|_2^2 - \|\tilde\vr^{(t)} - 2[\vr^{(t)}]_+\|_2^2 \le 2 \|[\vr^{(t)}]_+\|_1^2.
\end{align*}
Since $f^{(t)}$ is nondecreasing, we obtain
\begin{align}
    \|[\vr^{(t)}]_+\|_1 \ge \sqrt{\frac{f^{(t)}}2} \ge \sqrt{\frac{f^{(t_0)}}2} > 0.
    \label{eq:prediction_lower_hprm}
\end{align}
Consequently, we obtain
\begin{align*}
    \sum_{t=t_0-1}^{T} \frac{f^{(t+1)} - f^{(t)}}{2\|[\vr^{(t+1)}]_+\|_1}
    &\le \frac1{\sqrt2} \sum_{t=t_0-1}^{T} \frac{f^{(t+1)} - f^{(t)}}{\sqrt{f^{(t+1)}}} \le \sqrt2 \sum_{t=t_0-1}^{T} \left(\sqrt{f^{(t+1)}} - \sqrt{f^{(t)}}\right) = \sqrt{2f^{(T+1)}} \le \sqrt2 \|\tilde\vr^{(T+1)}\|_2.
\end{align*}
Combining the utility-improvement bound \eqref{eq:util_improvement_hprm} with \eqref{eq:alt_reg_bound_simple}, we therefore get
\begin{align}
    \operatorname{Alt-Reg}(T)
    &\le (2+\sqrt2)\|\tilde\vr^{(T+1)}\|_2 - \frac14 \sum_{t=t_0-1}^{T} \frac{\|\tilde\vr^{(t+1)} - \tilde\vr^{(t)}\|_2^2}{\|[\vr^{(t+1)}]_+\|_1} \notag \\
    &\le 4\max_{1\le s\le T+1}\|\tilde\vr^{(s)}\|_2 - \frac14 \sum_{t=t_0-1}^{T} \frac{\|\tilde\vr^{(t+1)} - \tilde\vr^{(t)}\|_2^2}{\|[\vr^{(t+1)}]_+\|_1}.
    \label{eq:hprm_alt_reg_corrections}
\end{align}
We next relate the negative sum to strategy movement. The prediction identity from Lemma~\ref{lem:half_bound_r_tilde} gives
\begin{align*}
    \|[\vr^{(t+1)}]_+ - [\vr^{(t)}]_+\|_2^2
    &\le \left\|\frac32\left(\tilde\vr^{(t+1)} - \tilde\vr^{(t)}\right) - \frac12\left(\tilde\vr^{(t)} - \tilde\vr^{(t-1)}\right)\right\|_2^2 \\
    &\le 3\|\tilde\vr^{(t+1)} - \tilde\vr^{(t)}\|_2^2 + \|\tilde\vr^{(t)} - \tilde\vr^{(t-1)}\|_2^2.
\end{align*}
Using the normalization bound \eqref{eq:normalization_bound} and the prediction norm bound \eqref{eq:prediction_lower_hprm}, for $t \ge t_0$ we obtain
\begin{align*}
    \sqrt{\frac{f^{(t_0)}}2}\|\vx^{(t+1)} - \vx^{(t)}\|_2^2
    &\le (1+\sqrt d)^2 \left(\frac{3\|\tilde\vr^{(t+1)} - \tilde\vr^{(t)}\|_2^2}{\|[\vr^{(t+1)}]_+\|_1} + \frac{\|\tilde\vr^{(t)} - \tilde\vr^{(t-1)}\|_2^2}{\|[\vr^{(t)}]_+\|_1}\right).
\end{align*}
Summing and using the stated choice of $\gamma$, we get
\begin{align}
    \sum_{t=t_0-1}^{T} \frac{\|\tilde\vr^{(t+1)} - \tilde\vr^{(t)}\|_2^2}{\|[\vr^{(t+1)}]_+\|_1}
    &\ge \frac{\sqrt{f^{(t_0)}}}{4\sqrt2(1+\sqrt d)^2} \sum_{t=t_0}^{T} \|\vx^{(t+1)} - \vx^{(t)}\|_2^2 \notag \\
    &= 16\gamma \sum_{t=t_0}^{T} \|\vx^{(t+1)} - \vx^{(t)}\|_2^2 \notag \\
    &\ge 16\gamma \sum_{t=1}^{T} \|\vx^{(t+1)} - \vx^{(t)}\|_2^2 - 32\gamma,
    \label{eq:hprm_corrections_movement}
\end{align}
where the last inequality uses that movements before $t_0-1$ vanish and $\|\vx^{(t_0)} - \vx^{(t_0-1)}\|_2^2 \le 2$.

It remains to absorb the positive maximum norm in \eqref{eq:hprm_alt_reg_corrections}. Since $\tilde\vr^{(t)} \ge \vec0$, the prediction identity also gives
\begin{align*}
    [\vr^{(t+1)}]_+ \le \frac32\tilde\vr^{(t+1)}, \; \|[\vr^{(t+1)}]_+\|_1 \le 2\sqrt d\max_{1\le s\le T+1}\|\tilde\vr^{(s)}\|_2.
\end{align*}
Therefore, we obtain
\begin{align*}
    \sum_{t=2}^{T}\|\tilde\vr^{(t)} - \tilde\vr^{(t-1)}\|_2^2
    &\le 2\sqrt d \left(\max_{1\le s\le T+1}\|\tilde\vr^{(s)}\|_2\right) \sum_{t=t_0-1}^{T} \frac{\|\tilde\vr^{(t+1)} - \tilde\vr^{(t)}\|_2^2}{\|[\vr^{(t+1)}]_+\|_1}.
\end{align*}
Substituting this bound into \eqref{eq:tilde_r_s_bound_half} with $\varepsilon=1/(64\sqrt d)$ and applying Young's inequality, we obtain
\begin{align*}
    \max_{1\le s\le T+1}\|\tilde\vr^{(s)}\|_2^2
    &\le \|\mA\|_2^2 + (1+16\sqrt d)\left(2(1+\sqrt d)^2\sum_{t=2}^{T}\|\vu^{(t)}-\vu^{(t-1)}\|_2^2 + 2d\|\mA\|_2^2\sum_{t=1}^{T}\|\vx^{(t+1)}-\vx^{(t)}\|_2^2\right) \\
    & + \frac1{32}\left(\max_{1\le s\le T+1}\|\tilde\vr^{(s)}\|_2\right) \sum_{t=t_0-1}^{T} \frac{\|\tilde\vr^{(t+1)} - \tilde\vr^{(t)}\|_2^2}{\|[\vr^{(t+1)}]_+\|_1} \\
    &\le \|\mA\|_2^2 + (1+16\sqrt d)\left(2(1+\sqrt d)^2\sum_{t=2}^{T}\|\vu^{(t)}-\vu^{(t-1)}\|_2^2 + 2d\|\mA\|_2^2\sum_{t=1}^{T}\|\vx^{(t+1)}-\vx^{(t)}\|_2^2\right) \\
    & + \frac12\max_{1\le s\le T+1}\|\tilde\vr^{(s)}\|_2^2 + \frac1{2048}\left(\sum_{t=t_0-1}^{T} \frac{\|\tilde\vr^{(t+1)} - \tilde\vr^{(t)}\|_2^2}{\|[\vr^{(t+1)}]_+\|_1}\right)^2.
\end{align*}
Rearranging and taking square roots, using $\sqrt{2+32\sqrt d}\le 6d^{1/4}$, we get
\begin{align*}
    \max_{1\le s\le T+1}\|\tilde\vr^{(s)}\|_2
    &\le 2\|\mA\|_2 + 6d^{1/4}\left(2(1+\sqrt d)^2\sum_{t=2}^{T}\|\vu^{(t)}-\vu^{(t-1)}\|_2^2 + 2d\|\mA\|_2^2\sum_{t=1}^{T}\|\vx^{(t+1)}-\vx^{(t)}\|_2^2\right)^{1/2} \\
    & + \frac1{32}\sum_{t=t_0-1}^{T} \frac{\|\tilde\vr^{(t+1)} - \tilde\vr^{(t)}\|_2^2}{\|[\vr^{(t+1)}]_+\|_1}.
\end{align*}
Substituting into \eqref{eq:hprm_alt_reg_corrections} and then applying \eqref{eq:hprm_corrections_movement}, we obtain
\begin{align*}
    \operatorname{Alt-Reg}(T)
    &\le 8\|\mA\|_2 + 4\gamma + 24d^{1/4}\left(2(1+\sqrt d)^2\sum_{t=2}^{T}\|\vu^{(t)}-\vu^{(t-1)}\|_2^2 + 2d\|\mA\|_2^2\sum_{t=1}^{T}\|\vx^{(t+1)}-\vx^{(t)}\|_2^2\right)^{1/2} \\
    & - 2\gamma\sum_{t=1}^{T}\|\vx^{(t+1)}-\vx^{(t)}\|_2^2 \\
    &\le 8\|\mA\|_2 + 4\gamma + \frac{288\sqrt d(1+\sqrt d)^2}{\beta} + \frac{288d^{3/2}\|\mA\|_2^2}{\gamma} \\
    &+ \beta\sum_{t=2}^{T}\|\vu^{(t)}-\vu^{(t-1)}\|_2^2 - \gamma\sum_{t=1}^{T}\|\vx^{(t+1)}-\vx^{(t)}\|_2^2,
\end{align*}
where the last inequality follows by splitting the square root and applying Young's inequality to each term. Thus, one admissible $\alpha$ choice is
\begin{align*}
    \alpha = 8\|\mA\|_2 + 4\gamma + \frac{288\sqrt d(1+\sqrt d)^2}{\beta} + \frac{288d^{3/2}\|\mA\|_2^2}{\gamma},
\end{align*}
which is independent of $T$. We hence obtain the desired Alt-RVU bound for $\HPRMplus$. 
\end{proof}

\subsection{Proofs from Subsection \ref{sec:necessity_alt_rvu}}

\halfStandardRvuFailure*
\begin{proof}
We construct a sequence whose standard regret and squared utility variation remain bounded, while its squared strategy movement diverges. We use the same blocks as in Proposition~\ref{prop:failure_above_half_explicit}, but increase their lengths and decrease their utility coefficients. This will keep the accumulated norm increase bounded while allowing each block to produce a nonvanishing strategy change.

Use the same three initial utilities $\vu^{(1)} = (1,-1)$, $\vu^{(2)} = (0,1)$, and $\vu^{(3)} = \vec0$, so that $\tilde\vr^{(4)} = \vr^{(4)} = (1,1)$. For block $k \ge 1$, choose
\begin{align*}
    N_k = k + 2, \; \varepsilon_k = \frac{1}{12N_k}.
\end{align*}
The block consists of $N_k-1$ utilities $\varepsilon_k(-r_2^{(t)},r_1^{(t)})$, one zero utility, $N_k-1$ utilities $\varepsilon_k(r_2^{(t)},-r_1^{(t)})$, and one zero utility.

The positivity argument and polynomial calculation in the proof of Proposition~\ref{prop:failure_above_half_explicit} remain valid at $\lambda = 1/2$: they require $N_k \ge 3$ and $\varepsilon_k = 1/[8N_k(1+\lambda)]$, but do not require the strict norm decrease. Thus, if block $k$ starts at time $t$ with $\tilde\vr^{(t)} = \vr^{(t)} = R_k(1,1)$, every coordinate of both vectors remains in $[R_k/2,3R_k/2]$, and the block ends with
\begin{align*}
    \tilde\vr^{(t+2N_k)} = \vr^{(t+2N_k)} = R_{k+1}(1,1), \; R_{k+1} = \rho_k R_k,
\end{align*}
where $R_1 = 1$ and $\rho_k$ is the forward half's squared-norm ratio. Throughout the block, clipping is inactive and
\begin{align*}
    \ip{\vu^{(s)}, \vx^{(s)}} = 0, \; \tilde\vr^{(s+1)} = \tilde\vr^{(s)} + \vu^{(s)}, \; \vr^{(s+1)} = \tilde\vr^{(s+1)} + \frac12 \vu^{(s)}.
\end{align*}

We first bound the regret-vector norms. Substituting $\lambda = 1/2$ into \eqref{eq:failure_above_half_half_energy}, the utility-square term vanishes. Using the forward-half variation estimate \eqref{eq:failure_above_half_variation_part}, scaled by $R_k^2$, we obtain
\begin{align*}
    0 \le \rho_k - 1 = \frac{\|\tilde\vr^{(t+N_k)}\|_2^2 - 2R_k^2}{2R_k^2} = \frac{1}{4R_k^2} \sum_{s=t}^{t+N_k-1} \|\vu^{(s)} - \vu^{(s-1)}\|_2^2 < \frac74 \varepsilon_k^2.
\end{align*}
Consequently,
\begin{align*}
    1 \le R_k = \prod_{j=1}^{k-1} \rho_j \le \exp\left(\frac74 \sum_{j=1}^{\infty} \varepsilon_j^2\right) = \exp\left(\frac7{576} \sum_{n=3}^{\infty} \frac1{n^2}\right) \le \exp\left(\frac7{1152}\right) < 2,
\end{align*}
where we use $\sum_{n=3}^{\infty} n^{-2} \le \sum_{n=3}^{\infty} [n(n-1)]^{-1} = 1/2$. In particular, every utility after initialization satisfies
\begin{align*}
    \|\vu^{(s)}\|_\infty \le \frac32 R_k \varepsilon_k < 3\varepsilon_k \le \frac1{12}.
\end{align*}
Together with the three initial utilities, this proves that the entire sequence lies in $[-1,1]^2$.

We next bound squared utility variation. Each block has four changes to or from zero, each contributing at most $9R_k^2\varepsilon_k^2/2$. For two consecutive nonzero utilities within the same half, the update identity gives
\begin{align*}
    \|\vu^{(s)} - \vu^{(s-1)}\|_2 = \varepsilon_k \|\vr^{(s)} - \vr^{(s-1)}\|_2 = \varepsilon_k \left\|\frac32 \vu^{(s-1)} - \frac12 \vu^{(s-2)}\right\|_2 \le 3\sqrt2 R_k\varepsilon_k^2.
\end{align*}
There are $2(N_k-2)$ such interior changes. We therefore obtain
\begin{align*}
    \sum_{s=t}^{t+2N_k-1} \|\vu^{(s)} - \vu^{(s-1)}\|_2^2 \le 18R_k^2\varepsilon_k^2 + 36(N_k-2)R_k^2\varepsilon_k^4 = R_k^2\varepsilon_k^2 \left(18 + \frac{N_k-2}{4N_k^2}\right) < 76\varepsilon_k^2.
\end{align*}
This sum includes the first change from the preceding zero utility. The initialization contributes $\|\vu^{(2)}-\vu^{(1)}\|_2^2 + \|\vu^{(3)}-\vu^{(2)}\|_2^2 = 6$, so
\begin{align*}
    \sum_{s=2}^{\infty} \|\vu^{(s)} - \vu^{(s-1)}\|_2^2 \le 6 + 76 \sum_{k=1}^{\infty} \varepsilon_k^2 < \infty.
\end{align*}

We now show that each forward half changes the strategy by an amount bounded away from zero. At the beginning of block $k$, we have $\vx^{(t)} = (1/2,1/2)$. Summing its $N_k-1$ nonzero forward utilities, we get
\begin{align*}
    \tilde r_2^{(t+N_k)} - \tilde r_1^{(t+N_k)} = \varepsilon_k \sum_{s=t}^{t+N_k-2} \left(r_1^{(s)} + r_2^{(s)}\right) \ge R_k(N_k-1)\varepsilon_k.
\end{align*}
The final zero utility in this half gives $\vr^{(t+N_k)} = \tilde\vr^{(t+N_k)}$. Since its coordinate sum is at most $3R_k$, we obtain
\begin{align*}
    \left|x_1^{(t+N_k)} - \frac12\right| = \frac{\tilde r_2^{(t+N_k)} - \tilde r_1^{(t+N_k)}}{2\|\tilde\vr^{(t+N_k)}\|_1} \ge \frac{(N_k-1)\varepsilon_k}{6} \ge \frac1{108}.
\end{align*}
Applying Cauchy--Schwarz to these $N_k$ strategy changes, we further obtain
\begin{align*}
    \sum_{s=t}^{t+N_k-1} \|\vx^{(s+1)} - \vx^{(s)}\|_2^2 \ge \frac{\|\vx^{(t+N_k)} - \vx^{(t)}\|_2^2}{N_k} = \frac{2}{N_k}\left|x_1^{(t+N_k)} - \frac12\right|^2 \ge \frac1{5832N_k}.
\end{align*}
The forward halves are disjoint, and $\sum_{k=1}^{\infty} 1/N_k = \infty$. Hence,
\begin{align*}
    \sum_{s=2}^{\infty} \|\vx^{(s)} - \vx^{(s-1)}\|_2^2 = \infty.
\end{align*}

It remains to bound standard regret. Every played payoff is zero, including during initialization. Moreover, the regret-vector updates telescope from time $4$ onward. Thus, for every $T \ge 3$,
\begin{align*}
    \sum_{s=1}^{T} \vu^{(s)} = (1,0) + \sum_{s=4}^{T} \left(\tilde\vr^{(s+1)} - \tilde\vr^{(s)}\right) = \tilde\vr^{(T+1)} - (0,1),
\end{align*}
and therefore
\begin{align*}
    \operatorname{Reg}(T) = \max\left\{\tilde r_1^{(T+1)}, \tilde r_2^{(T+1)} - 1\right\}.
\end{align*}
Positivity and the coordinate bounds imply $0 \le \operatorname{Reg}(T) \le 3$. The first two horizons both have regret $1$.

For any horizon-independent constants $\alpha, \beta, \gamma > 0$, we have
\begin{align*}
    \alpha + \beta \sum_{s=2}^{T} \|\vu^{(s)} - \vu^{(s-1)}\|_2^2 - \gamma \sum_{s=2}^{T} \|\vx^{(s)} - \vx^{(s-1)}\|_2^2 \longrightarrow -\infty,
\end{align*}
whereas $\operatorname{Reg}(T) \ge 0$. This contradicts standard RVU, even when the constants depend on the entire sequence. The same conclusion holds for any fixed pair of primal-dual norms by norm equivalence in dimension two. Finally, the common matrix-game realization of utilities in $[-1,1]^2$ shows that this sequence can be generated by an opponent under alternating updates.
\end{proof}

\subsection{Proofs from Section \ref{sec:no_lambda_except_half}}

\altRvuFailureAboveHalf*
\begin{proof}
We work with two actions throughout. Recall the scaled-prediction updates
\begin{align*}
    \tilde\vr^{(t+1)} = \left[\tilde\vr^{(t)} + \vu^{(t)} - \ip{\vu^{(t)}, \vx^{(t)}} \vec1\right]_+, \; \vr^{(t+1)} = \tilde\vr^{(t+1)} + \lambda \left(\vu^{(t)} - \ip{\vu^{(t)}, \vx^{(t)}} \vec1\right),
\end{align*}
with $\vx^{(t)} = [\vr^{(t)}]_+ / \|[\vr^{(t)}]_+\|_1$ and the uniform strategy when the denominator is zero.
\paragraph{Construction.} Starting from the initialization $\tilde\vr^{(1)} = \vr^{(1)} = \vec0$, we set the first three utilities
\begin{align*}
    \vu^{(1)} = (1,-1), \; \vu^{(2)} = (0,1), \; \vu^{(3)} = \vec0.
\end{align*}
The first three observations are
\begin{align*}
    \tilde\vr^{(2)} = (1,0), \; \vr^{(2)} = (1+\lambda,-\lambda), \; \vx^{(2)} = (1,0),
\end{align*}
and
\begin{align*}
    \tilde\vr^{(3)} = (1,1), \; \vr^{(3)} = (1,1+\lambda), \; \vx^{(3)} = \left(\frac{1}{2+\lambda}, \frac{1+\lambda}{2+\lambda}\right),
\end{align*}
and
\begin{align}
    \tilde\vr^{(4)} = \vr^{(4)} = (1,1), \; \vx^{(4)} = \left( \frac12, \frac12 \right).
    \label{eq:failure_above_half_start}
\end{align}
We seek a block that returns both regret vectors to this direction at a smaller scale. A sufficiently small $\varepsilon$ will keep every coordinate positive throughout the block, while sufficiently many nonzero utilities will make the regret vector norm decrease. We make the following fixed choices:
\begin{align}
    N = 1 + \left\lceil \frac{16\lambda}{2\lambda - 1} \right\rceil, \; \varepsilon = \frac{1}{8N(1+\lambda)}.
    \label{eq:failure_above_half_parameters}
\end{align}
We now prescribe the utilities
\begin{align}
    \vu^{(t)} =
    \begin{cases}
        \varepsilon (-r_2^{(t)}, r_1^{(t)}), & 4 \le t \le N+2,\\
        \vec0, & t = N+3,\\
        \varepsilon (r_2^{(t)}, -r_1^{(t)}), & N+4 \le t \le 2N+2,\\
        \vec0, & t = 2N+3.
    \end{cases}
    \label{eq:failure_above_half_block}
\end{align}
Each half contains $N-1$ nonzero utilities followed by one zero utility, so the full block has $2N$ observations. 
\paragraph{No clipping in a block.} First, we prove by induction that every coordinate of both regret vectors lies in $[1/2,3/2]$ throughout the block, and that the actual regret vectors updates satisfy
\begin{align}
    \tilde\vr^{(t+1)} = \tilde\vr^{(t)} + \vu^{(t)}, \; \vr^{(t+1)} = \tilde\vr^{(t+1)} + \lambda \vu^{(t)}.
    \label{eq:failure_above_half_linear_updates}
\end{align}
At $t=4$, both vectors equal $(1,1)$, and there are no preceding updates within the block. Fix an iterate $t$ where $4 \le t \le 2N+3$. We suppose that $\vr^{(s)}, \tilde \vr^{(s)} \in [1/2,3/2]^2$ for every $4 \le s \le t$, and that \eqref{eq:failure_above_half_linear_updates} holds for $4 \le s < t$.

For every $4 \le s \le t$, the induction hypothesis and the utility rule give
\begin{align*}
    \ip{\vu^{(s)}, \vx^{(s)}} = \frac{\ip{\vu^{(s)}, \vr^{(s)}}}{\|\vr^{(s)}\|_1} = 0, \; \|\vu^{(s)}\|_\infty \le \varepsilon \max\{|r_2^{(s)}|,|r_1^{(s)}|\} \le \frac32 \varepsilon.
\end{align*}
Summing the previous update identities gives
\begin{align*}
    \tilde{\vr}^{(t)} - (1,1) = \tilde{\vr}^{(t)} - \tilde{\vr}^{(4)} = \sum_{s=4}^{t-1}\left(\tilde{\vr}^{(s+1)}-\tilde{\vr}^{(s)}\right) = \sum_{s=4}^{t-1}\vu^{(s)},
\end{align*}
which implies 
\begin{align*}
    \tilde{\vr}^{(t)} + \vu^{(t)} - (1,1)  = \sum_{s=4}^{t}\vu^{(s)}.
\end{align*}
Since there are at most $2N-2$ nonzero utilities in the full block, we obtain
\begin{align*}
    \left\|\tilde\vr^{(t)} + \vu^{(t)} - (1,1)\right\|_\infty = \left\| \sum_{s=4}^{t}\vu^{(s)}\right\|_\infty \le \frac32(2N-2)\varepsilon \le \frac32(2N-2+\lambda)\varepsilon \le \frac38,
\end{align*}
where the last inequality follows from $2N-2+\lambda \le 2N(1+\lambda)$ and \eqref{eq:failure_above_half_parameters}. In particular, $\tilde\vr^{(t)} + \vu^{(t)} \in [5/8,11/8]^2$, so clipping is inactive and $\tilde\vr^{(t+1)} = \tilde\vr^{(t)} + \vu^{(t)}$. For the subsequent prediction regret vector, we similarly get
\begin{align*}
    \left\|\vr^{(t+1)} - (1,1)\right\|_\infty = \left\|\sum_{s=4}^{t}\vu^{(s)} + \lambda\vu^{(t)}\right\|_\infty \le \frac32(2N-2+\lambda)\varepsilon \le \frac38.
\end{align*}
Consequently, for $i = 1,2$, we obtain
\begin{align*}
    \frac58 \le \tilde r_i^{(t+1)}, r_i^{(t+1)}  \le \frac{11}{8}.
\end{align*}
Thus, both vectors at time $t+1$ belong to $[5/8,11/8]^2 \subset [1/2,3/2]^2$, and \eqref{eq:failure_above_half_linear_updates} holds at time $t$. This completes the induction through time $2N+4$.

In particular, truncation is inactive throughout the block. For $4 \le t \le 2N+4$, we also have
\begin{align*}
    \frac12 = 2\left(\frac12\right)^2 \le \|\vr^{(t)}\|_2^2 \le 2\left(\frac32\right)^2 = \frac92,
\end{align*}
which implies, at every nonzero-utility observation,
\begin{align}
    \frac{\varepsilon^2}{2} \le \|\vu^{(t)}\|_2^2 \le \frac{9\varepsilon^2}{2}, \text{ where } t \in \{4,\ldots,N+2\} \cup \{N+4,\ldots,2N+2\}.
    \label{eq:failure_above_half_positive_bounds}
\end{align}
The upper bound also holds at the two zero-utility observations. Moreover, for $4 \le t \le 2N+3$, every utility in the block satisfies
\begin{align*}
    \|\vu^{(t)}\|_\infty \le \frac32\varepsilon = \frac{3}{16N(1+\lambda)} < 1.
\end{align*}
\paragraph{Regret norm decrease over the first half block.} We now show that the first half block decreases the regret-vector norm. Since $\vr^{(t)} = \tilde\vr^{(t)} + \lambda\vu^{(t-1)}$ and $\ip{\vu^{(t)}, \vr^{(t)}} = 0$, we obtain
\begin{align*}
    \|\tilde\vr^{(t+1)}\|_2^2 - \|\tilde\vr^{(t)}\|_2^2 = 2 \ip{\vu^{(t)}, \tilde\vr^{(t)}} + \|\vu^{(t)}\|_2^2 = \|\vu^{(t)}\|_2^2 - 2\lambda \ip{\vu^{(t-1)}, \vu^{(t)}}.
\end{align*}
Using $2\ip{\vu^{(t-1)}, \vu^{(t)}} = \|\vu^{(t-1)}\|_2^2 + \|\vu^{(t)}\|_2^2 - \|\vu^{(t)} - \vu^{(t-1)}\|_2^2$ and summing over $t=4,\ldots,N+3$, we get
\begin{align}
    \|\tilde\vr^{(N+4)}\|_2^2 -  \|\tilde\vr^{(4)}\|_2^2 = -(2\lambda-1)\sum_{t=4}^{N+2}\|\vu^{(t)}\|_2^2 + \lambda\sum_{t=4}^{N+3}\|\vu^{(t)} - \vu^{(t-1)}\|_2^2,
    \label{eq:failure_above_half_half_energy}
\end{align}
where we use $\vu^{(3)} = \vu^{(N+3)} = \vec0$.

Each of the $N-1$ nonzero utilities contributes at least $\varepsilon^2/2$ to the first sum by \eqref{eq:failure_above_half_positive_bounds}. The choice of $N$ therefore gives
\begin{align}
    (2\lambda-1)\sum_{t=4}^{N+2}\|\vu^{(t)}\|_2^2 \ge \frac{(2\lambda-1)(N-1)}{2}\varepsilon^2 \ge 8\lambda\varepsilon^2.
    \label{eq:failure_above_half_negative_part}
\end{align}
For the utility variation sum, the initial and final changes contribute $2\varepsilon^2$ and at most $9\varepsilon^2/2$, respectively. This is because
\begin{align*}
    \|\vu^{(4)} - \vu^{(3)}\|_2^2 &= \|\vu^{(4)}\|_2^2 = 2 \varepsilon^2, \\
     \|\vu^{(N+3)} - \vu^{(N+2)}\|_2^2 &=  \|\vu^{(N+2)}\|_2^2 \overset{\eqref{eq:failure_above_half_positive_bounds}} {\le}\frac{9 \varepsilon^2}{2}.
\end{align*}
To bound the interior terms for $5 \le t \le N+2$, \eqref{eq:failure_above_half_linear_updates} gives
\begin{align*}
    \vr^{(t)} - \vr^{(t-1)} = (1+\lambda)\vu^{(t-1)} - \lambda\vu^{(t-2)} .
\end{align*}
The utilities satisfy $\|\vu^{(t)}\|_2 \le 3\varepsilon/\sqrt2$ from \eqref{eq:failure_above_half_positive_bounds}, so we have, for $5 \le t \le N+2$, 
\begin{align*}
    \|\vu^{(t)} - \vu^{(t-1)}\|_2 = \varepsilon\|\vr^{(t)} - \vr^{(t-1)}\|_2 \le \frac{3}{\sqrt2}(1+2\lambda)\varepsilon^2.
\end{align*}
Consequently, we obtain
\begin{align}
    \sum_{t=4}^{N+3}\|\vu^{(t)} - \vu^{(t-1)}\|_2^2 \le 2\varepsilon^2 + \frac{9\varepsilon^2}{2} + \frac92(N-2)(1+2\lambda)^2\varepsilon^4 \le \frac{13 \varepsilon^2}{2} + \frac{9(N-2)\varepsilon^2}{32N^2} < 7\varepsilon^2,
    \label{eq:failure_above_half_variation_part}
\end{align}
where we substitute \eqref{eq:failure_above_half_parameters} and use $1+2\lambda \le 2(1+\lambda)$ and $N \ge 3$. Substituting \eqref{eq:failure_above_half_negative_part} and \eqref{eq:failure_above_half_variation_part} into \eqref{eq:failure_above_half_half_energy}, we obtain
\begin{align}
    \|\tilde\vr^{(N+4)}\|_2^2 - \|\tilde\vr^{(4)}\|_2^2 \le -8\lambda\varepsilon^2 + \lambda \sum_{t=4}^{N+3}\|\vu^{(t)} - \vu^{(t-1)}\|_2^2 < -8\lambda\varepsilon^2 + 7\lambda\varepsilon^2 = -\lambda\varepsilon^2 < 0.
    \label{eq:failure_above_half_decrease}
\end{align}
Hence, the regret vector norm at the end of the forward half block is strictly smaller than at its beginning.
\paragraph{Regret vector to original direction over the second half block.} It remains to prove that the second half block returns the regret vector to its original direction. We denote the skew symmetric matrix $\mJ$ as
\begin{align*}
    \mJ = \begin{pmatrix}0&-1\\1&0\end{pmatrix}.
\end{align*}
It straightforwardly satisfy $\mJ^2 = -\mI, \mJ^\top = -\mJ$, where $\mI$ is the identity matrix. The first update of the block can be written in terms of $\mJ$ as $\tilde\vr^{(5)}=(I+\varepsilon J)\tilde\vr^{(4)}$. For $5 \le t \le N+2$, we have $\vu^{(t-1)}=\tilde\vr^{(t)}-\tilde\vr^{(t-1)}$, so
\begin{align}
    \tilde\vr^{(t+1)} = \left(\mI + (1 + \lambda)\varepsilon \mJ\right)\tilde\vr^{(t)} - \lambda\varepsilon \mJ\tilde\vr^{(t-1)}.
    \label{eq:failure_above_half_recurrence}
\end{align}
We can define the real polynomials that represent this recurrence:
\begin{align*}
    p_0(z)=1, \; p_1(z)=1+z, \; p_{k+1}(z)=\left(1+(1+\lambda)z\right)p_k(z)-\lambda z p_{k-1}(z),
\end{align*}
for $k \ge 1$. We now aim to prove by induction on $k$ that
\begin{align*}
    \tilde\vr^{(4+k)} = p_k(\varepsilon\mJ)\tilde\vr^{(4)},
\end{align*}
for $ 0 \le k \le N-1$.  For $k=0$, we have $p_0(\varepsilon\mJ)=\mI$, so the identity holds. For $k=1$, the first update gives
\begin{align*}
    \tilde\vr^{(5)} = (\mI+\varepsilon\mJ)\tilde\vr^{(4)} = p_1(\varepsilon\mJ)\tilde\vr^{(4)}.
\end{align*}
Now fix $1 \le k \le N-2$ and suppose the identity holds for $k-1$ and $k$, namely,
\begin{align*}
    \tilde\vr^{(3+k)} = p_{k-1}(\varepsilon\mJ)\tilde\vr^{(4)}, \; \tilde\vr^{(4+k)} = p_k(\varepsilon\mJ)\tilde\vr^{(4)}.
\end{align*}
Since $5 \le 4+k \le N+2$, we can apply \eqref{eq:failure_above_half_recurrence} at time $t=4+k$ to obtain
\begin{align*}
    \tilde\vr^{(5+k)} = \left(\mI+(1+\lambda)\varepsilon\mJ\right)\tilde\vr^{(4+k)} - \lambda\varepsilon\mJ\tilde\vr^{(3+k)}.
\end{align*}
Substituting the two induction hypotheses and using the definition of $p_{k+1}$, we get
\begin{align*}
    \tilde\vr^{(5+k)} = \left[\left(\mI+(1+\lambda)\varepsilon\mJ\right)p_k(\varepsilon\mJ) - \lambda\varepsilon\mJ p_{k-1}(\varepsilon\mJ)\right]\tilde\vr^{(4)} = p_{k+1}(\varepsilon\mJ)\tilde\vr^{(4)}.
\end{align*}
This proves the identity for $k+1$ and completes the induction.

Taking $k=N-1$ in the preceding identity gives $\tilde\vr^{(N+3)} = p_{N-1}(\varepsilon\mJ)\tilde\vr^{(4)}$. Since $\vu^{(N+3)}=\vec0$, the next update satisfies
\begin{align*}
    \tilde\vr^{(N+4)} = \tilde\vr^{(N+3)} + \vu^{(N+3)} = \tilde\vr^{(N+3)}, \; \vr^{(N+4)} = \tilde\vr^{(N+4)} + \lambda\vu^{(N+3)} = \tilde\vr^{(N+4)}.
\end{align*}
Consequently, we obtain
\begin{align}
    \tilde\vr^{(N+4)} = \vr^{(N+4)} = p_{N-1}(\varepsilon\mJ)\tilde\vr^{(4)}.
    \label{eq:failure_above_half_forward_map}
\end{align}

The second half block uses $\vu^{(t)} = -\varepsilon\mJ\vr^{(t)}$ for $N+4 \le t \le 2N+2$. Since $\vr^{(N+4)} = \tilde\vr^{(N+4)}$, its first update is
\begin{align*}
    \tilde\vr^{(N+5)} = \tilde\vr^{(N+4)} - \varepsilon\mJ\vr^{(N+4)} = (\mI-\varepsilon\mJ)\tilde\vr^{(N+4)} = p_1(-\varepsilon\mJ)\tilde\vr^{(N+4)}.
\end{align*}
For the remaining nonzero updates, the same derivation as in \eqref{eq:failure_above_half_recurrence} gives
\begin{align*}
    \tilde\vr^{(t+1)} = \left(\mI-(1+\lambda)\varepsilon\mJ\right)\tilde\vr^{(t)} + \lambda\varepsilon\mJ\tilde\vr^{(t-1)},
\end{align*}
for $N+5 \le t \le 2N+2.$ Thus, both the initial update and the recurrence are those of the first half with $\varepsilon\mJ$ replaced by $-\varepsilon\mJ$. Applying the similar induction, we obtain
\begin{align*}
    \tilde\vr^{(N+4+k)} = p_k(-\varepsilon\mJ)\tilde\vr^{(N+4)}, 
\end{align*}
for $0 \le k \le N-1$. In particular, at $k=N-1$, we have
\begin{align*}
    \tilde\vr^{(2N+3)} = p_{N-1}(-\varepsilon\mJ)\tilde\vr^{(N+4)}.
\end{align*}
Finally, $\vu^{(2N+3)}=\vec0$ gives
\begin{align*}
    \tilde\vr^{(2N+4)} = \tilde\vr^{(2N+3)}, \; \vr^{(2N+4)} = \tilde\vr^{(2N+4)}.
\end{align*}
Substituting the preceding expression and \eqref{eq:failure_above_half_forward_map}, we obtain
\begin{align}
    \tilde\vr^{(2N+4)} = \vr^{(2N+4)} = p_{N-1}(-\varepsilon\mJ)p_{N-1}(\varepsilon\mJ)\tilde\vr^{(4)}.
    \label{eq:failure_above_half_full_map}
\end{align}
We first show that the matrix multiplying $\tilde\vr^{(4)}$ in \eqref{eq:failure_above_half_full_map} is a scalar multiple of $\mI$. Replacing $z$ by $-z$ in the polynomial $p_{N-1}(-z)p_{N-1}(z)$ gives
\begin{align*}
    p_{N-1}(z)p_{N-1}(-z) = p_{N-1}(-z)p_{N-1}(z).
\end{align*}
Thus, this replacement leaves the polynomial unchanged. Since replacing $z$ by $-z$ reverses the sign of every odd power, all coefficients of odd powers must be zero. The product therefore contains only powers $z^{2j}$. After substituting $z=\varepsilon\mJ$, each such power satisfies
\begin{align*}
    (\varepsilon\mJ)^{2j} = \varepsilon^{2j}(\mJ^2)^j = (-1)^j\varepsilon^{2j}\mI.
\end{align*}
Every term in the product is consequently a scalar multiple of $\mI$, so there is a real scalar $\rho$ such that
\begin{align*}
    p_{N-1}(-\varepsilon\mJ)p_{N-1}(\varepsilon\mJ) = \rho\mI.
\end{align*}

We next identify $\rho$ using the norm at the end of the forward half block. Since the polynomial coefficients are real, transposing its matrix argument gives
\begin{align*}
    p_{N-1}(\varepsilon\mJ)^\top = p_{N-1}((\varepsilon\mJ)^\top) = p_{N-1}(-\varepsilon\mJ),
\end{align*}
where we use $\mJ^\top=-\mJ$. Hence,
\begin{align*}
    p_{N-1}(\varepsilon\mJ)^\top p_{N-1}(\varepsilon\mJ) = \rho\mI.
\end{align*}
Taking the inner product with $\tilde\vr^{(4)}$ on both sides, we obtain
\begin{align*}
    \|p_{N-1}(\varepsilon\mJ)\tilde\vr^{(4)}\|_2^2 = \ip{\tilde\vr^{(4)}, p_{N-1}(\varepsilon\mJ)^\top p_{N-1}(\varepsilon\mJ)\tilde\vr^{(4)}} = \rho\|\tilde\vr^{(4)}\|_2^2.
\end{align*}
By \eqref{eq:failure_above_half_forward_map}, the vector inside the norm on the left is $\tilde\vr^{(N+4)}$. Since $\tilde\vr^{(4)}=(1,1)$, we therefore get
\begin{align*}
    \rho = \frac{\|\tilde\vr^{(N+4)}\|_2^2}{\|\tilde\vr^{(4)}\|_2^2} = \frac{\|\tilde\vr^{(N+4)}\|_2^2}{2}.
\end{align*}
The vector $\tilde\vr^{(N+4)}$ has positive coordinates, and \eqref{eq:failure_above_half_decrease} gives $\|\tilde\vr^{(N+4)}\|_2^2 < \|\tilde\vr^{(4)}\|_2^2 = 2$. Consequently, $0<\rho<1$. Substituting the scalar matrix into \eqref{eq:failure_above_half_full_map}, we conclude that
\begin{align}
    \tilde\vr^{(2N+4)} = \vr^{(2N+4)} = \rho\tilde\vr^{(4)} = \rho(1,1),
    \label{eq:failure_above_half_return}
\end{align}
where $0 <\rho <1$. Thus, the full block returns both regret vectors to their {\em original directions} with {\em strictly smaller norms}.
\paragraph{Repeating the block to violate Alt-RVU.}
We now repeat the utility rule \eqref{eq:failure_above_half_block}, shifting its indices by $2N$ after each block. By \eqref{eq:failure_above_half_return}, the second block starts from $\tilde\vr^{(2N+4)} = \vr^{(2N+4)} = \rho(1,1)$. Multiplying a positive prediction vector by $\rho>0$ leaves its strategy unchanged, since
\begin{align*}
    \frac{\rho\vr^{(t)}}{\|\rho\vr^{(t)}\|_1} = \frac{\vr^{(t)}}{\|\vr^{(t)}\|_1} = \vx^{(t)}.
\end{align*}
The utility rule therefore multiplies the corresponding utility by $\rho$. Since the payoff remains zero, the two regret vectors updates give
\begin{align*}
    [\rho\tilde\vr^{(t)} + \rho\vu^{(t)}]_+ = \rho\tilde\vr^{(t+1)}, \; \rho\tilde\vr^{(t+1)} + \lambda\rho\vu^{(t)} = \rho\vr^{(t+1)}.
\end{align*}
Thus, every regret vector and utility in the second block is $\rho$ times its counterpart in the first block, whereas the strategies are identical. Positivity is preserved because $\rho>0$. Repeating this argument over successive blocks, we obtain
\begin{align*}
    \tilde\vr^{(t+2N)} = \rho\tilde\vr^{(t)}, \; \vr^{(t+2N)} = \rho\vr^{(t)}, \; \vu^{(t+2N)} = \rho\vu^{(t)}, \; \vx^{(t+2N)} = \vx^{(t)},
\end{align*}
for $t \ge 4$. The repeated strategies are not constant. Indeed, $\vu^{(4)}=\varepsilon(-1,1)$ and $\tilde\vr^{(4)}=\vr^{(4)}=(1,1)$ give
\begin{align*}
    \vr^{(5)} = \left(1-(1+\lambda)\varepsilon,1+(1+\lambda)\varepsilon\right), \; \vx^{(5)} = \left(\frac{1-(1+\lambda)\varepsilon}{2},\frac{1+(1+\lambda)\varepsilon}{2}\right).
\end{align*}
Since $\vx^{(4)}=(1/2,1/2)$, we obtain
\begin{align*}
    \|\vx^{(5)}-\vx^{(4)}\|_2^2 = \frac{(1+\lambda)^2\varepsilon^2}{2}>0.
\end{align*}
Every complete block contains this same strategy movement. There are $\lfloor (T-3)/(2N)\rfloor$ complete blocks among observations $4,\ldots,T$, so, for $T\ge3$, we get
\begin{align*}
    \sum_{t=1}^{T}\|\vx^{(t+1)}-\vx^{(t)}\|_2^2 \ge \left\lfloor\frac{T-3}{2N}\right\rfloor \frac{(1+\lambda)^2\varepsilon^2}{2}.
\end{align*}
On the other hand, $\|\vx^{(t+1)}-\vx^{(t)}\|_2^2 \le 2$ because both strategies belong to $\Delta(2)$. Consequently,
\begin{align*}
    \sum_{t=1}^{T}\|\vx^{(t+1)}-\vx^{(t)}\|_2^2 = \Theta(T).
\end{align*}

In contrast, the utilities decrease geometrically across blocks. Since $0<\rho<1$, we have
\begin{align*}
    \sum_{t=1}^{\infty}\|\vu^{(t)}\|_2 = \sum_{t=1}^{3}\|\vu^{(t)}\|_2 + \sum_{k=0}^{\infty}\rho^k \sum_{t=4}^{2N+3}\|\vu^{(t)}\|_2 = \sum_{t=1}^{3}\|\vu^{(t)}\|_2 + \frac{1}{1-\rho}\sum_{t=4}^{2N+3}\|\vu^{(t)}\|_2 < \infty.
\end{align*}
All utilities lie in $[-1,1]^2$, since this holds for the initialization and the first block, as subsequent blocks only decrease their magnitudes. Moreover, we note that
\begin{align*}
    \sum_{t=2}^{\infty}\|\vu^{(t)}-\vu^{(t-1)}\|_2^2 \le 2\sum_{t=2}^{\infty}\left(\|\vu^{(t)}\|_2^2+\|\vu^{(t-1)}\|_2^2\right) \le 4\sum_{t=1}^{\infty}\|\vu^{(t)}\|_2^2 \le 4\left(\sum_{t=1}^{\infty}\|\vu^{(t)}\|_2\right)^2 < \infty.
\end{align*}
Finally, for any $\vz\in\Delta(2)$, the triangle inequality gives $\|2\vz-\vx^{(t)}-\vx^{(t+1)}\|_2 \le 4$. Hence, alternating regret remains bounded in absolute value:
\begin{align*}
    |\operatorname{Alt-Reg}(T)| \le \max_{\vz\in\Delta(2)}\sum_{t=1}^{T}\left|\ip{\vu^{(t)},2\vz-\vx^{(t)}-\vx^{(t+1)}}\right| \le 4\sum_{t=1}^{T}\|\vu^{(t)}\|_2 \le 4\sum_{t=1}^{\infty}\|\vu^{(t)}\|_2 < \infty.
\end{align*}
For any horizon-independent constants $\alpha,\beta,\gamma>0$, we therefore have
\begin{align*}
    \alpha + \beta\sum_{t=2}^{T}\|\vu^{(t)}-\vu^{(t-1)}\|_2^2 - \gamma\sum_{t=1}^{T}\|\vx^{(t+1)}-\vx^{(t)}\|_2^2 \longrightarrow -\infty,
\end{align*}
whereas $\operatorname{Alt-Reg}(T)$ is bounded below. Thus, the Alt-RVU inequality cannot hold for every horizon on this sequence, even when its constants depend on the entire sequence. By norm equivalence in dimension two, the same contradiction holds for any fixed pair of primal-dual norms. Since refined Alt-RVU adds a nonnegative term to the left-hand side, it cannot hold either.
\end{proof}
\altRvuFailureBelowHalf*
\begin{proof}
We adapt the forward and reverse half blocks from the proof of Proposition~\ref{prop:failure_above_half_explicit}. For $\lambda<1/2$, these blocks increase the regret vector norm. We let their lengths increase with this norm and vary the utility coefficients smoothly within each half block to control the squared utility variation.

\paragraph{Construction.} We choose the initial direction
\begin{align*}
    \vv = \left(1,\frac1{1+\sqrt d},\ldots,\frac1{1+\sqrt d}\right), \; \|\vv\|_1 = \sqrt d, \; \|\vv\|_2^2 = \frac{2\sqrt d}{1+\sqrt d}.
\end{align*}
The assumption on $d$ gives
\begin{align*}
    \frac{\|\vv\|_2^2}{(1-2\lambda)\|\vv\|_1} = \frac{2}{(1-2\lambda)(1+\sqrt d)} < 1.
\end{align*}
We can therefore fix $\varepsilon>0$ sufficiently small that
\begin{align}
    \varepsilon < \frac1{2(1+\sqrt d)}, \; (1-\varepsilon)^2 > \frac{\|\vv\|_2^2}{(1-2\lambda)\|\vv\|_1}.
    \label{eq:failure_below_half_parameters}
\end{align}
We use the skew-symmetric matrix
\begin{align*}
    \mJ = \frac{(\ve_2-\ve_3)\vv^\top-\vv(\ve_2-\ve_3)^\top}{\sqrt2\|\vv\|_2},
\end{align*}
where $\ve_i$ is the $i$-th coordinate vector. Since $\ip{\vv,\ve_2-\ve_3}=0$, this matrix satisfies $\mJ^\top=-\mJ$ and $\mJ^2=-\mI$ on $\operatorname{span}\{\vv,\ve_2-\ve_3\}$. In particular, $\mJ$ preserves the Euclidean norm on this plane.

Starting from the standard initialization, we set
\begin{align*}
    \vu^{(1)} = \left(1,-\frac1{d-1},\ldots,-\frac1{d-1}\right), \; \vu^{(2)} = \left(0,\frac1{1+\sqrt d},\ldots,\frac1{1+\sqrt d}\right), \; \vu^{(3)} = \vec0.
\end{align*}
The first observation gives $\tilde\vr^{(2)}=\ve_1$ and $\vx^{(2)}=\ve_1$. The second gives $\tilde\vr^{(3)}=\vv$, and the third resets the prediction, so
\begin{align*}
    \tilde\vr^{(4)} = \vr^{(4)} = \vv.
\end{align*}
All three played utilities are zero.

We construct successive blocks. Let $\tau_k$ be the first observation of block $k$, where both regret vectors equal $s_k\vv$, with $\tau_1=4$ and $s_1=1$. We choose
\begin{align*}
    N_k = \max\{2,\lceil s_k\rceil\}, \; \tau_{k+1} = \tau_k+2N_k,
\end{align*}
and, for $1 \le j \le N_k$, prescribe
\begin{align}
    \vu^{(\tau_k+j-1)} &= \frac{\varepsilon\sin(\pi j/N_k)}{4(1+\lambda)\|\vv\|_2N_k}\mJ\vr^{(\tau_k+j-1)}, \nonumber\\
    \vu^{(\tau_k+N_k+j-1)} &= -\frac{\varepsilon\sin(\pi j/N_k)}{4(1+\lambda)\|\vv\|_2N_k}\mJ\vr^{(\tau_k+N_k+j-1)}.
    \label{eq:failure_below_half_block}
\end{align}
Each half ends with a zero utility. We will show that the full block returns both regret vectors to the direction $\vv$, thereby defining $s_{k+1}$ and allowing the construction to continue. All implicit constants below depend only on $\lambda$, $d$, and the fixed choice of $\varepsilon$.

\paragraph{No clipping in a block.} We follow the induction used in the preceding proof. Suppose that the prediction regret vectors up to the current observation satisfy $\|\vr^{(t)}\|_2 \le 2s_k\|\vv\|_2$ and have positive coordinates. Skew symmetry gives $\ip{\vu^{(t)},\vx^{(t)}}=0$, while the utility rule gives
\begin{align*}
    \|\vu^{(t)}\|_2 \le \frac{\varepsilon s_k}{2(1+\lambda)N_k}.
\end{align*}
Summing the previous linear updates, the candidate observation update satisfies
\begin{align*}
    \|\tilde\vr^{(t)}+\vu^{(t)}-s_k\vv\|_2 \le 2N_k\frac{\varepsilon s_k}{2(1+\lambda)N_k} \le \varepsilon s_k.
\end{align*}
Every coordinate of $s_k\vv$ is at least $s_k/(1+\sqrt d)$, so this candidate remains positive by~\eqref{eq:failure_below_half_parameters}. Thus, clipping is inactive and~\eqref{eq:failure_above_half_linear_updates} holds. The next prediction regret vector satisfies
\begin{align*}
    \|\vr^{(t+1)}-s_k\vv\|_2 \le (2N_k+\lambda)\frac{\varepsilon s_k}{2(1+\lambda)N_k} \le \varepsilon s_k.
\end{align*}
It also has positive coordinates and norm at most $2s_k\|\vv\|_2$, completing the induction.

Consequently, throughout the block, including its endpoints,
\begin{align}
    \|\tilde\vr^{(t)}-s_k\vv\|_2 \le \varepsilon s_k, \; \|\vr^{(t)}-s_k\vv\|_2 \le \varepsilon s_k.
    \label{eq:failure_below_half_state_bounds}
\end{align}
In particular,
\begin{align}
    \frac{s_k\|\vv\|_2}{2} \le \|\vr^{(t)}\|_2 \le 2s_k\|\vv\|_2, \; \|\vr^{(t)}\|_1 \ge (1-\varepsilon)s_k\|\vv\|_1, \; \tilde r_1^{(t)} \ge (1-\varepsilon)s_k.
    \label{eq:failure_below_half_norm_bounds}
\end{align}
For the $\ell_1$ bound, we use $\|\vv\|_1=\sqrt d$ and $\|\vr^{(t)}-s_k\vv\|_1 \le \sqrt d\|\vr^{(t)}-s_k\vv\|_2$. Moreover, since $N_k \ge s_k$,
\begin{align*}
    \|\vu^{(t)}\|_\infty \le \|\vu^{(t)}\|_2 \le \frac{\varepsilon s_k}{2(1+\lambda)N_k} \le \frac{\varepsilon}{2(1+\lambda)} < 1.
\end{align*}
Thus, every utility belongs to $[-1,1]^d$.

\paragraph{Utility bounds within a block.} All regret vectors remain in $\operatorname{span}\{\vv,\ve_2-\ve_3\}$, where $\mJ$ preserves their Euclidean norms. Combining~\eqref{eq:failure_below_half_norm_bounds} with
\begin{align*}
    \sum_{j=1}^{N_k}\sin^2(\pi j/N_k) = \frac{N_k}{2},
\end{align*}
we obtain
\begin{align*}
    \sum_{t=\tau_k}^{\tau_k+N_k-1}\|\vu^{(t)}\|_2^2 = \Theta\left(\frac{s_k^2}{N_k}\right).
\end{align*}
The same estimate holds for the second half.

To bound utility variation, we use
\begin{align*}
    |\sin(\pi j/N_k)-\sin(\pi(j-1)/N_k)| \le \frac{\pi}{N_k}.
\end{align*}
At every observation after the first in either half block, the linear updates give
\begin{align*}
    \vr^{(t)}-\vr^{(t-1)} = (1+\lambda)\vu^{(t-1)}-\lambda\vu^{(t-2)}.
\end{align*}
Therefore,
\begin{align*}
    \|\vu^{(t)}-\vu^{(t-1)}\|_2 \le \frac{\pi\varepsilon\|\vr^{(t-1)}\|_2}{4(1+\lambda)\|\vv\|_2N_k^2} + \frac{\varepsilon\|\vr^{(t)}-\vr^{(t-1)}\|_2}{4(1+\lambda)\|\vv\|_2N_k} = O\left(\frac{s_k}{N_k^2}\right).
\end{align*}
At the first observation of either half, the preceding utility is zero, and the same bound follows directly from $\sin(\pi/N_k)\le\pi/N_k$. Hence,
\begin{align}
    \sum_{t=\tau_k}^{\tau_{k+1}-1}\|\vu^{(t)}\|_2^2 &= \Theta\left(\frac{s_k^2}{N_k}\right), \nonumber\\
    \sum_{t=\tau_k}^{\tau_{k+1}-1}\|\vu^{(t)}-\vu^{(t-1)}\|_2^2 &= O\left(\frac{s_k^2}{N_k^3}\right).
    \label{eq:failure_below_half_block_bounds}
\end{align}
These estimates also hold when the sums are restricted to either half block.

\paragraph{Return to the original direction and norm growth.} The recurrence~\eqref{eq:failure_above_half_recurrence}, with the prescribed coefficient at each observation, expresses the first half map as a real polynomial in $\mJ$. The second half uses the same coefficients with $\mJ$ replaced by $-\mJ$. Both halves begin with equal regret and prediction regret vectors because the preceding utility is zero. Thus, the polynomial argument leading to~\eqref{eq:failure_above_half_full_map} applies on $\operatorname{span}\{\vv,\ve_2-\ve_3\}$ and gives
\begin{align*}
    \tilde\vr^{(\tau_{k+1})} = \vr^{(\tau_{k+1})} = s_{k+1}\vv, \; \frac{s_{k+1}}{s_k} = \frac{\|\tilde\vr^{(\tau_k+N_k)}\|_2^2}{s_k^2\|\vv\|_2^2}.
\end{align*}
This defines the next block's scale.

The identity~\eqref{eq:failure_above_half_half_energy} does not require a constant utility coefficient. Applying it to the first half block, whose preceding and final utilities are zero, yields
\begin{align*}
    s_k\|\vv\|_2^2(s_{k+1}-s_k) = (1-2\lambda)\sum_{t=\tau_k}^{\tau_k+N_k-1}\|\vu^{(t)}\|_2^2 + \lambda\sum_{t=\tau_k}^{\tau_k+N_k-1}\|\vu^{(t)}-\vu^{(t-1)}\|_2^2.
\end{align*}
Since $1-2\lambda>0$, the preceding bounds imply
\begin{align*}
    s_{k+1}-s_k = \Theta\left(\frac{s_k}{N_k}\right) > 0.
\end{align*}
Starting from $s_1=1$, we therefore have $s_k \ge 1$ and $s_k \le N_k \le 2s_k$. Consequently,
\begin{align}
    s_{k+1}-s_k = \Theta(1), \; s_k = \Theta(k), \; N_k = \Theta(k), \; \tau_k = \Theta(k^2).
    \label{eq:failure_below_half_scale}
\end{align}

For a horizon $T$ in block $K$, we have $T=\Theta(K^2)$. Summing the utility variation bound in~\eqref{eq:failure_below_half_block_bounds} gives
\begin{align*}
    \sum_{t=2}^{T}\|\vu^{(t)}-\vu^{(t-1)}\|_2^2 \le O(1) + \sum_{k=1}^{K}O\left(\frac{s_k^2}{N_k^3}\right) = O\left(\sum_{k=1}^{K}\frac1k\right) = O(\log T).
\end{align*}

\paragraph{Alternating regret.} It remains to show that alternating regret grows as $\sqrt T$. We first bound cumulative utility improvement. Since $\ip{\vu^{(t)},\vx^{(t)}}=0$, the linear updates give
\begin{align*}
    \ip{\vu^{(t)},\vx^{(t+1)}-\vx^{(t)}} = \frac{\|\vu^{(t)}\|_2^2+\lambda\ip{\vu^{(t)},\vu^{(t)}-\vu^{(t-1)}}}{\|\vr^{(t+1)}\|_1}.
\end{align*}
Using~\eqref{eq:failure_below_half_norm_bounds}, Cauchy--Schwarz, and~\eqref{eq:failure_below_half_block_bounds}, we obtain
\begin{align*}
    \sum_{t=\tau_k}^{\tau_{k+1}-1}\left|\ip{\vu^{(t)},\vx^{(t+1)}-\vx^{(t)}}\right| \le \frac{\sum_{t=\tau_k}^{\tau_{k+1}-1}\|\vu^{(t)}\|_2^2}{(1-\varepsilon)s_k\|\vv\|_1} + O\left(\frac{s_k}{N_k^2}\right).
\end{align*}
Applying~\eqref{eq:failure_above_half_half_energy} over the full block also gives
\begin{align*}
    \|\vv\|_2^2(s_{k+1}^2-s_k^2) = (1-2\lambda)\sum_{t=\tau_k}^{\tau_{k+1}-1}\|\vu^{(t)}\|_2^2 + \lambda\sum_{t=\tau_k}^{\tau_{k+1}-1}\|\vu^{(t)}-\vu^{(t-1)}\|_2^2.
\end{align*}
The last sum is nonnegative. Substituting the resulting upper bound on the squared utilities, and using $s_{k+1}-s_k=\Theta(1)$, yields
\begin{align*}
    \sum_{t=\tau_k}^{\tau_{k+1}-1}\left|\ip{\vu^{(t)},\vx^{(t+1)}-\vx^{(t)}}\right| &\le \frac{\|\vv\|_2^2}{(1-2\lambda)(1-\varepsilon)\|\vv\|_1}\frac{s_{k+1}^2-s_k^2}{s_k} + O\left(\frac1{s_k}\right)\\
    &= \frac{2\|\vv\|_2^2}{(1-2\lambda)(1-\varepsilon)\|\vv\|_1}(s_{k+1}-s_k) + O\left(\frac1{s_k}\right).
\end{align*}
In particular, the sum of absolute utility improvements within any one block is $O(1)$. We can therefore sum over the complete blocks and include any incomplete final block to obtain
\begin{align}
    \left|\sum_{t=1}^{T}\ip{\vu^{(t)},\vx^{(t+1)}-\vx^{(t)}}\right| \le \frac{2\|\vv\|_2^2}{(1-2\lambda)(1-\varepsilon)\|\vv\|_1}s_K + O(\log K).
    \label{eq:failure_below_half_total_improvement}
\end{align}

Every played utility is zero. Moreover, the initialization and the subsequent linear updates imply, for $T \ge 3$,
\begin{align*}
    \sum_{t=1}^{T}\vu^{(t)} = \tilde\vr^{(T+1)}-\left(0,\frac1{d-1},\ldots,\frac1{d-1}\right).
\end{align*}
Thus, for $T$ in block $K$, \eqref{eq:failure_below_half_state_bounds} gives
\begin{align*}
    (1-\varepsilon)s_K \le \tilde r_1^{(T+1)} \le \operatorname{Reg}(T) \le \|\tilde\vr^{(T+1)}\|_\infty \le (1+\varepsilon)s_K.
\end{align*}
Combining this with~\eqref{eq:failure_below_half_total_improvement} and the identity between ordinary and alternating regret, we obtain
\begin{align*}
    \operatorname{Alt-Reg}(T) &= 2\operatorname{Reg}(T)-\sum_{t=1}^{T}\ip{\vu^{(t)},\vx^{(t+1)}-\vx^{(t)}}\\
    &\ge 2\left(1-\varepsilon-\frac{\|\vv\|_2^2}{(1-2\lambda)(1-\varepsilon)\|\vv\|_1}\right)s_K-O(\log K).
\end{align*}
The coefficient of $s_K$ is strictly positive by~\eqref{eq:failure_below_half_parameters}. Since $s_K=\Theta(K)=\Theta(\sqrt T)$, this proves $\operatorname{Alt-Reg}(T)=\Omega(\sqrt T)$. The upper bounds on ordinary regret and absolute cumulative utility improvement give $\operatorname{Alt-Reg}(T)=O(\sqrt T)$ as well.

We have constructed a single bounded infinite utility sequence satisfying
\begin{align*}
    \operatorname{Alt-Reg}(T) = \Theta(\sqrt T), \; \sum_{t=2}^{T}\|\vu^{(t)}-\vu^{(t-1)}\|_2^2 = O(\log T).
\end{align*}
For any horizon-independent constants $\alpha,\beta>0$, $\operatorname{Alt-Reg}$ eventually exceeds $\alpha+\beta\sum_{t=2}^{T}\|\vu^{(t)}-\vu^{(t-1)}\|_2^2$. Subtracting a nonnegative strategy-movement term only decreases this upper bound. Hence, Alt-RVU fails on this sequence for every choice of horizon-independent positive constants.
\end{proof}

\subsection{Proofs from Section \ref{sec:best_last_nfg}}

\refinedAltRvuConvergence*
\begin{proof}
    From the bound \eqref{eq:alt_payoff_telescope} for $\operatorname{Alt-Reg}$, we can further obtain
\begin{align}
    \operatorname{Alt-Reg}_{\cal X} (T) + \operatorname{Alt-Reg}_{\cal Y} (T) &= 2T \operatorname{Gap}(\bar{\vx}^{(T)}, \bar{\vy}^{(T)}) + (\vx^{(1)})^\top \mA \vy^{(0)} - (\vx^{(T+1)})^\top \mA \vy^{(T)} \notag \\
    &\ge (\vx^{(1)})^\top \mA \vy^{(0)} - (\vx^{(T+1)})^\top \mA \vy^{(T)} \notag \\
    &\ge - \operatorname{range}(\mA),
    \label{eq:sum_alt_reg_refined_bound}
\end{align}
where the first inequality follows from $\operatorname{Gap}(\bar{\vx}^{(T)}, \bar{\vy}^{(T)}) \ge 0$, and the second inequality holds because $(\vx^{(1)})^\top \mA \vy^{(0)} \ge \min_{i,j} A_{ij}$ and $(\vx^{(T+1)})^\top \mA \vy^{(T)} \le \max_{i,j} A_{ij}$.

Assume both $\vx$ and $\vy$ players satisfy the refined Alt-RVU bound, we can add the two bounds and absorb utility variation into the negative strategy movement 
\begin{align*}
&\zeta \sqrt{\sum_{t=1}^{T}
\left[\max_{\vz\in\mathcal X}
\left\langle A\vy^{(t)}, 2\vz-\vx^{(t)} -\vx^{(t+1)}\right\rangle\right]^2} + \zeta \sqrt{\sum_{t=1}^{T}
\left[\max_{\vz\in\mathcal Y}
\left\langle - A^\top\vx^{(t)}, 2\vz-\vy^{(t-1)}-\vy^{(t)} \right\rangle\right]^2}
\\
& \le 2 \alpha - \operatorname{Alt-Reg}_{\cal X} (T) - \operatorname{Alt-Reg}_{\cal Y} (T) \\
& \overset{\eqref{eq:sum_alt_reg_refined_bound}}{\le} 2 \alpha + \operatorname{range}(\mA).
\end{align*}
Combining this inequality with the pointwise upper bound on the Nash gap preceding \eqref{eq:square_gap_bound} and applying the triangle inequality, we obtain
\begin{align*}
    &\sqrt{\sum_{t=1}^{T} \operatorname{Gap}(\vx^{(t)}, \vy^{(t)})^2}
    \\
    &\le \sqrt{\sum_{t=1}^{T} \left[\max_{\vz \in \mathcal X} \ip{\mA\vy^{(t)}, 2\vz - \vx^{(t)} - \vx^{(t+1)}} + \max_{\vz \in \mathcal Y} \ip{-\mA^\top\vx^{(t)}, 2\vz - \vy^{(t-1)} - \vy^{(t)}}\right]^2} \\
    &\le \sqrt{\sum_{t=1}^{T} \left[\max_{\vz \in \mathcal X} \ip{\mA\vy^{(t)}, 2\vz - \vx^{(t)} - \vx^{(t+1)}}\right]^2} + \sqrt{\sum_{t=1}^{T} \left[\max_{\vz \in \mathcal Y} \ip{-\mA^\top\vx^{(t)}, 2\vz - \vy^{(t-1)} - \vy^{(t)}}\right]^2} \\
     &\le \frac{2\alpha + \operatorname{range}(\mA)}{\zeta}.
\end{align*}

Consequently, we obtain
\begin{align*}
    \min_{1 \le t \le T} \operatorname{Gap}(\vx^{(t)}, \vy^{(t)}) \le \left(\frac1T \sum_{t=1}^{T} \operatorname{Gap}(\vx^{(t)}, \vy^{(t)})^2\right)^{1/2} \le \frac{2\alpha + \operatorname{range}(\mA)}{\zeta\sqrt T},
\end{align*}
which implies an $O(T^{-1/2})$ best-iterate rate. Returning to the bound on $\sqrt{\sum_{t=1}^{T} \operatorname{Gap}(\vx^{(t)}, \vy^{(t)})^2}$, squaring both sides and letting $T \to \infty$, we obtain
\begin{align*}
    \sum_{t=1}^{\infty} \operatorname{Gap}(\vx^{(t)}, \vy^{(t)})^2 \le \left(\frac{2\alpha + \operatorname{range}(\mA)}{\zeta}\right)^2 < \infty,
\end{align*}
which implies $\operatorname{Gap}(\vx^{(t)}, \vy^{(t)}) \to 0$. Since the strategy space is compact and the Nash gap is continuous and vanishes exactly at Nash equilibria, the last iterates converge to the Nash set.
\end{proof}
\refineAltRvuErrors*
\begin{proof}
For $\IRPRMplus$, the regret vector update \eqref{eq:correction}, rewritten using $c^{(t)}$, gives
\begin{align*}
    \max_{\vz \in \Delta(d)} \ip{\vu^{(t)}, \vz - \vx^{(t)}} + c^{(t)}
    \le \max_{\vz \in \Delta(d)} \ip{\tilde\vr^{(t+1)} - \tilde\vr^{(t)}, \vz}
    \le \|\tilde\vr^{(t+1)} - \tilde\vr^{(t)}\|_2.
\end{align*}
Combining this with \eqref{eq:util_improvement_ir}, we obtain
\begin{align*}
    \max_{\vz \in \Delta(d)} \ip{\vu^{(t)}, 2\vz - \vx^{(t)} - \vx^{(t+1)}}
    &= 2\max_{\vz \in \Delta(d)} \ip{\vu^{(t)}, \vz - \vx^{(t)}} - \ip{\vu^{(t)}, \vx^{(t+1)} - \vx^{(t)}} \\
    &\le 2\|\tilde\vr^{(t+1)} - \tilde\vr^{(t)}\|_2 - c^{(t)} - c^{(t+1)} \\
    &\le 2\|\tilde\vr^{(t+1)} - \tilde\vr^{(t)}\|_2,
\end{align*}
where the last inequality uses $c^{(t)}, c^{(t+1)} \ge 0$.

For $\HPRMplus$, the coordinatewise inequality \eqref{eq:ineq_tilde_r} remains valid because its regret vector update is unchanged. Suppose first that $[\vr^{(t+1)}]_+ \ne \vec0$. The prediction identity established in the proof of Lemma~\ref{lem:half_bound_r_tilde} implies
\begin{align*}
    \tilde\vr_i^{(t+1)} = 0 \; \Longrightarrow \; [\vr_i^{(t+1)}]_+ = \left[ -\frac12 \tilde\vr_i^{(t)} \right]_+ = 0.
\end{align*}
Thus, the regret vector update is untruncated on the support of $\vx^{(t+1)}$, and we obtain
\begin{align*}
    \ip{\vu^{(t)}, \vx^{(t+1)} - \vx^{(t)}}
    = \ip{\vu^{(t)} - \ip{\vu^{(t)}, \vx^{(t)}}\vec1, \vx^{(t+1)}}
    = \ip{\tilde\vr^{(t+1)} - \tilde\vr^{(t)}, \vx^{(t+1)}}.
\end{align*}
Using this identity and \eqref{eq:ineq_tilde_r}, we get
\begin{align*}
    \max_{\vz \in \Delta(d)} \ip{\vu^{(t)}, 2\vz - \vx^{(t)} - \vx^{(t+1)}} \le \max_{\vz \in \Delta(d)} \ip{\tilde\vr^{(t+1)} - \tilde\vr^{(t)}, 2\vz - \vx^{(t+1)}} \le \sqrt5\|\tilde\vr^{(t+1)} - \tilde\vr^{(t)}\|_2,
\end{align*}
where the last inequality follows from Cauchy-Schwarz and
\begin{align*}
    \|2\vz - \vx^{(t+1)}\|_2^2 = 4\|\vz\|_2^2 + \|\vx^{(t+1)}\|_2^2 - 4\ip{\vz, \vx^{(t+1)}} \le 5.
\end{align*}
Finally, if $[\vr^{(t+1)}]_+ = \vec0$, \eqref{eq:prediction_lower_hprm} implies that activation has not occurred. Hence $\tilde\vr^{(t)} = \tilde\vr^{(t+1)} = \vec0$; uniform play and the regret vector update force the instantaneous regret vector to be zero. Both sides of \eqref{eq:errors} then vanish.
\end{proof}
\refinedAltRvuAlgorithms*
\begin{proof}
If $t_0 = \infty$, the instantaneous regret vectors and strategy movements are zero, so \eqref{eq:refined_rvu} holds with any positive constants. Otherwise, it suffices to consider $T \ge t_0 - 1$, since alternating regret and the best-response errors vanish for earlier horizons.

\paragraph{$\IRPRMplus$.}
We first retain the terms involving $c^{(t)}$ in the alternating-regret analysis. The coordinatewise regret vector update gives, for every $\vz \in \Delta(d)$,
\begin{align*}
    \sum_{t = 1}^{T} \ip{\vu^{(t)}, \vz - \vx^{(t)}} \le \ip{\tilde \vr^{(T + 1)}, \vz} - \sum_{t = 1}^{T} c^{(t)}.
\end{align*}
Using the identity between alternating regret and cumulative utility improvement, we obtain
\begin{align*}
    \operatorname{Alt-Reg}(T) \le 2 \| \tilde \vr^{(T + 1)} \|_2 - 2 \sum_{t = 1}^{T} c^{(t)} - \sum_{t = 1}^{T} \ip{\vu^{(t)}, \vx^{(t + 1)} - \vx^{(t)}}.
\end{align*}
Set $\gamma = \| \tilde \vr^{(t_0)} \|_2 / (4d) > 0$. Summing \eqref{eq:util_improvement_ir}, using the nondecreasing regret-vector norms and $c^{(1)} = 0$, gives
\begin{align*}
    \sum_{t = 1}^{T} \ip{\vu^{(t)}, \vx^{(t + 1)} - \vx^{(t)}} \ge \gamma \sum_{t = 1}^{T} \| \vx^{(t + 1)} - \vx^{(t)} \|_2^2 + c^{(T + 1)}.
\end{align*}
Consequently, we obtain
\begin{align}
    \operatorname{Alt-Reg}(T) \le 2 \| \tilde \vr^{(T + 1)} \|_2 - 2 \sum_{t = 1}^{T} c^{(t)} - \gamma \sum_{t = 1}^{T} \| \vx^{(t + 1)} - \vx^{(t)} \|_2^2.
    \label{eq:refined_ir_credit}
\end{align}

We next use the retained sum of $c^{(t)}$ to control the additional best-response errors. Norm preservation and the definition of $c^{(t)}$ imply
\begin{align*}
    c^{(t)} \| [\vr^{(t)}]_+ \|_1 = \| \tilde \vr^{(t)} \|_2^2 - \ip{\tilde \vr^{(t)}, [\vr^{(t)}]_+} = \frac12 \| [\vr^{(t)}]_+ - \tilde \vr^{(t)} \|_2^2.
\end{align*}
Moreover, $[\vr^{(t)}]_+ = \| [\vr^{(t)}]_+ \|_1 \vx^{(t)}$ gives
\begin{align*}
    \ip{\vu^{(t)} - \ip{\vu^{(t)}, \vx^{(t)}} \vec1, [\vr^{(t)}]_+} = 0.
\end{align*}
Taking the inner product of the coordinatewise regret vector update with $[\vr^{(t)}]_+$, we therefore obtain
\begin{align*}
    \ip{\tilde \vr^{(t + 1)}, [\vr^{(t)}]_+} \ge \ip{\tilde \vr^{(t)}, [\vr^{(t)}]_+} + c^{(t)} \| [\vr^{(t)}]_+ \|_1 = \| [\vr^{(t)}]_+ \|_2^2.
\end{align*}
Expanding the square and using norm preservation, we get
\begin{align*}
    \| \tilde \vr^{(t + 1)} - [\vr^{(t)}]_+ \|_2^2 \le \| \tilde \vr^{(t + 1)} \|_2^2 - \| \tilde \vr^{(t)} \|_2^2.
\end{align*}
We now decompose each regret-vector increment around $[\vr^{(t)}]_+$ and sum to obtain
\begin{align*}
    \sum_{t = 1}^{T} \| \tilde \vr^{(t + 1)} - \tilde \vr^{(t)} \|_2^2 \le 2 \sum_{t = 1}^{T} \| \tilde \vr^{(t + 1)} - [\vr^{(t)}]_+ \|_2^2 + 2 \sum_{t = 1}^{T} \| [\vr^{(t)}]_+ - \tilde \vr^{(t)} \|_2^2.
\end{align*}
Substituting the preceding bounds and using $\tilde \vr^{(1)} = \vec0$, we obtain
\begin{align*}
    \sum_{t = 1}^{T} \| \tilde \vr^{(t + 1)} - \tilde \vr^{(t)} \|_2^2 \le 2 \| \tilde \vr^{(T + 1)} \|_2^2 + 4 \sum_{t = 1}^{T} c^{(t)} \| [\vr^{(t)}]_+ \|_1 \le 2 \| \tilde \vr^{(T + 1)} \|_2^2 + 4 \sqrt d \, \| \tilde \vr^{(T + 1)} \|_2 \sum_{t = 1}^{T} c^{(t)}.
\end{align*}
The last inequality follows from $\| [\vr^{(t)}]_+ \|_1 \le \sqrt d \, \| \tilde \vr^{(t)} \|_2$ and the nondecreasing regret-vector norms.

Applying Lemma~\ref{lem:refine_alt_rvu_errors} with its constant $2$ for $\IRPRMplus$, we get
\begin{align*}
    \sqrt{\sum_{t = 1}^{T} \left[ \max_{\vz \in \Delta(d)} \ip{\vu^{(t)}, 2 \vz - \vx^{(t)} - \vx^{(t + 1)}} \right]^2} \le 2 \sqrt2 \, \| \tilde \vr^{(T + 1)} \|_2 + 4 d^{1/4} \sqrt{\| \tilde \vr^{(T + 1)} \|_2 \sum_{t = 1}^{T} c^{(t)}}.
\end{align*}
Young's inequality gives
\begin{align*}
    4 d^{1/4} \sqrt{\| \tilde \vr^{(T + 1)} \|_2 \sum_{t = 1}^{T} c^{(t)}} \le 4 \sqrt d \, \| \tilde \vr^{(T + 1)} \|_2 + \sum_{t = 1}^{T} c^{(t)}.
\end{align*}
Adding these bounds to \eqref{eq:refined_ir_credit} and discarding the remaining negative sum of $c^{(t)}$, we obtain
\begin{align*}
    \operatorname{Alt-Reg}(T) + \sqrt{\sum_{t = 1}^{T} \left[ \max_{\vz \in \Delta(d)} \ip{\vu^{(t)}, 2 \vz - \vx^{(t)} - \vx^{(t + 1)}} \right]^2} \le (2 + 2 \sqrt2 + 4 \sqrt d) \| \tilde \vr^{(T + 1)} \|_2 - \gamma \sum_{t = 1}^{T} \| \vx^{(t + 1)} - \vx^{(t)} \|_2^2.
\end{align*}
The norm estimate established in the proof of Proposition~\ref{prop:alt_rvu_IR} gives
\begin{align*}
    \| \tilde \vr^{(T + 1)} \|_2 \le \| \mA \|_2 + (1 + \sqrt d) \sqrt{\sum_{t = 2}^{T} \| \vu^{(t)} - \vu^{(t - 1)} \|_2^2}.
\end{align*}
Substituting this estimate and applying Young's inequality proves \eqref{eq:refined_rvu} with
\begin{align*}
    \zeta = 1, \text{ and } \alpha = (2 + 2 \sqrt2 + 4 \sqrt d) \| \mA \|_2 + \frac{(2 + 2 \sqrt2 + 4 \sqrt d)^2 (1 + \sqrt d)^2}{4 \beta}.
\end{align*}

\paragraph{$\HPRMplus$.}
For this learner, we retain part of the negative weighted regret-vector increment sum in \eqref{eq:hprm_alt_reg_corrections}. The prediction bound from the preceding Alt-RVU proof gives
\begin{align*}
    \| [\vr^{(t + 1)}]_+ \|_1 \le 2 \sqrt d \, \max_{1 \le s \le T + 1} \| \tilde \vr^{(s)} \|_2.
\end{align*}
Since the regret-vector increments vanish before $t_0 - 1$, we obtain
\begin{align*}
    \sum_{t = 1}^{T} \| \tilde \vr^{(t + 1)} - \tilde \vr^{(t)} \|_2^2 \le 2 \sqrt d \left( \max_{1 \le s \le T + 1} \| \tilde \vr^{(s)} \|_2 \right) \sum_{t = t_0 - 1}^{T} \frac{\| \tilde \vr^{(t + 1)} - \tilde \vr^{(t)} \|_2^2}{\| [\vr^{(t + 1)}]_+ \|_1}.
\end{align*}
Set $\zeta = 1 / (4 \sqrt5 \, d^{1/4})$. Applying Lemma~\ref{lem:refine_alt_rvu_errors} and Young's inequality, we get
\begin{align}
    \zeta \sqrt{\sum_{t = 1}^{T} \left[ \max_{\vz \in \Delta(d)} \ip{\vu^{(t)}, 2 \vz - \vx^{(t)} - \vx^{(t + 1)}} \right]^2} \le \max_{1 \le s \le T + 1} \| \tilde \vr^{(s)} \|_2 + \frac1{32} \sum_{t = t_0 - 1}^{T} \frac{\| \tilde \vr^{(t + 1)} - \tilde \vr^{(t)} \|_2^2}{\| [\vr^{(t + 1)}]_+ \|_1}.
    \label{eq:refined_hprm_errors}
\end{align}

Recall that \eqref{eq:hprm_alt_reg_corrections} gives
\begin{align*}
    \operatorname{Alt-Reg}(T) \le 4 \max_{1 \le s \le T + 1} \| \tilde \vr^{(s)} \|_2 - \frac14 \sum_{t = t_0 - 1}^{T} \frac{\| \tilde \vr^{(t + 1)} - \tilde \vr^{(t)} \|_2^2}{\| [\vr^{(t + 1)}]_+ \|_1}.
\end{align*}
Adding \eqref{eq:refined_hprm_errors}, we obtain
\begin{align*}
    \operatorname{Alt-Reg}(T) + \zeta \sqrt{\sum_{t = 1}^{T} \left[ \max_{\vz \in \Delta(d)} \ip{\vu^{(t)}, 2 \vz - \vx^{(t)} - \vx^{(t + 1)}} \right]^2} \le 5 \max_{1 \le s \le T + 1} \| \tilde \vr^{(s)} \|_2 - \frac7{32} \sum_{t = t_0 - 1}^{T} \frac{\| \tilde \vr^{(t + 1)} - \tilde \vr^{(t)} \|_2^2}{\| [\vr^{(t + 1)}]_+ \|_1}.
\end{align*}
The maximum-norm estimate established in the preceding Alt-RVU proof gives
\begin{align*}
    \max_{1 \le s \le T + 1} \| \tilde \vr^{(s)} \|_2 &\le 2 \| \mA \|_2 + 6 d^{1/4} \left( 2 (1 + \sqrt d)^2 \sum_{t = 2}^{T} \| \vu^{(t)} - \vu^{(t - 1)} \|_2^2 + 2 d \| \mA \|_2^2 \sum_{t = 1}^{T} \| \vx^{(t + 1)} - \vx^{(t)} \|_2^2 \right)^{1/2} \\
    &+ \frac1{32} \sum_{t = t_0 - 1}^{T} \frac{\| \tilde \vr^{(t + 1)} - \tilde \vr^{(t)} \|_2^2}{\| [\vr^{(t + 1)}]_+ \|_1}.
\end{align*}
Substituting this estimate, the coefficient of the weighted increment sum becomes $5/32 - 7/32 = -1/16$. We therefore obtain
\begin{align*}
    &\operatorname{Alt-Reg}(T) + \zeta \sqrt{\sum_{t = 1}^{T} \left[ \max_{\vz \in \Delta(d)} \ip{\vu^{(t)}, 2 \vz - \vx^{(t)} - \vx^{(t + 1)}} \right]^2} \\
    &\le 10 \| \mA \|_2 + 30 d^{1/4} \left( 2 (1 + \sqrt d)^2 \sum_{t = 2}^{T} \| \vu^{(t)} - \vu^{(t - 1)} \|_2^2 + 2 d \| \mA \|_2^2 \sum_{t = 1}^{T} \| \vx^{(t + 1)} - \vx^{(t)} \|_2^2 \right)^{1/2} \\
    &- \frac1{16} \sum_{t = t_0 - 1}^{T} \frac{\| \tilde \vr^{(t + 1)} - \tilde \vr^{(t)} \|_2^2}{\| [\vr^{(t + 1)}]_+ \|_1}.
\end{align*}
Choose $\gamma = \sqrt{f^{(t_0)}} / (128 \sqrt2 (1 + \sqrt d)^2) > 0$, which is half the movement coefficient in the preceding Alt-RVU proposition. With this choice, \eqref{eq:hprm_corrections_movement} reads
\begin{align*}
    \sum_{t = t_0 - 1}^{T} \frac{\| \tilde \vr^{(t + 1)} - \tilde \vr^{(t)} \|_2^2}{\| [\vr^{(t + 1)}]_+ \|_1} \ge 32 \gamma \sum_{t = 1}^{T} \| \vx^{(t + 1)} - \vx^{(t)} \|_2^2 - 64 \gamma.
\end{align*}
Consequently, we obtain
\begin{align*}
    &\operatorname{Alt-Reg}(T) + \zeta \sqrt{\sum_{t = 1}^{T} \left[ \max_{\vz \in \Delta(d)} \ip{\vu^{(t)}, 2 \vz - \vx^{(t)} - \vx^{(t + 1)}} \right]^2} \\
    &\le 10 \| \mA \|_2 + 4 \gamma + 30 d^{1/4} \left( 2 (1 + \sqrt d)^2 \sum_{t = 2}^{T} \| \vu^{(t)} - \vu^{(t - 1)} \|_2^2 + 2 d \| \mA \|_2^2 \sum_{t = 1}^{T} \| \vx^{(t + 1)} - \vx^{(t)} \|_2^2 \right)^{1/2} \\
    &- 2 \gamma \sum_{t = 1}^{T} \| \vx^{(t + 1)} - \vx^{(t)} \|_2^2.
\end{align*}
Splitting the square root and applying Young's inequality to each term, we obtain
\begin{align*}
    \operatorname{Alt-Reg}(T) + \zeta \sqrt{\sum_{t = 1}^{T} \left[ \max_{\vz \in \Delta(d)} \ip{\vu^{(t)}, 2 \vz - \vx^{(t)} - \vx^{(t + 1)}} \right]^2} \le \alpha + \beta \sum_{t = 2}^{T} \| \vu^{(t)} - \vu^{(t - 1)} \|_2^2 - \gamma \sum_{t = 1}^{T} \| \vx^{(t + 1)} - \vx^{(t)} \|_2^2,
\end{align*}
where
\begin{align*}
    \alpha = 10 \| \mA \|_2 + 4 \gamma + \frac{450 \sqrt d (1 + \sqrt d)^2}{\beta} + \frac{450 d^{3/2} \| \mA \|_2^2}{\gamma}.
\end{align*}
For both learners, $\gamma$ and $\zeta$ are positive and independent of $\beta$, and all constants are independent of $T$.
\end{proof}
\nfgBestConvergence*
\begin{proof}
Applying Proposition~\ref{prop:refined_alt_rvu_convergence} to the bounds established in Proposition~\ref{prop:refined_alt_rvu_algorithms} gives the best-iterate bound and $\operatorname{Gap}(\vx^{(t)}, \vy^{(t)}) \to 0$. We now prove convergence to one equilibrium for both algorithms.

Because the preceding Alt-RVU bounds hold for every $\beta > 0$, we choose $\beta$ so that $\beta \|\mA\|_2^2$ is strictly smaller than both players' negative movement coefficients. Summing the two bounds and using \eqref{eq:sum_alt_reg_refined_bound}, we obtain
\begin{align*}
    \sum_{t=1}^{\infty} \left(\|\vx^{(t+1)} - \vx^{(t)}\|_2^2 + \|\vy^{(t)} - \vy^{(t-1)}\|_2^2\right) < \infty.
\end{align*}
Since utilities are linear in the opponent's strategy, both utility-variation sums are also finite.

We first establish the required bounds on the regret vectors. For $\IRPRMplus$, the norm estimate in the proof of Proposition~\ref{prop:alt_rvu_IR} gives
\begin{align*}
    \sup_{t \ge 1} \|\tilde\vr^{(t)}\|_2^2 \le \|\mA\|_2^2 + (1 + \sqrt d)^2 \sum_{t=2}^{\infty} \|\vu^{(t)} - \vu^{(t-1)}\|_2^2 < \infty.
\end{align*}
The IR mechanism makes these norms nondecreasing, so they converge to a positive limit whenever the learner becomes nonzero.

For $\HPRMplus$, first consider a learner with $t_0<\infty$. For $T \ge t_0-1$, the maximum-norm estimate in the proof of its Alt-RVU bound gives
\begin{align*}
    \max_{1 \le s \le T+1} \|\tilde\vr^{(s)}\|_2 \le O(1) + \frac1{32} \sum_{t=t_0-1}^{T} \frac{\|\tilde\vr^{(t+1)} - \tilde\vr^{(t)}\|_2^2}{\|[\vr^{(t+1)}]_+\|_1},
\end{align*}
where $O(1)$ is independent of $T$, because the utility-variation and strategy-movement sums are bounded. Multiplying this inequality by $5$ and combining it with \eqref{eq:hprm_alt_reg_corrections}, we obtain
\begin{align*}
    \operatorname{Alt-Reg}(T) + \max_{1 \le s \le T+1} \|\tilde\vr^{(s)}\|_2 + \frac1{16} \sum_{t=t_0-1}^{T} \frac{\|\tilde\vr^{(t+1)} - \tilde\vr^{(t)}\|_2^2}{\|[\vr^{(t+1)}]_+\|_1} \le O(1).
\end{align*}
For a learner that never becomes nonzero, its alternating regret and regret vectors are zero, and we omit its weighted increment sum. For sufficiently large $T$, summing over the two players and using \eqref{eq:sum_alt_reg_refined_bound}, we conclude that both maximum norms and both nonnegative sums are bounded. Since $[\vr^{(t + 1)}]_+ \le 3 \tilde \vr^{(t + 1)} / 2$, the denominators are bounded above, and therefore
\begin{align*}
    \sum_{t=1}^{\infty} \|\tilde\vr^{(t+1)} - \tilde\vr^{(t)}\|_2^2 < \infty.
\end{align*}
Moreover, $f^{(t)}$ is nondecreasing and bounded. Consequently, we have
\begin{align*}
    \|\tilde\vr^{(t)}\|_2^2 = f^{(t)} + \frac12 \|\tilde\vr^{(t)} - \tilde\vr^{(t-1)}\|_2^2
\end{align*}
converges to a positive limit whenever the learner becomes nonzero.

We next show that the sum of the two regret-vector inner products converges for every Nash equilibrium. Fix any Nash equilibrium $(\vx^*, \vy^*)$. The column learner's local round $t$ starts with the pending strategy $\vy^{(t-1)}$ and ends with $\vy^{(t)}$. For $\HPRMplus$, the coordinatewise regret vector updates give
\begin{align*}
    &\ip{\tilde\vr_{\mathcal X}^{(t+1)} - \tilde\vr_{\mathcal X}^{(t)}, \vx^*} + \ip{\tilde\vr_{\mathcal Y}^{(t+1)} - \tilde\vr_{\mathcal Y}^{(t)}, \vy^*} \\
    &\ge (\vx^*)^\top \mA \vy^{(t)} - (\vx^{(t)})^\top \mA \vy^* + \ip{-\mA^\top \vx^{(t)}, \vy^{(t)} - \vy^{(t-1)}} \\
    &\ge \ip{-\mA^\top \vx^{(t)}, \vy^{(t)} - \vy^{(t-1)}},
\end{align*}
where the last inequality follows from the saddle-point inequalities. For $\IRPRMplus$, the same calculation retains the additional terms $c_{\mathcal X}^{(t)} + c_{\mathcal Y}^{(t)}$.

For $\IRPRMplus$, applying the column player's utility-improvement bound therefore gives
\begin{align*}
    \ip{\tilde\vr_{\mathcal X}^{(t+1)} - \tilde\vr_{\mathcal X}^{(t)}, \vx^*} + \ip{\tilde\vr_{\mathcal Y}^{(t+1)} - \tilde\vr_{\mathcal Y}^{(t)}, \vy^*} \ge c_{\mathcal X}^{(t)} + c_{\mathcal Y}^{(t+1)} \ge 0.
\end{align*}
Thus $\ip{\tilde\vr_{\mathcal X}^{(t)}, \vx^*} + \ip{\tilde\vr_{\mathcal Y}^{(t)}, \vy^*}$ is nondecreasing and bounded, hence convergent. The same inequality also implies that both credit sums are finite.

For $\HPRMplus$, whenever $\tilde \vr_{\mathcal Y}^{(t)} \ne \vec0$, \eqref{eq:util_improvement_hprm} and \eqref{eq:prediction_lower_hprm} give
\begin{align*}
    \ip{-\mA^\top \vx^{(t)}, \vy^{(t)} - \vy^{(t-1)}} \ge -\frac{f_{\mathcal Y}^{(t+1)} - f_{\mathcal Y}^{(t)}}{2\|[\vr_{\mathcal Y}^{(t+1)}]_+\|_1} \ge -\frac{f_{\mathcal Y}^{(t+1)} - f_{\mathcal Y}^{(t)}}{\sqrt{2f_{\mathcal Y}^{(t+1)}}} \ge -\sqrt2 \left(\sqrt{f_{\mathcal Y}^{(t+1)}} - \sqrt{f_{\mathcal Y}^{(t)}}\right).
\end{align*}
If $\tilde \vr_{\mathcal Y}^{(t)} = \tilde \vr_{\mathcal Y}^{(t + 1)} = \vec0$, the utility improvement and the $f$'s difference are both zero. At the first nonzero update, \eqref{eq:t_0_1_sum} gives a positive utility improvement, and $f_{\mathcal Y}^{(t)} = 0 < f_{\mathcal Y}^{(t + 1)}$. It follows that
\begin{align*}
    \ip{\tilde\vr_{\mathcal X}^{(t)}, \vx^*} + \ip{\tilde\vr_{\mathcal Y}^{(t)}, \vy^*} + \sqrt{2f_{\mathcal Y}^{(t)}}
\end{align*}
is nondecreasing and bounded. Since $f_{\mathcal Y}^{(t)}$ converges, the sum of the two inner products also converges.

We now relate the regret vectors to the played strategies. For $\HPRMplus$, \eqref{eq:bound_in_hprm_1} gives
\begin{align*}
    \|[\vr^{(t)}]_+ - \tilde\vr^{(t)}\|_2 \le \frac12 \|\tilde\vr^{(t)} - \tilde\vr^{(t-1)}\|_2 \to 0.
\end{align*}
For $\IRPRMplus$, norm preservation and the definition of $c^{(t)}$ give
\begin{align*}
    c^{(t)} \|[\vr^{(t)}]_+\|_1
    &= \ip{\vr^{(t)} - \tilde\vr^{(t)}, [\vr^{(t)}]_+} \\
    &= \|[\vr^{(t)}]_+\|_2^2 - \ip{\tilde\vr^{(t)}, [\vr^{(t)}]_+}
    = \frac12 \|[\vr^{(t)}]_+ - \tilde\vr^{(t)}\|_2^2.
\end{align*}
The credit sums are finite and the prediction norms are bounded, so this discrepancy also tends to zero. For either algorithm, the positive limiting regret norm and the normalization bound imply the corresponding limit below for each learner that becomes nonzero:
\begin{align*}
    \left\|\vx^{(t)} - \frac{\tilde\vr_{\mathcal X}^{(t)}}{\|\tilde\vr_{\mathcal X}^{(t)}\|_1}\right\|_2 \to 0, \;
    \left\|\vy^{(t)} - \frac{\tilde\vr_{\mathcal Y}^{(t)}}{\|\tilde\vr_{\mathcal Y}^{(t)}\|_1}\right\|_2 \to 0.
\end{align*}
For the column player, normalization initially gives $\vy^{(t-1)}$; the displayed conclusion follows because $\|\vy^{(t)} - \vy^{(t-1)}\|_2 \to 0$. A learner that never becomes nonzero remains uniform and requires no normalization.

Finally, choose any accumulation point $(\vx^*, \vy^*)$ of the played strategies. It is a Nash equilibrium because the Nash gap tends to zero. The convergence of the regret norms and of the sum of their inner products with this equilibrium implies that
\begin{align*}
    \|\tilde\vr_{\mathcal X}^{(t)}\|_2 \|\vx^*\|_2 - \ip{\tilde\vr_{\mathcal X}^{(t)}, \vx^*} + \|\tilde\vr_{\mathcal Y}^{(t)}\|_2 \|\vy^*\|_2 - \ip{\tilde\vr_{\mathcal Y}^{(t)}, \vy^*}
\end{align*}
has a limit. Both summands are nonnegative by Cauchy--Schwarz. Along the subsequence converging to $(\vx^*, \vy^*)$, the preceding normalization limits show that this expression tends to zero. Its full limit is therefore zero, and each summand tends to zero.

For an activated row learner, we consequently obtain
\begin{align*}
    \left\|\frac{\tilde\vr_{\mathcal X}^{(t)}}{\|\tilde\vr_{\mathcal X}^{(t)}\|_2} - \frac{\vx^*}{\|\vx^*\|_2}\right\|_2^2
    = \frac{2\left(\|\tilde\vr_{\mathcal X}^{(t)}\|_2 \|\vx^*\|_2 - \ip{\tilde\vr_{\mathcal X}^{(t)}, \vx^*}\right)}{\|\tilde\vr_{\mathcal X}^{(t)}\|_2 \|\vx^*\|_2} \to 0.
\end{align*}
Normalizing by the coordinate sum and using the preceding strategy approximation, we obtain $\vx^{(t)} \to \vx^*$. The same argument gives $\vy^{(t)} \to \vy^*$. A never-activated learner is already constant. This proves convergence to the selected Nash equilibrium.
\end{proof}
\subsection{Proofs from Section \ref{sec:separation_alt_sim}}

\AltSimSeparation*
\begin{proof}
We first consider simultaneous updates and omit the subscript ${\rm Sim}$. Since $\mA^\top = -\mA$ and both players have the same initialization, their regret vectors and strategies coincide at every iteration. Consequently, $\vx^{(t)} = \vy^{(t)}$, $\vu^{(t)} = \mA\vx^{(t)}$, and $\ip{\vu^{(t)}, \vx^{(t)}} = 0$.

The construction starts from three-action {\em rock-paper-scissors (RPS)} and adds a strictly dominated fourth action. This action creates a nonuniform strategy on the first three actions through the actual updates. Without this action, a zero initialization in three-action RPS alone would remain at equilibrium. After the standard initialization at time $(1)$, the first observation gives
\begin{align*}
    \tilde\vr^{(2)} = (1,3/4,3/4,0), \; \vx^{(2)} = \vy^{(2)} = (2/5,3/10,3/10,0).
\end{align*}
Whenever $x_4^{(t)} = 0$, we have $(\mA\vx^{(t)})_4 = -4x_1^{(t)} - 3x_2^{(t)} - 3x_3^{(t)} < 0$. Thus $\tilde r_4^{(t)} = x_4^{(t)} = 0$ for every $t \ge 2$, and
\begin{align*}
    \left((\mA\vx^{(t)})_1,(\mA\vx^{(t)})_2,(\mA\vx^{(t)})_3\right) = \left(x_3^{(t)} - x_2^{(t)},x_1^{(t)} - x_3^{(t)},x_2^{(t)} - x_1^{(t)}\right).
\end{align*}
These three coordinates sum to zero. Therefore, for $t \ge 2$, we have the following two bounds.
\begin{align}
    \|\tilde\vr^{(t+1)}\|_1 &= \sum_{i=1}^3 \left[\tilde r_i^{(t)} + (\mA\vx^{(t)})_i\right]_+ \ge \sum_{i=1}^3 \left(\tilde r_i^{(t)} + (\mA\vx^{(t)})_i\right) = \|\tilde\vr^{(t)}\|_1 \ge \frac52, \label{eq:sim_half_mass}\\
    \|\tilde\vr^{(t+1)} - \tilde\vr^{(t)}\|_2^2 &\le \sum_{i=1}^3 (\mA\vx^{(t)})_i^2 = 3\|\vx^{(t)} - \vx^\ast\|_2^2 \le 2. \label{eq:sim_half_increment}
\end{align}
Here and below, strict truncation refers only to the first three coordinates. The first inequality is an equality whenever no coordinate is strictly truncated in the regret vector update. The second follows from nonexpansiveness of the positive-part operation.

Recall that, for $\HPRMplus$, we define
\begin{align*}
    f^{(t)} = \|\tilde\vr^{(t)}\|_2^2 - \frac12 \|\tilde\vr^{(t)} - \tilde\vr^{(t-1)}\|_2^2.
\end{align*}
We first strengthen its bound from Lemma~\ref{lem:half_bound_r_tilde}. The identity $[\vr^{(t)}]_+ = [\tilde\vr^{(t)} + (\tilde\vr^{(t)} - \tilde\vr^{(t-1)})/2]_+$ gives
\begin{align*}
    \ip{\tilde\vr^{(t)} - \tilde\vr^{(t-1)}, [\vr^{(t)}]_+ - \tilde\vr^{(t)}} \ge 2\|[\vr^{(t)}]_+ - \tilde\vr^{(t)}\|_2^2.
\end{align*}
Applying Young's inequality and then this bound, we obtain
\begin{align*}
    2\ip{\tilde\vr^{(t+1)} - \tilde\vr^{(t)}, [\vr^{(t)}]_+ - \tilde\vr^{(t)}} &= 2\ip{\tilde\vr^{(t+1)} - \tilde\vr^{(t-1)}, [\vr^{(t)}]_+ - \tilde\vr^{(t)}} - 2\ip{\tilde\vr^{(t)} - \tilde\vr^{(t-1)}, [\vr^{(t)}]_+ - \tilde\vr^{(t)}}\\
    &\le \frac14\|\tilde\vr^{(t+1)} - \tilde\vr^{(t-1)}\|_2^2 + 4\|[\vr^{(t)}]_+ - \tilde\vr^{(t)}\|_2^2 - 2\ip{\tilde\vr^{(t)} - \tilde\vr^{(t-1)}, [\vr^{(t)}]_+ - \tilde\vr^{(t)}}\\
    &\le \frac14\|\tilde\vr^{(t+1)} - \tilde\vr^{(t-1)}\|_2^2.
\end{align*}
Since $\ip{\tilde\vr^{(t+1)} - \tilde\vr^{(t)}, [\vr^{(t)}]_+} \ge 0$, as in Lemma~\ref{lem:half_bound_r_tilde}, we can expand the $f$'s difference to obtain
\begin{align}
    f^{(t+1)} - f^{(t)} &\ge \frac12\|\tilde\vr^{(t+1)} - \tilde\vr^{(t)}\|_2^2 + \frac12\|\tilde\vr^{(t)} - \tilde\vr^{(t-1)}\|_2^2 - 2\ip{\tilde\vr^{(t+1)} - \tilde\vr^{(t)}, [\vr^{(t)}]_+ - \tilde\vr^{(t)}} \notag\\
    &\ge \frac12\|\tilde\vr^{(t+1)} - \tilde\vr^{(t)}\|_2^2 + \frac12\|\tilde\vr^{(t)} - \tilde\vr^{(t-1)}\|_2^2 - \frac14\|\tilde\vr^{(t+1)} - \tilde\vr^{(t-1)}\|_2^2 \notag\\
    &= \frac14\|\tilde\vr^{(t+1)} - 2\tilde\vr^{(t)} + \tilde\vr^{(t-1)}\|_2^2. \label{eq:sim_half_potential}
\end{align}

We next show that the regret vectors remain separated from vectors with three equal positive coordinates. Direct calculation gives
\begin{align*}
    \tilde\vr^{(3)} = (1,17/20,13/20,0), \; f^{(3)} = \frac{427}{200}, \; f^{(3)} - \frac{\|\tilde\vr^{(3)}\|_1^2}{3} = \frac{31}{600} > \frac1{24}.
\end{align*}
If no coordinate is strictly truncated, $\|\tilde\vr^{(t+1)}\|_1 = \|\tilde\vr^{(t)}\|_1$ and $f^{(t+1)} \ge f^{(t)}$. If a coordinate is strictly truncated, $\tilde\vr^{(t+1)}$ has at most two positive coordinates. Using \eqref{eq:sim_half_mass} and \eqref{eq:sim_half_increment}, we then obtain
\begin{align*}
    f^{(t+1)} - \frac{\|\tilde\vr^{(t+1)}\|_1^2}{3} &\ge \frac{\|\tilde\vr^{(t+1)}\|_1^2}{2} - 1 - \frac{\|\tilde\vr^{(t+1)}\|_1^2}{3} = \frac{\|\tilde\vr^{(t+1)}\|_1^2}{6} - 1 \ge \frac1{24}.
\end{align*}
For $t \ge 3$, we get by induction that
\begin{align}
    f^{(t)} - \frac{\|\tilde\vr^{(t)}\|_1^2}{3} \ge \frac1{24}. \label{eq:sim_half_separation}
\end{align}
We now prove that $\|\tilde\vr^{(t)}\|_1 \to \infty$. Suppose instead that these norms are bounded. Then $f^{(t)}$ is bounded and nondecreasing, so \eqref{eq:sim_half_potential} implies that
\begin{align*}
    \|\tilde\vr^{(t+1)} - 2\tilde\vr^{(t)} + \tilde\vr^{(t-1)}\|_2 \to 0.
\end{align*}
For every fixed integer $k \ge 1$, we can sum the consecutive increments to obtain the following bounds
\begin{align*}
    \tilde\vr^{(t+k)} - \tilde\vr^{(t)} = k\left(\tilde\vr^{(t+1)} - \tilde\vr^{(t)}\right) + o(1), \; \limsup_{t\to\infty}\|\tilde\vr^{(t+1)} - \tilde\vr^{(t)}\|_2 \le \frac{2}{k}\sup_{s\ge2}\|\tilde\vr^{(s)}\|_2.
\end{align*}
Letting $k \to \infty$, we get $\tilde\vr^{(t+1)} - \tilde\vr^{(t)} \to \vec0$. The regret vector update also gives
\begin{align*}
    (\mA\vx^{(t)})_i \le \tilde r_i^{(t+1)} - \tilde r_i^{(t)}, \text{ for } i=1,2,3.
\end{align*}
Since the three utility coordinates sum to zero, they all tend to zero. Their explicit expressions above imply $\vx^{(t)} \to \vx^\ast$. Moreover, we have
\begin{align*}
    \|[\vr^{(t)}]_+ - \tilde\vr^{(t)}\|_2 \le \frac12\left(\sum_{i=1}^3(\mA\vx^{(t-1)})_i^2\right)^{1/2} \to 0.
\end{align*}
Together with boundedness of $\|\tilde\vr^{(t)}\|_1$, this yields $\tilde\vr^{(t)} - \|\tilde\vr^{(t)}\|_1\vx^\ast \to \vec0$. Consequently,
\begin{align*}
    f^{(t)} - \frac{\|\tilde\vr^{(t)}\|_1^2}{3} = \|\tilde\vr^{(t)} - \|\tilde\vr^{(t)}\|_1\vx^\ast\|_2^2 - \frac12\|\tilde\vr^{(t)} - \tilde\vr^{(t-1)}\|_2^2 \to 0,
\end{align*}
contradicting \eqref{eq:sim_half_separation}. Thus the nondecreasing sequence $\|\tilde\vr^{(t)}\|_1$ must diverge.

There is therefore a strict truncation, since otherwise \eqref{eq:sim_half_mass} would hold with equality at every iteration. Immediately after any strict truncation, \eqref{eq:sim_half_increment} implies that
\begin{align*}
    f^{(t)} \ge \frac{\|\tilde\vr^{(t)}\|_1^2}{2} - 1.
\end{align*}
Between strict truncations, the right-hand side is constant and $f^{(t)}$ is nondecreasing. Hence this inequality holds at every iteration after the first strict truncation. We now obtain
\begin{align*}
    \left\|\frac{\tilde\vr^{(t)}}{\|\tilde\vr^{(t)}\|_1} - \vx^\ast\right\|_2^2 = \frac{\|\tilde\vr^{(t)}\|_2^2}{\|\tilde\vr^{(t)}\|_1^2} - \frac13 \ge \frac16 - \frac1{\|\tilde\vr^{(t)}\|_1^2}.
\end{align*}
For $t \ge 3$, the half-prediction update also satisfies
\begin{align*}
    \|[\vr^{(t)}]_+ - \tilde\vr^{(t)}\|_1 \le \frac12\sum_{i=1}^3|(\mA\vx^{(t-1)})_i| \le 1, \; \left\|\vx^{(t)} - \frac{\tilde\vr^{(t)}}{\|\tilde\vr^{(t)}\|_1}\right\|_1 \le \frac{2}{\|\tilde\vr^{(t)}\|_1} \to 0.
\end{align*}
It follows that
\begin{align*}
    \liminf_{t\to\infty}\|\vx^{(t)} - \vx^\ast\|_2 \ge \frac1{\sqrt6}.
\end{align*}
Finally, from the skew symmetry and negativity of the fourth utility coordinate, we note that
\begin{align*}
    \operatorname{Gap}(\vx^{(t)},\vx^{(t)}) = 2\max_{1\le i\le3}(\mA\vx^{(t)})_i.
\end{align*}
For three real numbers with zero sum, their squared sum is at most six times the square of their maximum. Therefore, we get
\begin{align*}
    3\|\vx^{(t)} - \vx^\ast\|_2^2 = \sum_{i=1}^3(\mA\vx^{(t)})_i^2 \le 6\left(\max_{1\le i\le3}(\mA\vx^{(t)})_i\right)^2 = \frac32\operatorname{Gap}(\vx^{(t)},\vx^{(t)})^2.
\end{align*}
Combining the last two bounds, we conclude that
\begin{align*}
    \liminf_{t\to\infty}\operatorname{Gap}(\vx^{(t)},\vy^{(t)}) \ge \sqrt2\liminf_{t\to\infty}\|\vx^{(t)} - \vx^\ast\|_2 \ge \frac1{\sqrt3}.
\end{align*}
\end{proof}

\section{Additional Experimental Details and Results}
\label{sec:exp_add}

\subsection{Construction and Validation of the Numerical Separations}
\label{sec:separation_avg}

The two block games in Figure~\ref{fig:average_last_separation} share the unscaled integer kernel
\begin{align*}
    \mA_0 = \begin{pmatrix} -1 & 0 & 2 \\ 19 & -1 & -19 \\ -19 & 2 & -19 \end{pmatrix}.
\end{align*}
We construct $\mA(a)\in\R^{(a+2)\times(a+2)}$ by replacing the first row of $\mA_0$ with $a$ identical copies and its second column with $a$ identical copies, retaining one copy of each remaining row and column. We use $a=11000$ for $\lambda=10^{-2}$ and $a=177700$ for $\lambda=1/3$, giving games of dimensions $11002\times11002$ and $177702\times177702$, respectively. On each game, we compare the specified multiplier with half prediction in both uniform average-iterate and last-iterate gap.

Under standard initialization, identical copies of an action retain identical regret coordinates. We therefore simulate each game using three action types per player, storing the regret coordinates of one copy of each type. If $s_1,s_2,s_3$ are the positive parts of the row player's prediction regret coordinates, its aggregate strategy is
\begin{align*}
    \frac{(a s_1,s_2,s_3)}{a s_1+s_2+s_3},
\end{align*}
when the denominator is positive; the column player analogously weights its second action type by $a$. When the denominator is zero, uniform play over the expanded actions gives aggregate strategies $(a,1,1)/(a+2)$ and $(1,a,1)/(a+2)$ for the row and column players, respectively. We compute utilities from the kernel and the opponent's aggregate strategy, retaining the multiplicity weights at every normalization. Aggregation preserves payoffs and duality gaps and commutes with uniform averaging, so the reduction preserves both gap measures in Figure~\ref{fig:average_last_separation}. Duplication nevertheless changes the learners' aggregate initialization and normalization weights.

\paragraph{Choosing the kernel and multiplicities.}
We first varied $q$ and $a$ within the structured kernel family
\begin{align*}
    \mK(q) = \begin{pmatrix} -1 & 0 & 2 \\ q & -1 & -q \\ -q & 2 & -q \end{pmatrix}.
\end{align*}
We sought candidates with persistent last-iterate gap oscillations and growing $T\operatorname{Gap}(\bar{\vx}^{(T)},\bar{\vy}^{(T)})$ under non-half prediction, while half prediction showed faster average-iterate decay and decreasing last-iterate gaps. Initial screening used $10^6$ iterations, followed by longer runs to reject candidates whose apparent slow average-iterate decay was transient.

For $\lambda=1/3$, a local refinement with $q=19$ and $a\in\{165000,165100,\ldots,185000\}$ selected $a=177700$. For $\lambda=10^{-2}$, we fixed the same kernel $\mK(19)$ and screened $779$ distinct values of $a$ before selecting $a=11000$. Both selected games were then evaluated through $10^9$ iterations. These multiplicities were selected numerically rather than derived from an analytic threshold, and neither is claimed to be minimal.

\paragraph{Validation and observed behavior.}
All trajectories in Figure~\ref{fig:average_last_separation} use the same double-double implementation, with approximately $31$ decimal digits of precision, standard initialization, and no positive normalization floor. Every iteration contributes to the uniform average, and we record both gap measures at the same checkpoints through $10^9$ iterations.

\begin{figure}[t]
    \centering
    \includegraphics[width=\linewidth]{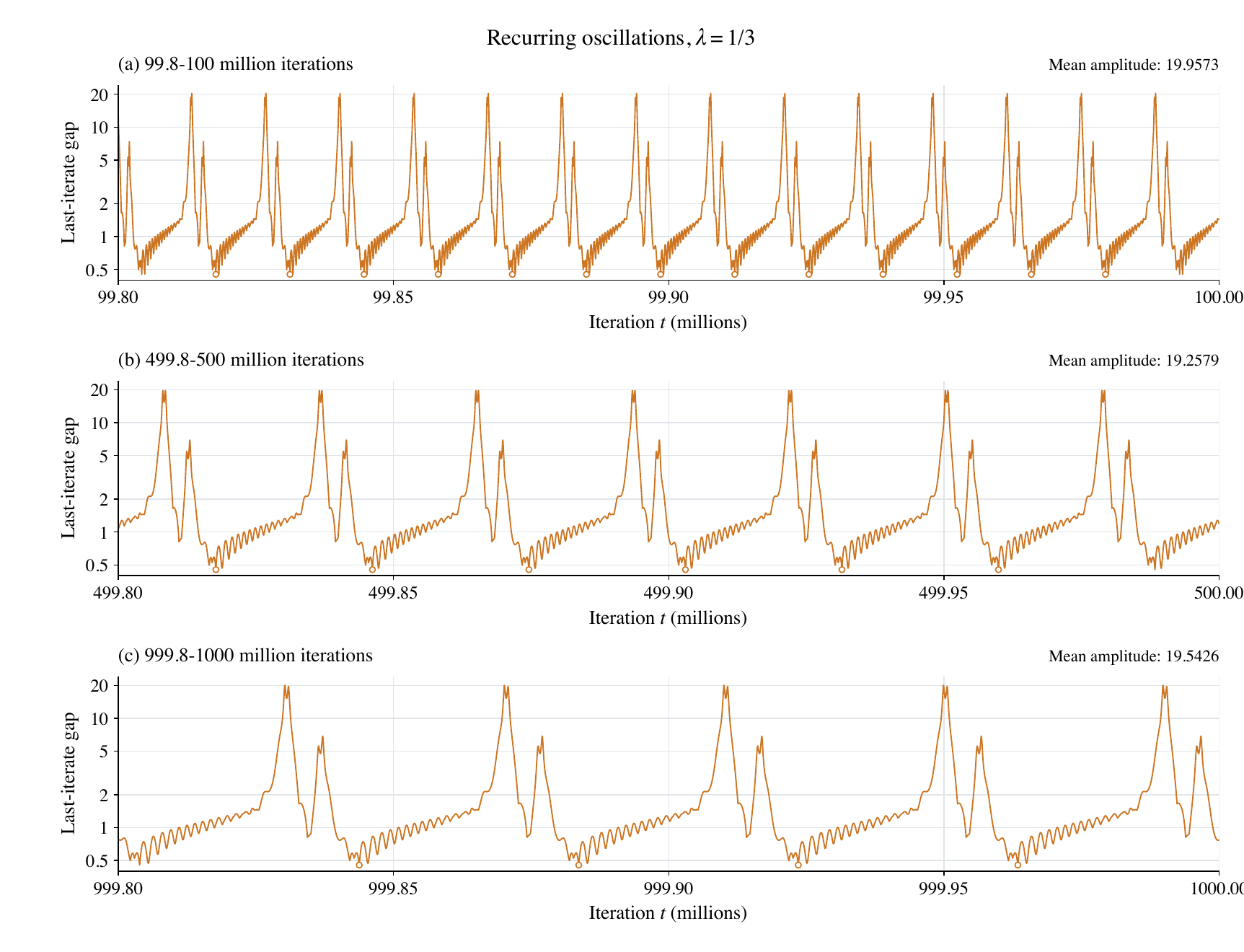}
    \caption{Last-iterate gap for $\lambda=1/3$ on the $177{,}702\times177{,}702$ block game over three late windows of $200{,}000$ iterations, with identical axis scales. Circles mark the minimum gap within each complete cycle, defined by consecutive upward crossings of gap $10$. Incomplete cycles at the window boundaries are displayed but excluded from cycle averages.}
    \label{fig:last_iterate_windows}
\end{figure}

For the two block games, the fitted average-iterate decay exponents over $10^8\le T\le10^9$, measured as the negative slopes of the log--log gap curves, are approximately $0.487$ for $\lambda=10^{-2}$ and $0.509$ for $\lambda=1/3$, compared with approximately $1$ for half prediction. These fits are consistent with $T^{-1/2}$ versus $T^{-1}$ average-iterate decay over the tested interval. In the same runs, the last-iterate gaps for the two non-half multipliers remain large and oscillatory, whereas half prediction reduces them to the numerical precision limit. To examine these oscillations more closely, Figure~\ref{fig:last_iterate_windows} shows every iterate in the final $200{,}000$ iterations before $10^8$, $5\times10^8$, and $10^9$ for $\lambda=1/3$ on the larger block game. The minimum gaps in these windows are $0.451095$, $0.453408$, and $0.453838$, respectively, increasing slightly rather than decreasing toward zero. From the first window to the last, the minimum gap increases by approximately $0.61\%$ and the mean cycle trough by $0.53\%$, while the mean peak-to-trough amplitude remains between $19.26$ and $19.96$. Individual cycle troughs fluctuate in both directions within each window. These observations strengthen the numerical evidence of persistent last-iterate oscillations, but do not rigorously establish a monotone trend or an asymptotic lower bound.

\subsection{Visualizations for the Separation of Alternating and Simultaneous Updates}

\begin{figure}[t]
    \centering
    \includegraphics[width=\linewidth]{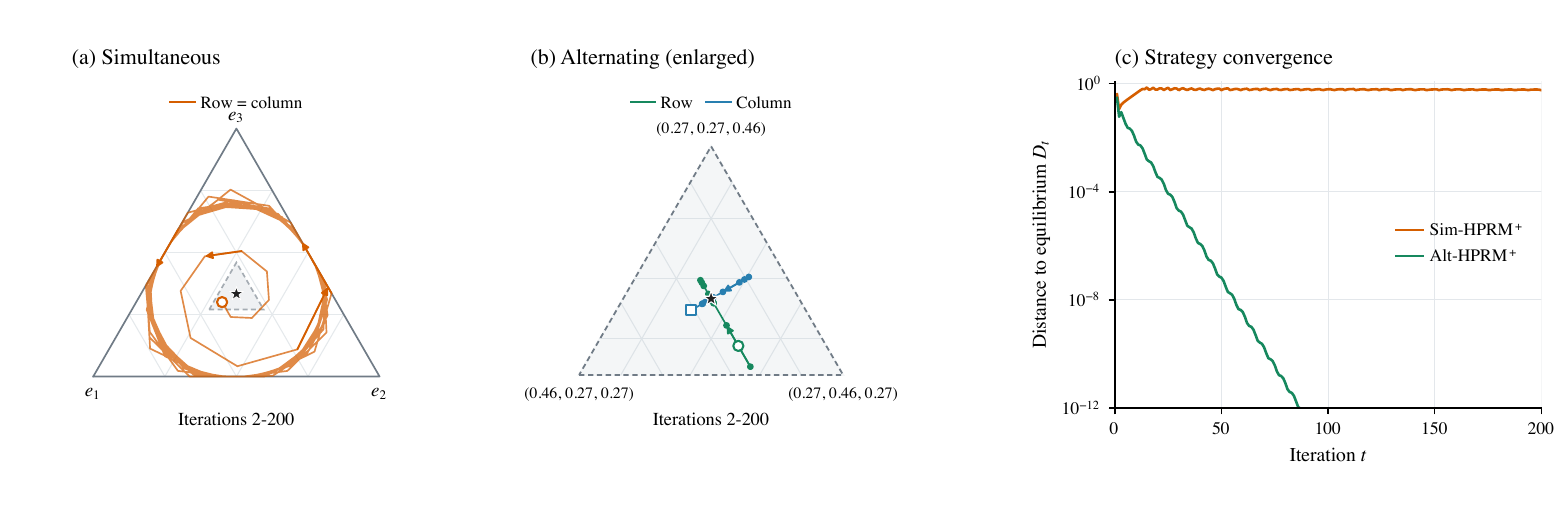}
    \caption{Half prediction under simultaneous and alternating updates (stars mark the equilibrium, and open markers denote iteration 2).
    (a) Simultaneous strategy trajectories.
    (b) Alternating trajectories, enlarging the dashed triangular region in (a).
    (c) Joint Euclidean distance to equilibrium.
    }
    \label{fig:separation}
\end{figure}

Figure~\ref{fig:separation} visualizes the trajectories of $\AHPRMplus$ and $\SHPRMplus$ on the separation game~\eqref{eq:separation_matrix}. The triangular representation is similar to that in \citet[Figure~2]{Nan26:On}, where the AltGDA trajectories settle into cycles while their averages converge. Here, the $\AHPRMplus$ trajectories in plot (b) instead exhibit damped oscillations toward equilibrium.

\subsection{Further Details for Performance on Extensive-Form Benchmarks}
\label{sec:exp_details}

All six methods are implemented in a common CPU codebase using Python 3.13.16, NumPy 2.5.4, OpenSpiel 2.0.2, and a modified LiteEFG 1.0.0 backend. Tree-based games use compiled exhaustive traversal, while HUNL uses a public-tree backend; within each game, all methods share the game representation, traversal, averaging, and evaluation routines. Computations use \texttt{float64}, zero initial regrets, and uniform initial strategies. Positive strategy weights are normalized without an $\varepsilon$ floor, with uniform fallback only at zero mass. $\APCFRplus$, $\AHPCFRplus$, and $\ASAPCFRplus$ share the same local update implementation with prediction coefficients $1$, $1/2$, and $1/3$, respectively. Following \citet{Xu24:Minimizing}, $\APDCFRplus$ and $\ADCFRplus$ use regret-discount exponents $\alpha=2.3$ and $\alpha=1.5$, respectively, multiplying past cumulative regrets by $(t-1)^\alpha/((t-1)^\alpha+1)$ before adding the current regret and clipping at zero. Neither method uses a negative-regret discount exponent $\beta$, since negative cumulative regrets are clipped. The paper's averaging exponents $\gamma=5$ and $\gamma=4$, respectively, assign iteration weights $t^\gamma$; we replace these output weights with the common averaging rules in Section~\ref{sec:exp}, while retaining the paper's local recurrences.

The iteration budgets are given in Table~\ref{tab:game-sizes}. Both averages accumulate post-iteration realization plans at every iteration, and HS30 uses the full declared run horizon, including when a figure displays a shorter interval. All three outputs are evaluated by exact best responses at identical checkpoints across methods: every iteration through $20$, every $5$ iterations through $100$, every $10$ through $500$, every $50$ through $5000$, and thereafter every $\max\{50,\lfloor T/1000\rfloor\}$ iterations, always including $T$. Runs do not stop early upon reaching a target. Target comparisons use the full saved horizons and the first saved crossing, without interpolation or a persistence requirement. Checkpoint exploitabilities and timers are saved, together with final policies, but full strategy histories and historical best-policy copies are not retained.

\begin{table*}
    \centering
    \setlength{\tabcolsep}{5pt}
    \begin{tabular}{@{}lrrrr@{}}
        \toprule
        Game
        & Histories & Information sets & Terminal histories & \shortstack{No. of iterations} \\
        \midrule
        Kuhn poker
        & 58 & 12 & 30 & 200 \\
        GoofspielImp (4 cards)
        & 1,077 & 162 & 576 & 500 \\
        Liar's Dice (4 sides)
        & 8,181 & 1,024 & 4,080 & 1000 \\
        Leduc poker
        & 9,457 & 936 & 5,520 & 50000\\
        GoofspielImp (5 cards)
        & 26,931 & 2,124 & 14,400 & 50000 \\
        Liar's Dice (5 sides)
        & 51,181 & 5,120 & 25,575 & 50000 \\
        Pursuit–evasion (5 turns)
        & 87,927 & 2,100 & 61,000 & 50000\\
        Liar's Dice (6 sides)
        & 294,883 & 24,576 & 147,420 & 100000\\
        Leduc poker (9 ranks)
        & 371,809 & 9,288 & 221,544 & 100000 \\
        Battleship ($3\times2$, 3 shots)
        & 732,607 & 81,027 & 552,132 & 10000 \\
        Leduc poker (13 ranks)
        & 1,179,777 & 19,656 & 704,600 & 100000 \\
        Liar's Dice (8 sides)
        & 8,388,553 & 524,288 & 4,194,240 & 5000 \\
        HUNL subgame 4
        & 244,005,483 & 43,240 & 158,388,120 & 100000 \\
        HUNL subgame 3
        & 398,112,843 & 69,184 & 261,126,360 & 100000 \\
        HUNL subgame 2
        & $\approx 3.5\times10^{10}$
        & $\approx 6.7\times10^{6}$
        & $\approx 2.2\times10^{10}$ & 10000\\
        \bottomrule
    \end{tabular}
    \caption{Benchmark games (the number of information sets counts those of both players).}
    \label{tab:game-sizes}
\end{table*}

The experiments are run on Apple M5 Pro with 18 CPU cores and 24\,GiB of memory; the run logs record arm64 macOS 26.6.2. Each learner uses one native thread, with numerical-library thread pools also limited to one. Each game--method pair contributes one run, without aggregation across repetitions. Learners run sequentially. Runtime convergence plots report cumulative learner wall time, including traversal, local updates, reach bookkeeping, both average accumulations, and checkpoint average-policy materialization. The runtime ratios in Figure~\ref{fig:heatmap} additionally include recorded setup and exploitability-evaluation time. Neither clock includes logging, final policy export, independent validation, or plotting; tree preparation and graph construction also precede the recorded setup timer. The timings are descriptive single-run measurements, so near-unit ratios should not be interpreted as reliable percent-level speed differences.

\subsection{Extra Plots for Benchmark Games}
\begin{figure}[!ht]
    \centering
    \includegraphics[width=0.9 \linewidth]{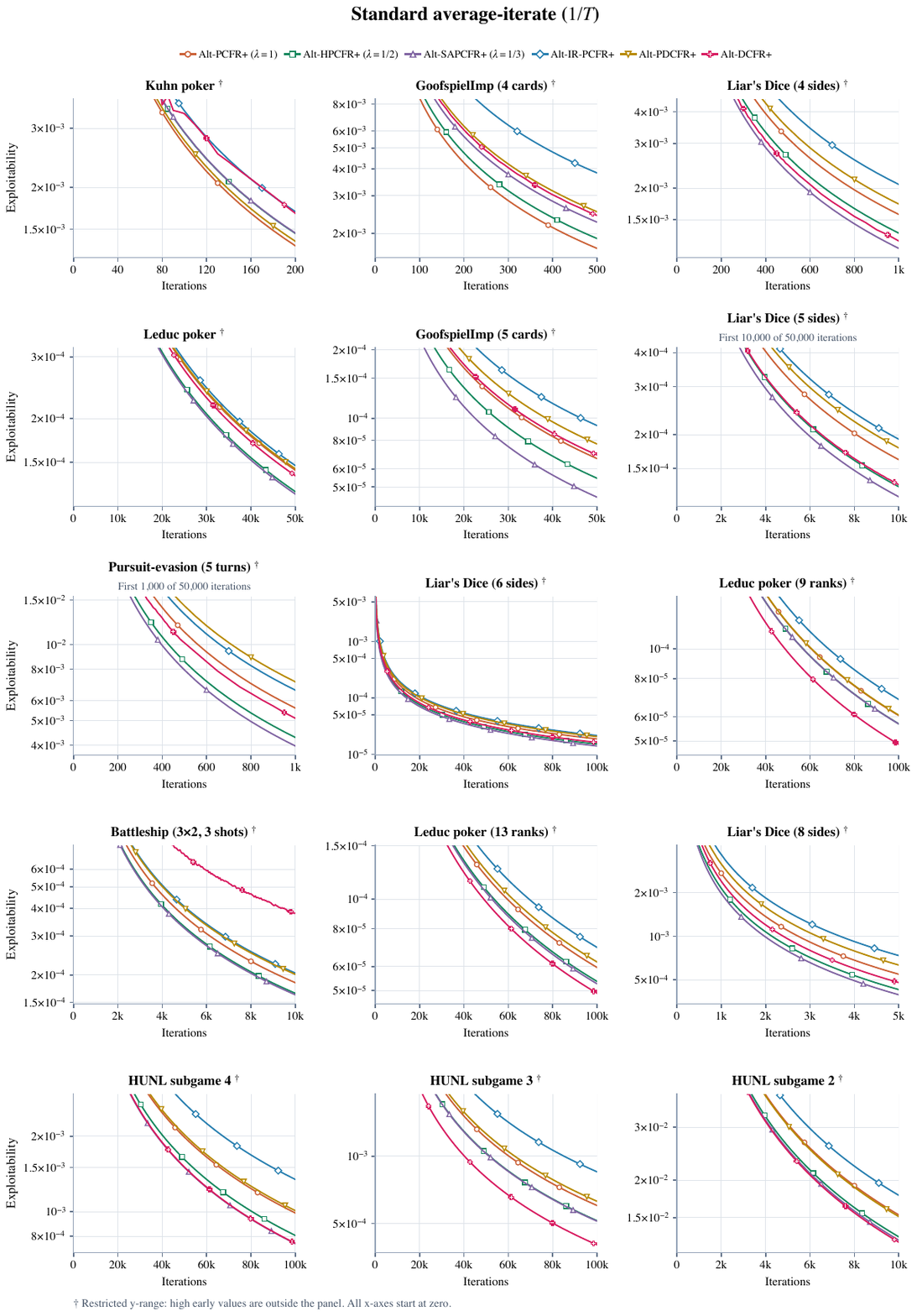}
    \caption{Average-iterate convergence against iteration count.}
    \label{fig:standard_iter}
\end{figure}

\begin{figure}[!ht]
    \centering
    \includegraphics[width=0.9 \linewidth]{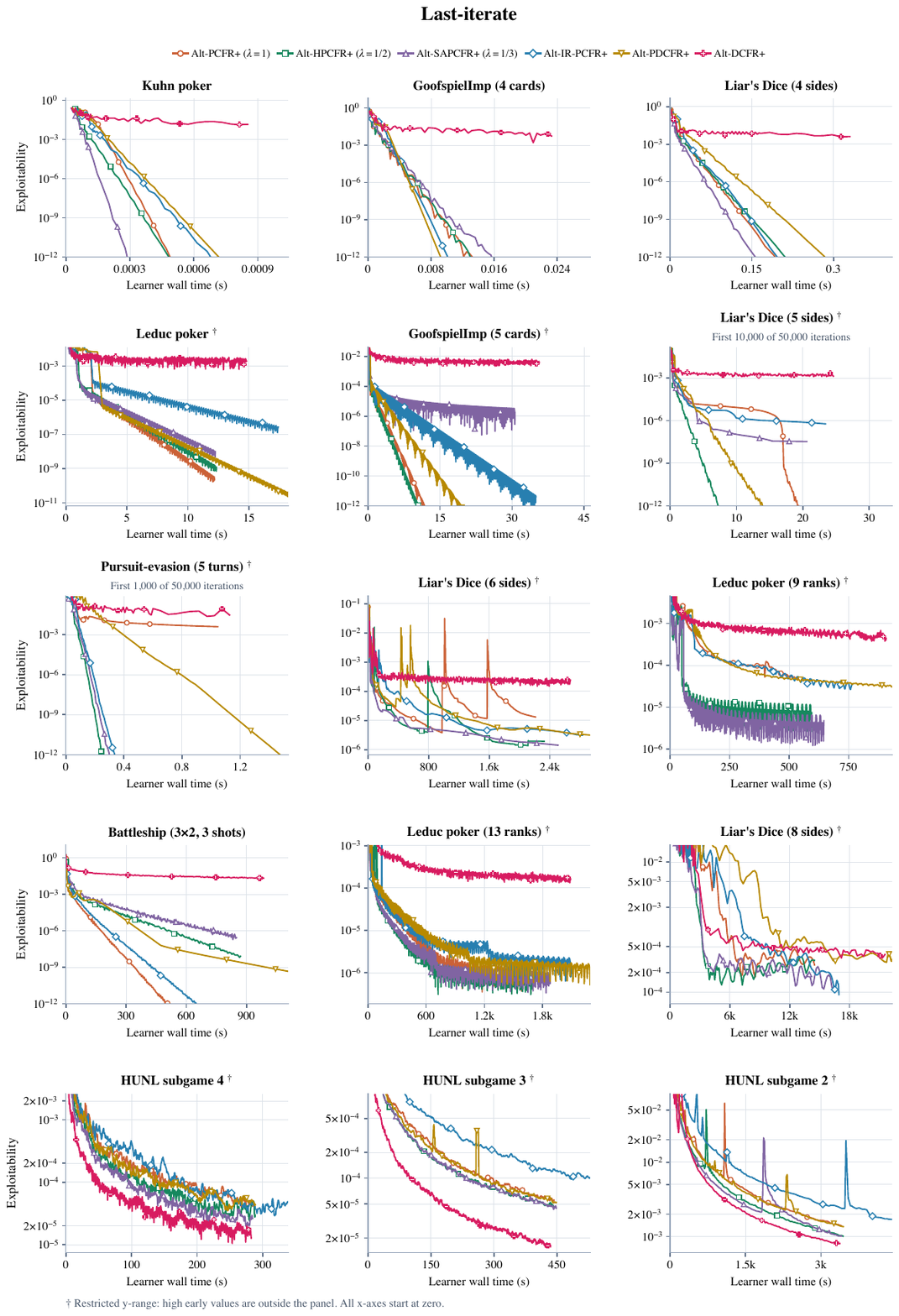}
    \caption{Last-iterate convergence against learner wall time.}
    \label{fig:last_runtime}
\end{figure}

\begin{figure}[!ht]
    \centering
    \includegraphics[width=0.9 \linewidth]{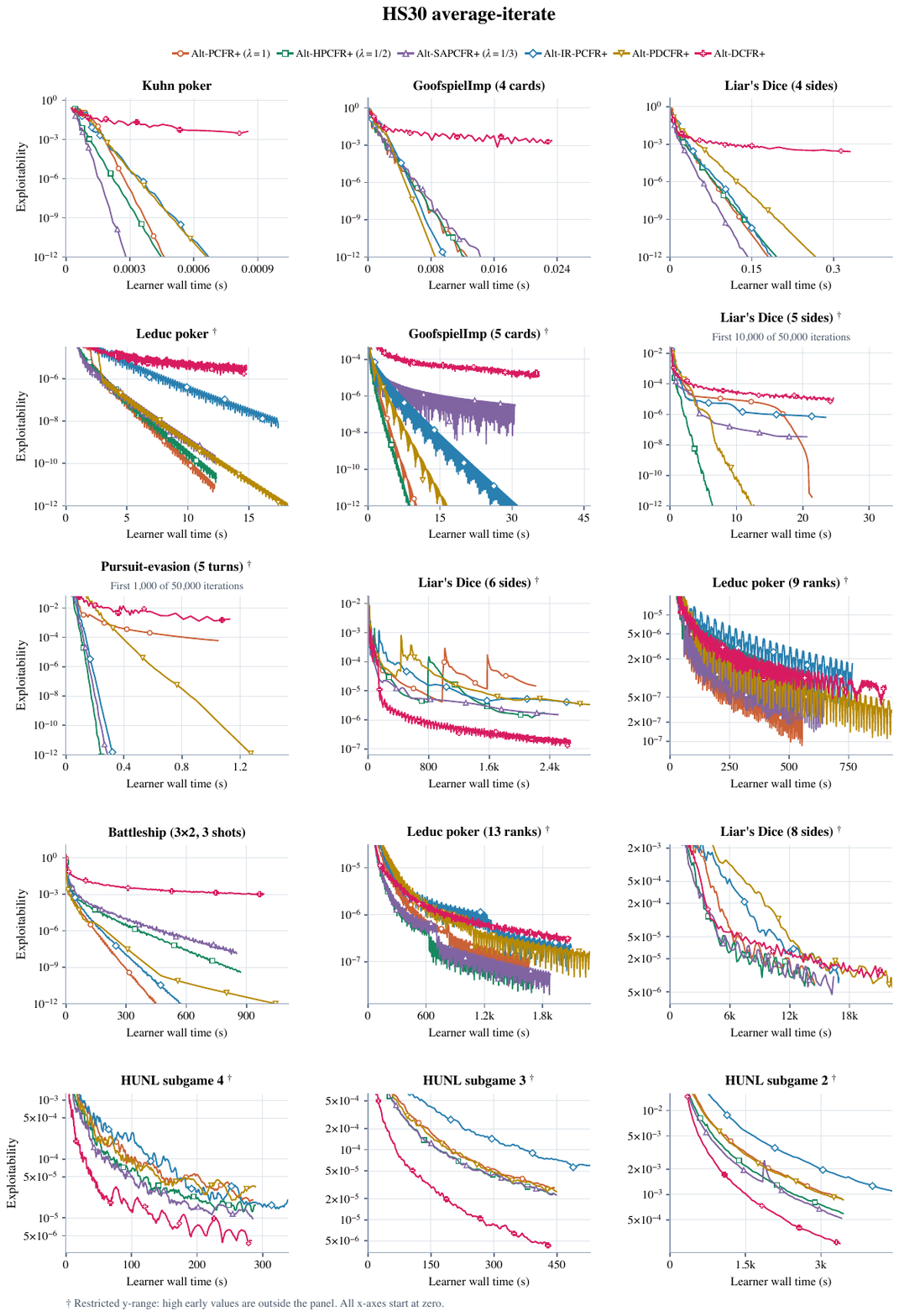}
    \caption{HS30 average convergence against learner wall time.}
    \label{fig:hs_runtime}
\end{figure}

\begin{figure}[!ht]
    \centering
    \includegraphics[width=0.9 \linewidth]{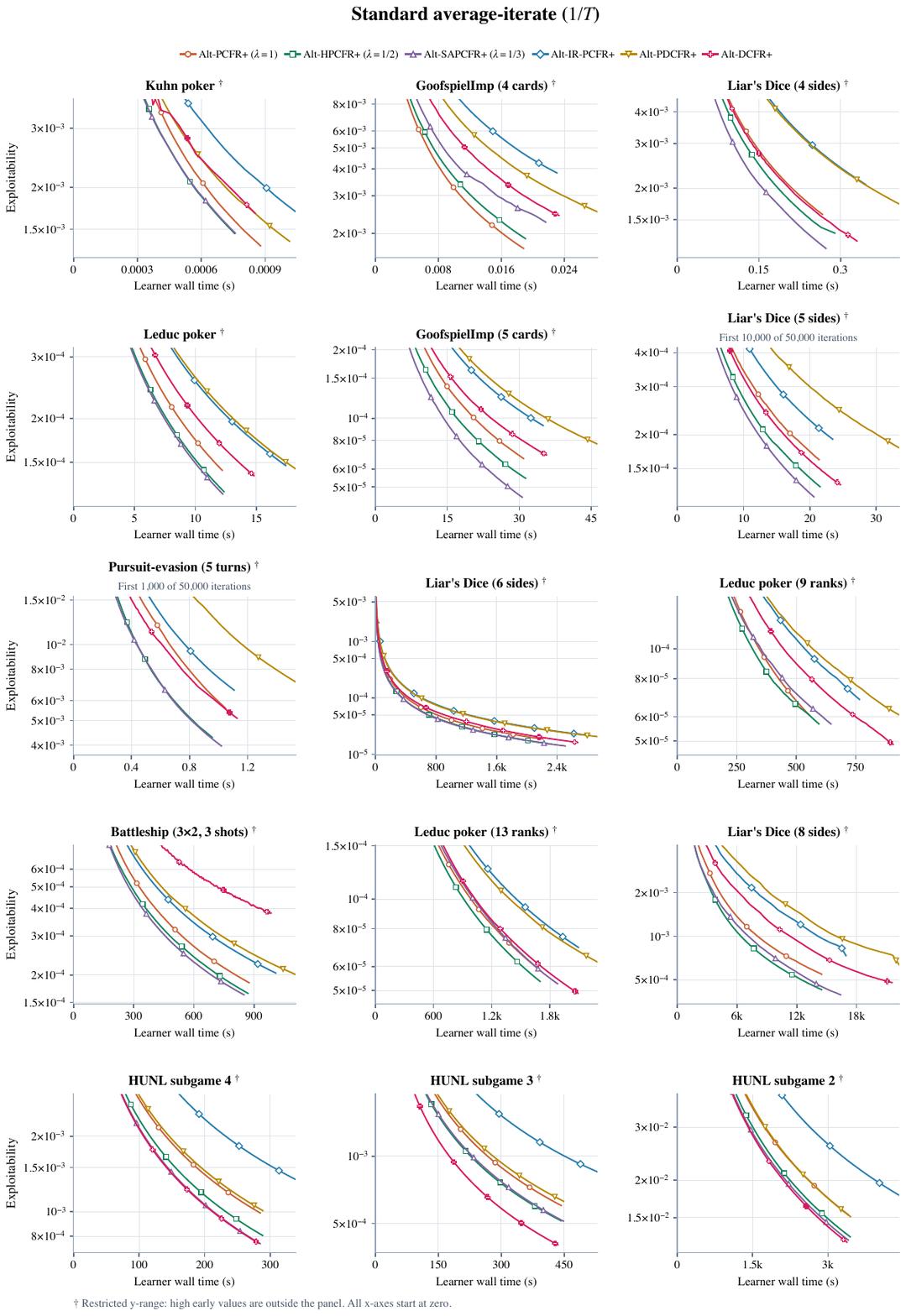}
    \caption{Average-iterate convergence against learner wall time.}
    \label{fig:unweighted_runtime}
\end{figure}

In the body of this paper, we showed the last-iterate convergence and HS30 average convergence plots. In Figure~\ref{fig:standard_iter}, we show the standard average-iterate convergence. We have also provided experimental plots showing exploitability as a function of iterations in Section \ref{sec:exp}. Here in Figures~\ref{fig:last_runtime}, \ref{fig:hs_runtime}, and~\ref{fig:unweighted_runtime}, we provide exploitability as a function of run time. 

\end{document}